\documentclass[reprint,
superscriptaddress,
amsmath,amssymb,
 aps,
prd,
floatfix]{revtex4-2}
\usepackage{xcolor}
\usepackage{comment}
\usepackage{graphicx}
\usepackage{dcolumn}
\usepackage{bm}
\usepackage{xcolor}
\usepackage{tabularx}
\usepackage{array}
\usepackage{titlesec}
\usepackage{hyperref}
\newcolumntype{Y}{>{\centering\arraybackslash}X}
\newcommand{\di}{\text{d}}
\newcommand{\pare}[1]{\left(#1\right)}
\newcommand{\parea}[1]{\left[#1\right]}
\newcommand{\pareb}[1]{\left\{#1\right\}}
\newcommand{\avg}[1]{\left<#1\right>}
\newcommand{\abs}[1]{\left|#1\right|}

\newcommand{\ket}[1]{\left|#1\right>}
\newcommand{\bra}[1]{\left<#1\right|}

\newcommand{\os}{\omega_0}
\renewcommand{\oc}{\omega_\text{LC}}

\newcommand{\bext}{B_\text{ext}}
\newcommand{\css}{\text{ECSS}}
\newcommand{\tz}[1]{\left.#1\right|_0}

\newcommand{\tsq}{\tau_\text{sq}}

\newcommand{\nocontentsline}[3]{}
\newcommand{\tocless}[2]{\bgroup\let\addcontentsline=\nocontentsline#1{#2}\egroup}

\begin{document}
\title{Superradiant Interactions for Relic Detection with Entangled
Nuclear Spins}
%\title{ Toward 48 dB Spin Squeezing and 96 dB Signal Magnification \\ for Cosmic Relic Searches with Nuclear Spins}

\author{Marios Galanis}
\email{mgalanis@pitp.edu}
\thanks{Author contributed equally to this work}
\affiliation{Perimeter Institute for Theoretical Physics, Waterloo, ON, N2L 2Y5, Canada}

\author{Onur Hosten}
\email{onur.hosten@ist.ac.at}
\thanks{Author contributed equally to this work}
\affiliation{Institute of Science and Technology Austria (ISTA), 3400 Klosterneuburg, Austria}

\author{Asimina Arvanitaki}
\email{aarvanitaki@pitp.ca}
\affiliation{Perimeter Institute for Theoretical Physics, Waterloo, ON, N2L 2Y5, Canada}

\author{Savas Dimopoulos}
\email{savas@stanford.edu}
\affiliation{Leinweber Institute for Theoretical Physics at Stanford,  \\382 Via Pueblo, Stanford, CA, 94305, USA}
\affiliation{Perimeter Institute for Theoretical Physics, Waterloo, ON, N2L 2Y5, Canada}

\date{\today}

\begin{abstract}

We recently  showed that macroscopic nuclear spin ensembles prepared in coherent spin states can dramatically enhance the interaction rates of weakly interacting cosmic relics—such as dark matter and the cosmic neutrino background —through collective quantum effects analogous to Dicke superradiance, where the de-excitation and excitation rates scale as the square of the number of spins, $N^2$. We thus coined these processes superradiant interactions.

In this paper, we propose a protocol to realize this enhancement and boost the discovery potential for such relics. We show how concepts from quantum optics can be adapted to nuclear spins coupled to superconducting circuits, enabling high-sensitivity systems. The spins are first initialized into a coherent spin state via a $\pi/2$ Rabi pulse from the ground state. When the circuit is sufficiently detuned from resonance, the spin-circuit interaction implements a squeezing Hamiltonian. Because squeezing must outpace spin relaxation and dephasing, the protocol favors macroscopic ensembles and high-quality superconducting circuits. During this squeezing phase, the standard quantum variance is reduced by up to 4.8 orders of magnitude—equivalent to 48~dB of squeezing—for circuits with quality factors $Q \sim 10^8$–$10^9$. The signal imprinted on the spins during the squeezing protocol can be magnified by further utilizing the squeezing interactions, easing the requirement for shot-noise-limited readout.

This protocol has the potential to significantly accelerate axion and dark photon dark matter searches and extend the reach of existing axion experiments to probe QCD axion–nuclear spin couplings. More broadly, it paves the way for detecting coherent inelastic interactions from other cosmic relics—most notably the cosmic neutrino background—and establishes nuclear-spin-based systems as a new class of quantum, ultra-low-threshold detectors.

\end{abstract}

\maketitle
\tableofcontents
\section{Introduction}
\label{sec:intro}

  One of the most exciting research directions in physics today is the search for cosmic relics, such as dark matter (DM) and the cosmic neutrino background (C$ \nu$B). These relics interact only very weakly with individual atoms or nuclei. To compensate for this, traditional experiments employ large detectors containing many particles, thereby enhancing the interaction rate linearly with the number of particles, $N$. In a recent paper, we showed that detectors composed of special quantum states of matter—such as product states (PS) or coherent spin states (CSS)—can exhibit interaction rates for inelastic processes that scale quadratically with particle number, as $N^2$ \cite{Arvanitaki:2024taq}. This coherent enhancement, which is analogous to Dicke superradiance, dramatically improves the discovery potential for cosmic relics. For instance, the C$\nu$B interacts with a 10~cm sphere of liquid- or solid-density nuclear spins prepared in a coherent spin state (CSS) at a rate of roughly one event per second, whereas its interaction rate with 10~cm of ordinary matter is less than one event over the entire age of the universe. Even more promising, interaction rates for dark matter candidates such as the QCD axion are similarly enhanced, reaching $\mathcal{O}(1~\text{Hz})$ for a detector just ten microns in size. In earlier work, we computed these interaction rates and identified possible observable signatures. Here, we go further by proposing a specific protocol capable of magnifying such signals by many orders of magnitude. 
  
Over the past decade, many-atom cavity QED systems have enabled the creation of highly entangled quantum states, consisting of half-a-million atoms squeezed by 20~dB \cite{Hosten2016,Cox2016}---for a review see \cite{Pezze18,Szigeti2021}. Building on this progress, we explore extending such techniques to nuclear spin systems coupled to electromagnetic circuits. Nuclear magnetic resonance (NMR) is already central to precision measurements and axion searches, see for example~\cite{Budker:2013hfa, Arvanitaki:2014dfa}. NMR setups are also a platform to probe superradiant interactions~\cite{Arvanitaki:2024taq}. These interactions offer the potential for ultra-sensitive, low-threshold detectors based on NMR.

As we will see, the coupling of nuclear spins to a circuit is described by the Dicke Hamiltonian, which reduces to the Tavis–Cummings Hamiltonian in the rotating wave approximation (RWA), as in atom-cavity QED. However, the smallness of the nuclear magnetic dipole moment leads to extremely weak single-spin coupling~\cite{rollano2022}, requiring large ensembles of spins to achieve meaningful squeezing. These macroscopic spin samples bring challenges such as spin dephasing and relaxation, not necessarily relevant in atomic systems. Additionally, radiation damping, the direct analogue of the Purcell effect in NMR, must be mitigated by operating in the far-detuned regime of a high-$Q$, cryogenic superconducting circuit.

We demonstrate that if an NMR setup satisfies the requirements of high-$Q$ and high sample densities, the induced Hamiltonian enables a one-axis twisting (OAT) interaction~\cite{Kitagawa93} that can generate up to 48~dB of spin squeezing---a value that could be achieved for a total number of spins ranging from $10^{7}$ to $10^{26}$ or greater. The envisioned interaction is similar to spin-exchange interactions realized in trapped neutral atomic systems with long-lived optical transitions~\cite{norcia2018} and complementary to cavity-feedback-based interactions realized in hyperfine atomic states~\cite{leroux2010,Hosten2016Science,li2023}, or collisionally realized couplings in Bose-Einstein condensates~\cite{strobel2014,berrada2013}.  The OAT interaction not only enhances the spin sample sensitivity but also supports efficient readout through a magnification protocol~\cite{Davis16,Hosten2016Science}, potentially overcoming the current limitations associated with technical challenges in achieving shot-noise-limited read-out sensitivity. Crucially, the squeezing and readout processes can be decoupled and optimized separately. We also consider the requirements on spin-circuit coupling homogeneity and time variation, the effects of finite polarization, and the effects of spin relaxation and dephasing, as well as imperfection of Rabi pulses. Due to these requirements, we find that hyperpolarized $^3 \text{He}$ is the best NMR candidate for achieving the  spin squeezing target. 

If successful, this approach could substantially advance dark matter searches and other cosmic relic searches, such as the cosmic neutrino background. First, it may allow probing QCD axion–nucleon spin couplings for axion DM{. The only other instance, to our knowledge, of quantum-enhanced DM searches have been achieved by the HAYSTAC collaboration~\cite{haystacsqueeze} for the axion-photon coupling. Second, an experiment based on our protocol would} have comparable sensitivity to the cosmic neutrino background with existing experiments like KATRIN~\cite{KATRIN:2022kkv}, while being much smaller in size. { We believe the qualitative properties of this approach are captured by Superradiant Interactions for Relic Detection with Entangled Nuclear Spins (SIREN).}

Our paper starts (sec.~\ref{sec:setup}) by describing the idealized experimental setup of the $^3\text{He}$ sample coupled to an LC circuit, deriving the OAT hamiltonian including corrections, when operating away from the RWA. In sec.~\ref{sec:magsqueeze}, we describe how in-circuit decays and spin relaxation limit the amount of squeezing to about 48~dB, and outline the Rabi pulse sequence needed for the subsequent magnification protocol. We also discuss a two-axis counter-twisting (TACT) \cite{Kitagawa93} protocol that could achieve up to an \emph{additional} $\sim12$~dB of squeezing at a shorter timescale. In sec.~\ref{sec:axion_and_dp}, we show how our protocol applied to a macroscopic $^3\text{He}$ sample consisting of $10^{26}$ spins has enhanced sensitivity to axion as well as dark photon dark matter. This enhancement can be enough to probe the GUT scale QCD-axion dark matter coupling to nuclear spins. In this section, we also comment on improving KATRIN's C$\nu$B sensitivity. In sec.~\ref{sec:tolerances} we calculate the systematic tolerances for the desired levels of squeezing to persist, and we finally conclude in sec.~\ref{sec:discussion}. Detailed computations of the effects presented in the main paper are included in Apps.~\ref{app:unitary}---\ref{app:tech_noise}.

%%%%%%%%%%%%%%%%%%%%%%
\section{System description}
\label{sec:setup} 

We consider $N$ identical nuclear spins that are coupled to a superconducting circuit with resonant frequency $\oc$ by enclosing the spin sample within the circuit's solenoid, as shown in Fig.~\ref{fig:setup}. For concreteness, we will consider gaseous or liquid $^3\text{He}$ as a benchmark system, but our results can be generalized to any NMR system, from gaseous to solid state form.  A magnetic field $\bext$ induces a splitting between the spin-up and spin-down states of magnitude $\omega_0=\gamma \bext$, where $\gamma=2 \times\mu $ is the $^3$He gyromagnetic ratio, with $\mu\approx-2.127\frac{e}{2m_p}$ for $^3$He, $e$ is the electron charge and $m_p$ the proton mass. For a $\bext$ up to a few Tesla, $\omega_0$ can be up to $10^{-6}$~eV or about $200$~MHz.

While it is easier to discuss the interactions in the rotating wave approximation (RWA) using the Tavis-Cummings Hamiltonian, we find that for large detunings, where the RWA no longer holds, the performance can continue to improve in terms of circuit decay and adverse effects of thermal excitations. Consequently we adopt the full Dicke model in the long wavelength limit as the appropriate model to describe the system (see Appendix~\ref{app:nonRWA}):

\begin{figure}[t!]
\centering
\includegraphics[width=.55\linewidth]{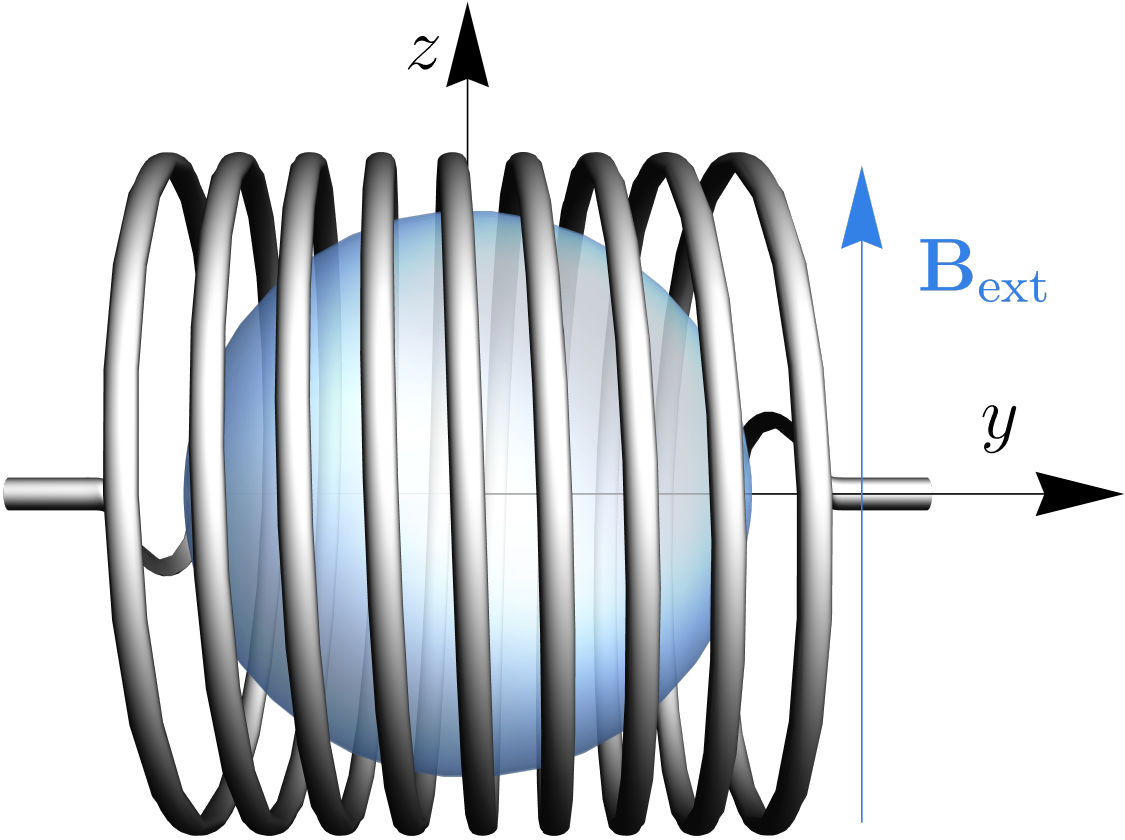}\hfill
\raisebox{-0.045\height}{\includegraphics[width=.43\linewidth]{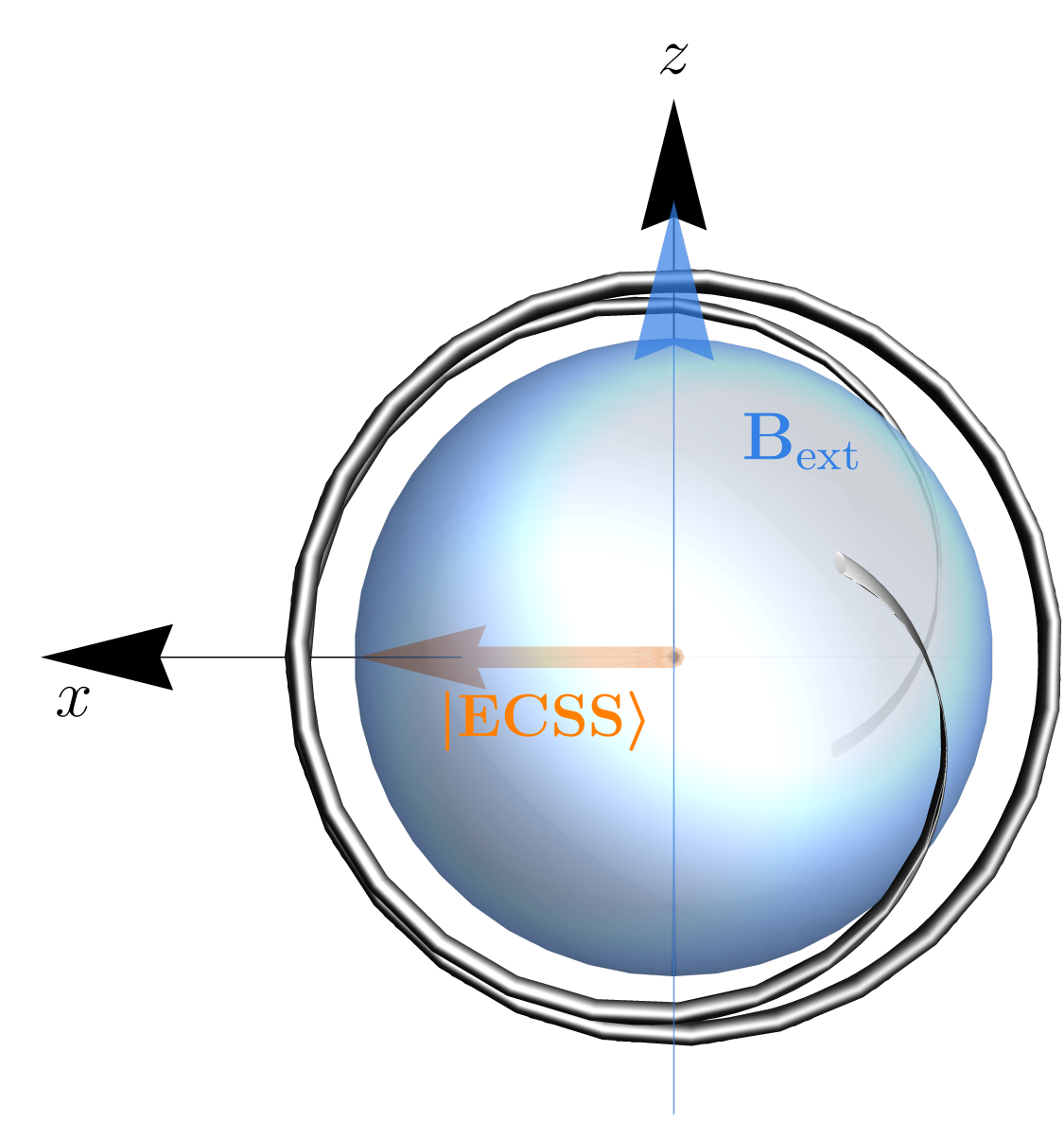}}
\caption{Abstract illustration of the squeezing setup. An NMR sample is inside a solenoid of a circuit. The whole setup is placed in a background magnetic field $\bext$ that determines the Larmor precession frequency. Rabi pulses are applied to the spins through the circuit. These pulses are off resonance to the circuit but on resonance with spins, and the squeezing protocol is realized through the spin-circuit coupling.}
\label{fig:setup}
\end{figure}

\begin{equation}
    H_\text{Dicke}/\hbar=\os J_z + \oc a^\dagger a + g\pare{a - a^\dagger}\pare{J_- - J_+},
    \label{eq:htc}
\end{equation}
where $a$ is the circuit mode annihilation operator, $J_i=\frac{1}{2}\sum_{i=1}^N \sigma_i$, $i\in\{x,y,z\}$ are the collective spin operators, $\sigma_i$ the Pauli matrices, $g=\mu\sqrt{\beta_N~\oc/(2V_\text{L})}$ is the coupling of the cavity to individual spins, and $V_\text{L}$ is the physical volume of the coil in the circuit. $\beta_N$ is an order one number that parametrizes the Nagaoka expression~\cite{1390853649533379968}. The collective raising and lowering operators are defined as $J_\pm\equiv J_x\pm i J_y$.

A squeezing interaction appears when the circuit is operated off-resonance from the spins, quantified by an amount set by the detuning $\Delta_-\equiv \os-\oc$. The circuit has a linewidth $\kappa=\oc/Q$, where $Q$ is its quality factor. The entire system is enclosed in a larger superconducting shield that controls systematics and suppresses the spontaneous decay of the spins to free space with a larger volume and a quality factor $Q_\text{shield}\gg Q$.

The spins are initialized in their ground state and an adiabatic $\pi/2$ Rabi pulse rotates them to the Equatorial Coherent Spin State (ECSS)
\begin{equation}
    \ket{\text{ECSS}}\equiv\prod_{i=1}^N\frac{1}{\sqrt{2}}\parea{\ket{g}_i+\ket{e}_i},
    \label{eq:cssvac}
\end{equation}
where $\ket{g}$ and $\ket{e}$ are the ground and excited states of each spin. 

In the limit $gN/|\Delta_-|\gg 1$ and ignoring decays, a mean circuit field builds up, with equilibrium mean occupation number $\approx (gN/\Delta_-)^2$ (see App.~\ref{app:nonRWA} for the exact expression). Expanding the circuit operators around this coherent state, perturbative diagonalization to lowest order in $g$ yields the effective Hamiltonian:
\begin{equation}
    H_\text{eff}/\hbar\approx\chi\pare{-J_z^2+2a^\dagger a J_z+J_z},
    \label{eq:hsqueeze}
\end{equation}
where $\chi\equiv \frac{2 g^2 \omega_\text{LC}}{\Delta_- \Delta_+}$, where $\Delta_+= \omega_0+\omega_\text{LC}$, and the circuit state is the vacuum $\ket{0}$ in this picture. Importantly, the above value of $\chi$ does not assume the RWA. This perturbative expansion is under control as long as $g\sqrt{N}\ll \Delta_-$. Since $g\propto \sqrt{V_\text{L}^{-1}}$, the perturbativity criterion sets a limit only on the effective density of the spins, $n_\text{eff}\equiv N/V_\text{L}$, and not directly on their total number. As we will see, the total number of spins is indirectly constrained by considerations beyond the squeezing protocol. A detailed proof of these statements is included in App.~\ref{app:unitary}.

The first term in Eq.~\ref{eq:hsqueeze} is dominant and is known in the literature as the one-axis twisting (OAT) Hamiltonian~\cite{Kitagawa93}. Since squeezing proceeds solely from the quantum fluctuations of the (displaced) circuit vacuum, this protocol is equivalent to vacuum spin squeezing~\cite{hu2017}. In addition to squeezing, the OAT Hamiltonian can also lead to magnification. We describe the full squeezing and magnification protocol next.

%%%%%%%%%%%%%%%%%%%%%%
\section{Squeeze and magnify protocol}
\label{sec:magsqueeze}

The OAT Hamiltonian of Eq.~\ref{eq:hsqueeze} reduces spin shot noise along the squeezing axis by orders of magnitude below the standard quantum limit (SQL), as we compute below. A protocol only implementing squeezing would require readout at this {\emph{reduced}} noise level. But for NMR systems readout even at the SQL is challenging. Thus, while not essential for increasing the sensitivity of the spin system achievable with the OAT Hamiltonian, a magnification protocol~\cite{Hosten2016Science}---see also \cite{Davis16,colombo2022,li2023}---is paramount for any realistic experimental implementation.

As we will see, squeezing and magnification are intertwined, as squeezing implies magnification to the same degree and vice versa.

\begin{figure*}[t!]
    \includegraphics[width=0.98\textwidth]{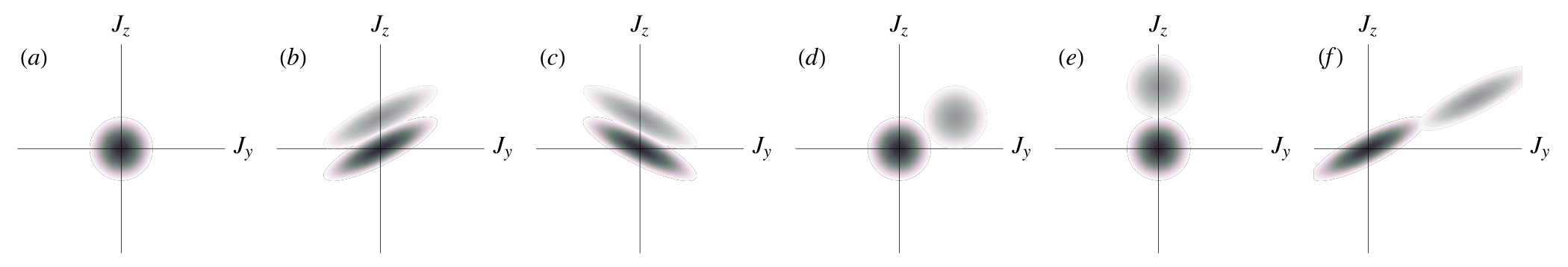}
    \caption{Qualitative depiction of the squeezing ({\it{(a)}} and {\it{(b)}} panels)and magnification ({\it{(c)}} to {\it{(f)}} panels) protocols when detecting a signal that changes $J_z$. The evolution of the state without a signal is shown in gray, and in light gray when a signal is present. {\it{(a)}} The phase space of spins prepared in a coherent state along the $J_x$ axis (out of the plane). {\it{(b)}} The state after it evolves for $\tsq$. The signal we would like to detect will appear as a shift of the state along the z direction by an amount $\delta J_z$. This is the measuring phase, when the effects of the signal are imprinted on the state. {\it{(c)}} A small rotation of size $\sim 2 \xi^{-1}$ is applied. This causes the state to evolve back to the original state that is shown in {\it{(d)}}. If there is a signal, then the new state is displaced in the $J_y$ direction by an amount $\xi\times \delta J_z$. {\it{(e) and (f)}} represent an additional magnification stage of the protocol. {\it{(e)}} The state after a $\pi/2-\xi^{-1}$ rotation. After $\tsq$, this causes the state to be squeezed in such a way that the signal is further displaced in the $J_y$ direction by a factor of $\xi$, shown in {\it{(f)}}. The noise is also enhanced along $J_y$, so that the fundamental SNR remains the same.}
    \label{fig:protocolnet}
\end{figure*}

\begin{figure*}[t!]
    \includegraphics[width=0.98\textwidth]{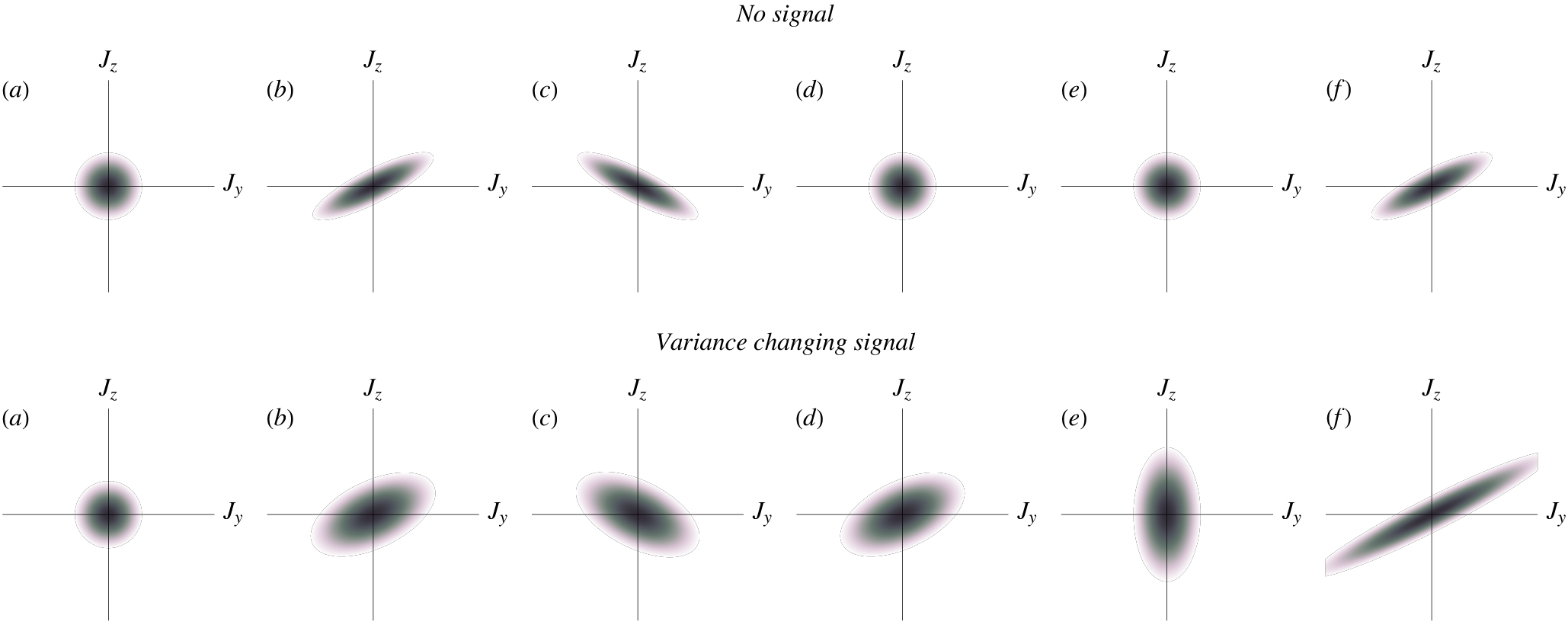}
    \caption{
     Qualitative depiction of the squeezing ({\it{(a)}} and {\it{(b)}} panels)and magnification ({\it{(c)}} to {\it{(f)}} panels) protocols when detecting a signal that causes a change in the variance $\text{var}(J_z')$ (see text for details). The top(bottom) {\it{(a)}} to {\it{(f)}} panels depict the evolution of the state when a signal is absent(present).  {\it{(a)}} The phase space of spins prepared in a coherent state along the $J_x$ axis (out of the plane). {\it{(b)}} The state after it evolves for $\tsq$. The signal we would like to detect will appear as an increase of the state uncertainty along the z direction. This is the measuring phase, when the effects of the signal are imprinted on the state. {\it{(c)}} A small rotation of size $\sim 2 \xi^{-1}$ is applied. This causes the state to evolve back to the original state that is shown in {\it{(d)}}. If there is a signal, then the new state has an enhanced uncertainty in the $J_y$ direction. {\it{(e) and (f)}} represent an additional magnification stage of the protocol. {\it{(e)}} The state after a $\pi/2-\xi^{-1}$ rotation. After $\tsq$, this causes the state to be squeezed in such a way that the signal caused uncertainty is further enhanced in the $J_y$ direction by a factor of $\xi$, shown in {\it{(f)}}. Note that the noise is also enhanced, so that the fundamental SNR remains the same.}
    \label{fig:protocoltot}
\end{figure*}

%%%%%%%%%%%%%%%%%%%%%%
\subsection{Vacuum spin squeezing}
\label{sec:vacsq}

The OAT Hamiltonian of Eq.~\eqref{eq:hsqueeze} can be thought of as a $J_z$-dependent energy shift $\chi J_z$. In the rotating frame, this causes a shearing of the phase space distribution as shown in panel {\it{(b)}} of Figs.~\ref{fig:protocolnet} and~\ref{fig:protocoltot}. This shearing results in reduced uncertainty, i.e. squeezing, along an axis $z'$ at a small angle $\theta\approx (N\chi t)^{-1}$ relative to the $z$-axis. 

The squeezing action of the OAT Hamiltonian competes with effects that increase the uncertainty. There are four main such sources:
\begin{itemize}

\item{ decay of spins through the circuit, parametrized by the single spin decay rate $\eta$,}
\item{ spin relaxation, parametrized by the timescale $T_1$,}
\item{ spin dephasing, parametrized by the timescale $T_\text{deph}$, and}
\item curvature of the Bloch sphere
\end{itemize}

The in-circuit spin decay is a collective effect, and its parametric dependence is determined by the physical mechanism that dominates circuit decay when the spin–circuit system operates far off resonance, requiring the full Dicke model, Eq.~\eqref{eq:htc}. If the linewidth of the circuit $\kappa$ is dominated by capacitive losses, the single-spin decay rate in the limit $|\Delta_-|\gg\kappa$ is $\eta = 4g^2\os^2\kappa/(\os^2-\oc^2)^2$. If, instead, inductive losses dominate,  $\eta = 4g^2\oc^2\kappa/(\os^2-\oc^2)^2$. This distinction is irrelevant within the RWA but leads to qualitatively different results in the far-off-resonant regime. The full computation is given in App.~\ref{app:spin_decays}. To our knowledge, the dependence of the spin decay rate on the type of circuit losses is a new effect. Since we operate in the limit $\oc\gg\os$, it may be worthwhile exploring whether circuits with $Q\gtrsim 10^8$ can be engineered, such that the linewidth is dominated by capacitive losses, with much more suppressed inductive losses. This will depend on a concrete experimental setup and is beyond the scope of this paper{, but will be considered in forthcoming work~\cite{upcomingRapidis}}. Instead, in Sec.~\ref{sec:sm_bsm} we present results for the two limiting cases in separate plots.

This rate describes the decay of $\avg{J_z}$. The variance of $J_z$ increases with a single-spin rate of $\eta(2\bar{n}+1)$, where $\bar{n}\equiv (e^{\os/T}-1)^{-1}$ the  thermal population of circuit quanta at the frequency of the spins. Close to resonance, the denominator of $\eta$  would instead be $\propto \kappa^2$, known at the Purcell enhancement, and the associated radiation damping effects would significantly back-react on the circuit and degrade the quality factor of the system. Nevertheless, when $|\Delta_-|\geq \kappa$ and the circuit quality factor Q is high, $\eta\propto \kappa$ and so it can be significantly suppressed by Q. If inductive losses dominate the circuit linewidth, $\eta$ can be further suppressed by the ratio $\omega_0^2/\oc^2$.

The spin relaxation time $T_1$ is a local phenomenon that parametrizes the thermalization of the nuclear spin system, due to spin-spin and spin-lattice interactions, among other effects. Its formal definition in terms of Lindblad operators, as well as its effects are described  in detail in App.~\ref{app:decoherence}. For liquid or gaseous $^3\text{He}$, this timescale  can be long \cite{PhysRevA.48.4411}: $T_1=1000~\text{s}~\frac{3\times 10^{22}~\text{cm}^3}{n_\text{S}}$, where $n_\text{S}$ is the nuclear spin sample density. As we will see, these uncertainty enhancing effects can be dominant if the in-circuit decays are sufficiently suppressed. 

The spin dephasing time $T_\text{deph}$ parametrizes any effect on the spins that affect their phase but not their energy. Similar to $T_1$ it parametrizes local phenomena and it is a major contributor to the better known $T_2$ timescale that parametrizes the decay of the transverse spin magnetization, which we define as $T_2\equiv\pare{(2T_1)^{-1}+T_\text{deph}^{-1}}^{-1}$. For systems where $T_1\gg T_2$, such as solid-state NMR, $T_2\approx T_\text{deph}$. The formal definitions of $T_2$ and $T_\text{deph}$ in terms of Lindblad operators and magnetization decays are described in detail in App.~\ref{app:decoherence}. For liquid or gaseous $^3\text{He}$, motional narrowing allows for $T_\text{deph}\sim T_1$ and this is what we assume. For reference, in solid NMR samples, $T_\text{deph}\approx 1~\text{ms}~\frac{3\times 10^{22}~\text{cm}^3}{n_\text{S}}$, assuming no external field gradients. The dephasing time can also be affected by magnetic field gradients, but as we will see, the OAT protocol is not as sensitive to its effects as one might naively expect.

In App.~\ref{app:decoherence}, we show that the above effects can be included in the evolution of the spin system under Eq.~\eqref{eq:hsqueeze}. The spin variance along the optimal axis $z'$, i.e. the spin projection direction in which the largest squeezing is observed, $\text{var}(J_z')\equiv \avg{J_z^{'2}}-\avg{J_z'}^2$, evolves as

\begin{eqnarray}\label{eq:Jzvar}
    \text{var}(J_z')\approx \frac{N}{4}\left[\frac{1+\frac{2t}{T_\text{deph}}}{(N \chi t)^2}+ N \eta(2 \bar{n}+1)t + \frac{2t}{3T_1}\right].
    \label{eq:jzprime}
\end{eqnarray}

The maximum achievable reduction with respect to the SQL can be calculated when the above quantity is minimized with respect to time, and it roughly occurs when the  squeezing term that scales as $t^{-2}$ becomes comparable to the terms proportional to $\eta$ or $T_1^{-1}$. It is customary to define this maximum achievable reduction with respect to the SQL as  $\text{var}(J_z')_\text{min}=\xi^{-2} N/4$, where $\xi^2\geq1$ is the squeezing parameter. This reduction can be achieved in time:
\begin{eqnarray}
\label{eq:tausq}
    \tau_\text{sq}& \approx \frac{2}{N}\pare{\frac{1}{[\eta(2\bar{n}+1)+2/(3NT_1)]\chi^2}}^{1/3}, 
\end{eqnarray}
and, as long as $\tau_\text{sq}\lesssim T_\text{deph}$, we find that $\xi^2$ is given by

\begin{equation}
    \begin{split}
         \xi^2&\approx \frac{2^{2/3}}{3}\parea{\frac{\chi}{\eta(2\bar{n}+1)+2/(3NT_1)}}^{2/3}.
    \end{split}
    \label{eq:xi2}
\end{equation}

If the circuit decay is inductive, then decoherence is dominated by $\eta(2\bar{n}+1)$ for all spin frequencies, and 
$\xi^2=\frac{2^{2/3}}{3}\parea{\frac{ |\Delta_-| \Delta+}{2 \oc \kappa (2\bar{n}+1)}}^{2/3}$. Operating far-off-resonance with $|\Delta_-|\gtrsim\oc$, this simplifies to $\xi^2\approx \frac{2^{2/3}}{3}\parea{\frac{Q}{2\bar{n}+1}}^{2/3}$, which is maximized for high $Q$ and low thermal quanta number $\bar{n}$. The maximum squeezing achievable at a 10~mK cryogenic environment, with $\os=10^{-6}$~eV and $\oc=1.3\times 10^{-6}$ eV is $\xi^2\approx 70000\pare{\frac{10^9}{Q}}^{2/3}$, or 48~dB (we remind the reader that the amount of squeezing in dB is $10\log_{10}\xi^2$).

If, instead, the circuit decay is capacitive, then spin decays dominate only for $Q\lesssim 10^7\pare{\frac{\os}{10^{-7}~\text{eV}}}$. In this case $\xi^2=\frac{2^{2/3}}{3}\parea{\frac{ |\Delta_-| \Delta+}{2 \oc \kappa (2\bar{n}+1)}\frac{\oc^2}{\os^2}}^{2/3}$. Thus for high-Q circuits, detunings comparable to $\oc$, and liquid samples of $^3\text{He}$, squeezing limitation is dominated by $T_1$, giving
$\xi^2\approx3^{-1/3}\left(\chi N T_1\right)^{2/3}$. As $T_1$ is a function of the density $n_\text{S}$ of the spin sample, so that $T_1= C n_\text{S}^{-1}$, the above limiting relation reduces to $\xi^2=3^{-1/3}\left(\frac{2 \mu^2 \omega_\text{LC}^2 n_\text{eff}}{\Delta_- \Delta_+ n_\text{S}~C}\right)^{2/3}\approx 3^{-1/3}\left(\frac{\mu^2n_\text{eff} }{n_\text{S} C}\right)^{2/3}$, where $n_\text{eff}\equiv N/V$ is the effective density of the spin sample as compared to the mode volume, and $C=(\mu^4 \sqrt{m^3 T})^{-1}$ below $1$~K~\cite{PhysRevA.48.4411}. This also shows that the maximal squeezing factor at large deturnings is independent of $\omega_0$ or $\oc$. In particular, for $^3\text{He}$ at 10~mK this gives $\xi^2\approx 76800\left(\frac{n_\text{eff}}{n_\text{S}}\right)^{2/3}$ or 48.8~dB. This amount of squeezing can be achieved within time $\tau_\text{sq} \approx0.6~\sec \frac{ 10^{21} \text{cm}^{-3}}{(n_\text{eff}^2 n_\text{S})^{1/3}}$. {The state of the art for resonator quality factors for lumped-element resonators at mircrowave frequencies is $Q\approx 2.1\times 10^6$~\cite{highq}, corresponding to 28~dB of achievable squeezing with our OAT protocol. Despite the fact that superconducting radio frequency (SRF) cavities routinely reach quality factors on the order of $\sim 10^9$, microwave lumped-element resonators remain relatively underdeveloped~\cite{highq}, and so similarly high quality factors may be achievable~\cite{kentprivate}}.

Two crucial observations stem from Eqs.~\eqref{eq:jzprime},~\eqref{eq:tausq}, and ~\eqref{eq:xi2}. First, as long as $\tau_\text{sq}\lesssim T_2$, squeezing is essentially unaffected by dephasing. Qualitatively, this is due to the fact that the squeezed observable $J_z'$ is almost aligned with $J_z$, and is thus not directly affected by energy-conserving processes. This effect is computed rigorously in App.~\ref{app:T1T2}. Second, the dynamics of squeezing are set by $n_\text{eff}$ and $n_\text{S}$, but not the total number of spins. As we will see in sec.~\ref{sec:axion_and_dp}, the signal from cosmic relics scales with $N^2$, so we can improve the signal to noise ratio by increasing $N$ while keeping the relevant densities fixed. Thus, a macroscopic number of spins can be used, increasing the SNR, as long as the effective volume of the LC mode and sample are appropriately scaled. 

Constraints on the number of spins come indirectly: the equilibrium population of modes $(gN/\Delta)^2$ may induce a current too large for the circuit to support, its decoherence can back-react on the spins (see App.~\ref{app:morenoise} and Sec.~\ref{sec:tolerances}), or the decay rate of spins in the circuit, $N^2\eta/4$, may heat up the circuit beyond the cooling power of the dilution fridge.  Given current technologies and assuming large detuning, $N$ as high as $10^{26}$ should be feasible, as long as the effective volume is also large, $(\mathcal{O}( \text{10 cm}^3))$. 

The curvature of the Bloch sphere sets the fundamental limit for the amount of squeezing achievable with the OAT protocol. For $N$ spins, the minimum variance $\text{var}(J_z')=\frac{3^{2/3}}{8} N^{1/3}$~\cite{Kitagawa93} (see also App.~\ref{app:QL}). Because this depends directly on $N$, curvature effects set a lower limit on the number of spins to achieve 48~dB of squeezing, $N\gtrsim  10^{7}$. In this case, $V_\text{L}$ as well as the physical sample volume need to be scaled accordingly so that squeezing occurs within reasonable time scales.

Perhaps the most severe current limitation comes from RF pulse uncertainties, which are relevant only for the magnification part of our protocol (see Sec.~\ref{sec:mag}). As we discuss in more detail in sec.~\ref{sec:tolerances}, RF pulse phase stability needs to be at the level of $1/\sqrt{N}$ for manipulating the state without introducing spurious signals. For $N=10^{26}$, this may be problematic in some range of frequencies, but this can be improved as atomic clock stability improves. Section~\ref{sec:tolerances} discusses several sources of technical noise, which could limit the level of squeezing, but any further considerations should be addressed in a concrete experimental realization of the protocol, and lie beyond the scope of this paper.

%%%%%%%%%%%%%%%%%%%%%%%%%%
\subsubsection*{Two-axis counter-twisting}
\label{sec:cmag}
In principle, it is simple to add a continuous spin rotation around the $x$ axis with the existing experimental setup. 
The unitary Hamiltonian $H=\frac{N\chi}{2}J_x + \chi J_z^2$ rotates the squeezed state of panel $(b)$ in figs.~\ref{fig:protocolnet}, and~\ref{fig:protocoltot} towards the $z$ axis. As the state is rotated continuously, $\langle J_z^2 \rangle$ becomes larger than $N$ and the squeezing is accelerated to the point that is becomes exponentially fast, $\xi^2 \approx e^{N \chi t}$ \cite{muessel2015,li2023}. The maximum squeezing is achieved along the diagonal of the $z-y$ plane at $\abs{\theta_\text{c}}=\pi/4$, and is has been shown to be equivalent to a two-axis counter-twisting (TACT) squeezing protocol~\cite{hu2017}. { The unitary evolution under the TACT Hamiltonian does not have a simple analytic form similar to the OAT evolution~\cite{hu2017}, and so a rigorous understanding of the full dynamics of the system under unitary evolution, relaxation, photon decay, dephasing and the cosmic relic signal, requires the development of different computational tools. Furthermore, the addition of the large coherent rotation $N\chi J_x/2$ may present further experimental challenges (see Sec.~\ref{sec:tolerances}) in terms of phase and amplitude stability of the drive. Finally, the standard treatment of dephasing as a Markovian dissipation term may be inadequate for a system in the ECSS~\cite{niu2025} For all these reasons, we leave a detailed understanding of the TACT protocol to future work. To motivate the importance of this study, however, in the rest of this subsection and in App.~\ref{app:tact}, we provide a heuristic derivation of the squeezing time $\tau_\text{sq}$ and squeezing factor $\xi^2$ that the TACT protocol may be able to achieve.}

As squeezing occurs along the diagonal, decoherence effects on $J_z$ and $J_{x,y}$ should be equally important. Thus, we expect that dephasing should now become relevant and a term $\sim (N/4) \times (2t/T_\text{deph})$ must be added to Eq.~\eqref{eq:jzprime}. For very long $T_\text{deph}$, TACT has a parametric enhancement $(\chi/\bar{\eta})^{1/3}/\log(\chi/\bar{\eta})$ compared to OAT, where $\bar{\eta}\equiv\eta(2\bar{n}+1)$, while the squeezing time is also significantly reduced, $\tau_\text{sq}\approx \log(\chi/\bar{\eta})/(N\chi)$.  For a $^3\text{He}$ system with $n_\text{eff}\approx n_\text{S}=10^{20}~\text{cm}^{-3}$, $\os=10^{-6}$~eV, $\oc=1.3\times 10^{-6}$~eV, $Q=10^9$, and $T_\text{deph}\approx 1000~\text{s}\frac{3\times 10^{22}~\text{cm}^{-3}}{n_\text{S}}$ cooled at 10~mK, TACT can achieve a $\sim 12.6$~dB boost to the maximum amount of squeezing, bringing it to a total of 60~dB. In addition, $\tau_\text{sq}$ is reduced by a factor of $\sim45$ compared to the OAT protocol. Since maximum squeezing is achieved for $\os\approx\oc$ close to the upper limit $10^{-6}$~eV, whether the circuit decay is capacitive or inductive is not important. 

For systems where $T_\text{deph}$ is reduced compared to the expectation from spin-spin interactions (see Table~\ref{tab:parameters}) due to technical noise, whether TACT is preferable to OAT depends on the frequency, the quality factor of the circuit and its dominant decay mode (capacitive or inductive). Nevertheless, it has recently been shown that certain dephasing effects contributing to $T_\text{deph}$ can be alleviated by the large gap $g \sqrt{N}$ needed to change the total dipole $J^2$~\cite{norcia2018,niu2025}. As a detailed understanding of this effect is crucial but beyond the scope of this work, we leave all details of a TACT protocol to future work.

%%%%%%%%%%%%%%%%%%%%%%%%%%
\subsection{Magnification}
\label{sec:mag}

The OAT Hamiltonian after time $\tau_\text{sq}$ results in a sheared state that has enhanced sensitivity to the effects that change $J_z'$ or $\text{var}(J_z')$ as shown in panel $(b)$ of Figs.~\ref{fig:protocolnet} and~\ref{fig:protocoltot}. If we were to read out the system at this stage, readout noise on the relevant magnetization component would have to be at the $\sqrt{N\xi^{-2}}$ level. However, even achieving the SQL, i.e., a readout noise at $\sqrt{N}$ level,  is challenging but recently demonstrated for macroscopic NMR systems~\cite{sushkovtalk}. We can get around this problem by observing that the OAT Hamiltonian can also {\emph{magnify}} both a change in the mean and the variance above the SQL by an amount parametrically equal to the achievable squeezing, thus ultimately requiring readout sensitivity less stringent than the SQL---as was first proposed and demonstrated in~\cite{Davis16,Hosten2016Science} in the context of manipulation of spin systems in the optical domain. The details of the computations that follow can be found in App.~\ref{app:magnification}. 

It is qualitatively shown in panels $(c)$ to $(f)$ of Figs.~\ref{fig:protocolnet}, and~\ref{fig:protocoltot}. First, the squeezed state is rotated by a small angle $\theta\approx 2 \xi^{-1}$ (panel {\it{c}}), equivalent in practice to flipping the sign of $\chi$, and $H_\text{eff}$ causes it to come back to the unsqueezed state in time $\tau_\text{sq}$. A signal $S$ that changed the mean, $\avg{J_z}$, before the rotation, will be mapped to a displacement $S\xi$ in the $J_y$-direction at the end of this process. Similarly, a signal $S'$ that changed the variance $\text{var}(J_z')$, before the rotation, will be mapped to an increase of the variance of $J_y$ compared to the SQL by $S'\xi^2$. This behavior is shown in panel $(d)$. The signal and noise have already been magnified by a factor of $\xi^2$.

The requirements on the readout can be further relaxed, as {\emph{both}} signal and shot noise can be further magnified. For this, a $\pi/2-\xi^{-1}$ pulse flips $J_y$ and $J_z$ ($(e)$ panels). From a practical standpoint, note that at this point in the protocol, the spin states are significantly less sensitive to imperfections in state rotations in comparison to the squeezed-state stage ($(b)$ panels).  After another $\tau_\text{sq}$, the OAT Hamiltonian further magnifies the mean displacement by a factor of $\xi$ or the variance by a factor of $\xi^2$. Thus, at the end of the magnification process we get schematically,
\begin{align}
    \avg{J_y}&\approx S\xi^2 \quad\text{and}\quad \text{var}(J_y) \approx \xi^2 N/4,\quad\text{or}\\
    \avg{J_y^2}&\approx \frac{N\xi^2}{4}\pare{1+\xi^2 S'} \quad\text{and}\quad \text{var}(J_y^2) \approx (\xi^2 N/4)^2
\end{align}
where we kept the leading order terms only. Here $S$ and $S'$ include photon decays, spin relaxation, as well as any signal from cosmic relics. In the end, the sensitivity to $S$ compared to spin shot-noise is enhanced by the squeezing factor $\xi^2$, exactly as described in the squeezing part of the protocol, but the total spin fluctuations are magnified by another factor of $\xi^2$. A possible signal changing $\langle J_z \rangle$, corresponds to a shift of the center of mass of the phase space distribution and it is also correspondingly magnified.

{ We emphasize that magnification does not affect the sensitivity of the squeezed state to the cosmic relic signal, as it amplifies both signal and noise to circumvent the potentially large noise of standard circuit readout schemes of NMR systems (e.g. with a SQUID). Given existing technologies, the initial $\approx 2\xi^{-1}$ of step (c) is most likely needed, as readout to $\sqrt{N}/\xi$ level may be unattainable, but the steps (d)$\to$(f) are not required if an SQL-limited readout exists. Such considerations are important for a future concrete experimental setup~\cite{upcomingRapidis}, because phase stability requirements of these unitary rotations may be too stringent (see Sec.~\ref{sec:tolerances}).}

%%%%%%%%%%%%%%%%%%%%%%
\subsubsection*{Readout}

This greatly relaxes the readout requirements, allowing the measurement to now be performed at a level $\xi$ {\emph{above}} the SQL. While the exact details of the readout setup are beyond the scope of this paper, a few remarks are in order. { First, the magnified spin fluctuations correspond to voltage fluctuations on the order of a few nV/$\sqrt{\text{Hz}}$ in the circuit— which appears to be within the detection capabilities of current technologies~\cite{DMRadio:2023igr, kentprivate}, especially given that SQL limited readout has been recently reported~\cite{sushkovtalk}.}  Second, a final $\pi/2$ pulse can be applied to rotate the state back to the bottom of the Bloch sphere, where the enhanced fluctuations are mapped onto the {\emph{transverse}} magnetization. This may be necessary, for example, if the readout needs to be performed closer to resonance, in which case the circuit would otherwise experience significant backaction due to collective spin decay. This highlights a key point: squeezing, magnification, and signal accumulation can—and should—be decoupled from the readout considerations, allowing both parts of the experiment to be optimized separately. In particular, squeezing and magnification can be carried out at the equator, far detuned from resonance, and at the highest possible $Q$, while the final measurement can take place at the bottom of the Bloch sphere, as a phase measurement (i.e., of $J_x$ or $J_y$), closer to resonance, and at a much lower $Q$.

Irrespective of readout, for a protocol that achieves 48 dB of squeezing, at the end of the magnification part, we have magnified the fluctuations of a very weak signal by 96 dB, or 9.6 orders of magnitude.

For the convenience of the reader, we add a table (Table~\ref{tab:parameters}) that summarizes relevant parameters and their fiducial values for the OAT protocol.

\begin{table*}[t]
     \centering
 \begin{tabularx}{0.8\textwidth}{X||c}\hline \hline
    Parameter &  Fiducial Value\\ \hline
    Number of spins $N$ & $10^{7}\lesssim N \lesssim 10^{26}$\\
    Circuit mode and spin frequencies $\omega_\mathrm{LC}$, $\os$ & $10^{-12}~\text{eV}-10^{-6}~\text{eV}$\\
    Circuit quality factor $Q$ & $10^6-10^9$\\
    Circuit line width $\kappa$ & $\omega_\text{LC}/Q$\\
     Circuit mode volume $V_\mathrm{LC}$ & $V_\text{L}\lesssim1 ~\text{m}^3$\\
      Circuit-spin coupling $g$ at $\omega_\mathrm{LC}=10^{-8}~\text{eV}$ & $\mu \sqrt{\frac{\omega_\text{LC}}{2 V_\text{L}}}\approx 2\times10^{-24}~\text{eV}\frac{\mu}{\mu_{^3\text{He}}}\sqrt{\frac{1~\text{m}^3}{V_\text{L}}}$\\
     Spin-circuit mode detuning $\Delta\equiv \omega_0-\omega_\text{LC}$ & $|\Delta|\approx\oc$\\
    Squeezing parameter $\chi$ at $|\Delta|\approx \omega_\text{LC}$ & $\frac{g^2}{\Delta}=4\times 10^{-38}$~eV\\
    Single-spin decay rate $\eta$ ($i=0$ capacitive, $i=\text{LC}$ inductive)& $4 \frac{g^2 \omega_i^2 \kappa}{(\omega_0^2-\omega_\text{LC}^2)^2 }\approx\frac{2\mu^2 \omega_i^2}{ V_\text{L}Q \oc^2}$\\
    Spin dephasing time $T_2$ & $1000~\text{s}\frac{3\times 10^{22}~\text{cm}^{-3}}{n_\text{S}}$\\
     Spin thermalization time $T_1$ & $1000~\text{s}\frac{3\times 10^{22}~\text{cm}^{-3}}{n_\text{S}}$\\
    Squeezing factor $\xi^2$ & $\frac{2^{2/3}}{3}\parea{\frac{\chi}{\eta(2\bar{n}+1)+2/(3 N T_1)}}^{2/3}$\\
    Squeezing time $\tsq$ & $\frac{2}{N}\pare{\frac{1}{\parea{\eta(2\bar{n}+1)+2/(3 N T_1)}\chi^2}}^{1/3}$\\\hline \hline
 \end{tabularx}
 \caption{Definitions and fiducial values of parameters used throughout the paper for the OAT protocol.}
 \label{tab:parameters}
 \end{table*}

%%%%%%%%%%%%%%%%%%%%%%%%%%
\section{Applying the protocol to new physics searches}
\label{sec:sm_bsm}

In order to demonstrate the power of the protocol presented above, we show how it can be applied to axion and dark photon DM, as well as C$\nu$B searches.

Axions are pseudoscalar particles that are predicted in extensions of the SM, like string theory. They can couple to nuclear or electron spins. They are excellent DM candidates as they can be produced cosmologically, e.g. through the misalignment mechanism. Their properties are parametrized by the axion decay constant $f_a$ and their mass $m_a$.

There is one axion that stands out among all those predicted in beyond the Standard Model (BSM) physics models, and that is the QCD axion, as it solves the outstanding puzzle of the missing electric dipole moment of the neutron~\cite{Weinberg:1977ma,Wilczek:1977pj,Dine:1982ah,Preskill:1982cy}. The QCD axion mass is uniquely determined by non-perturbative QCD effects and the axion decay constant, to be $m_a= 5.7~\mu \text{eV}\left(\frac{10^{12} ~\text{GeV}}{f_a}\right)$. 

Dark photons are another possible DM candidate that can be probed when looking for axions with a nuclear spin system. While they can be motivated by string theory similarly to axions, for masses below $10^{-5}~\text{eV}$ they do not have such a clean DM production mechanism such as misalignment. As a result, we do not consider that they should be the primary focus of searches in these low masses, but they are DM candidates more suited as a``parasitic"-type of experiment. 

The C$\nu$B is a relic of the big bang that was created when the universe was less than a second old, and is an elusive prediction of the standard models of particle physics and cosmology. There are about $56~\text{cm}^{-3}$ for every mass and helicity eigenstate. They are by far the most abundant and at the same time the least energetic species of neutrinos in the Universe. Their kinetic energy is just $E_\nu\sim 10^{-6}~\text{eV}\left(\frac{0.1~\text{eV}}{m_\nu}\right)$, which means that the extremely rare interactions that they have only deposit a minuscule amount of energy compared to traditional particle physics relic searches. A way to illustrate how challenging their detection is, the per nucleon elastic scattering cross-section is just $\mathcal{O}(10^{-64}~\text{cm}^2)$!

As a subset of the authors pointed out~\cite{Arvanitaki:2024taq}, superradiant interactions of these extremely weakly interacting cosmic relics can greatly enhance their discovery potential. The squeezed states that our proposed protocol creates have enhanced sensitivity to changes of $\langle J_z\rangle$ and $\langle J_z^2 \rangle$, or equivalently $\avg{J_z'}$ and $\text{var}(J_z')$, induced by superradiant interactions. Barring readout considerations, we present the reach  of the above protocol for each case below.

%%%%%%%%%%%%%%%%%%%%%%%%%%
\subsection{Axion and dark photon dark matter searches}
\label{sec:axion_and_dp}

DM which is lighter than $\mathcal{O}(1~\text{eV})$, can be thought of as a classical background field with finite spatial and time coherence of size $(m_\text{DM} \upsilon_\text{DM})^{-1}$ and $(m_\text{DM} \upsilon_\text{DM}^2)^{-1}$, respectively, where $m_\text{DM}$ is the DM mass, and $\upsilon_0\sim 10^{-3}$ is the DM velocity dispersion in the galaxy. Stimulated emission and absorption of bosonic DM, axion or dark photon, from the spins can be thought of as Markovian noise that induces changes in $\text{var}(J_z')$, when the running time of the experiment is longer than the coherence time~\cite{Arvanitaki:2024taq}. Such changes can be measured following the procedure outlined in Fig.~\ref{fig:protocoltot}. When the experiment running time is less than the bosonic DM coherence time, DM manifests as changes in $\langle J_z \rangle$, which can also be measured by what is shown in Fig.~\ref{fig:protocolnet}. We will now discuss the axion DM and dark photon DM cases separately.

%%%%%%%%%%%%%%%%%%%%%%%%%%
\subsubsection{Axion DM searches}
\label{sec:axion}

 \begin{figure*}[t!]
 \centering
    \includegraphics[width=1\textwidth]{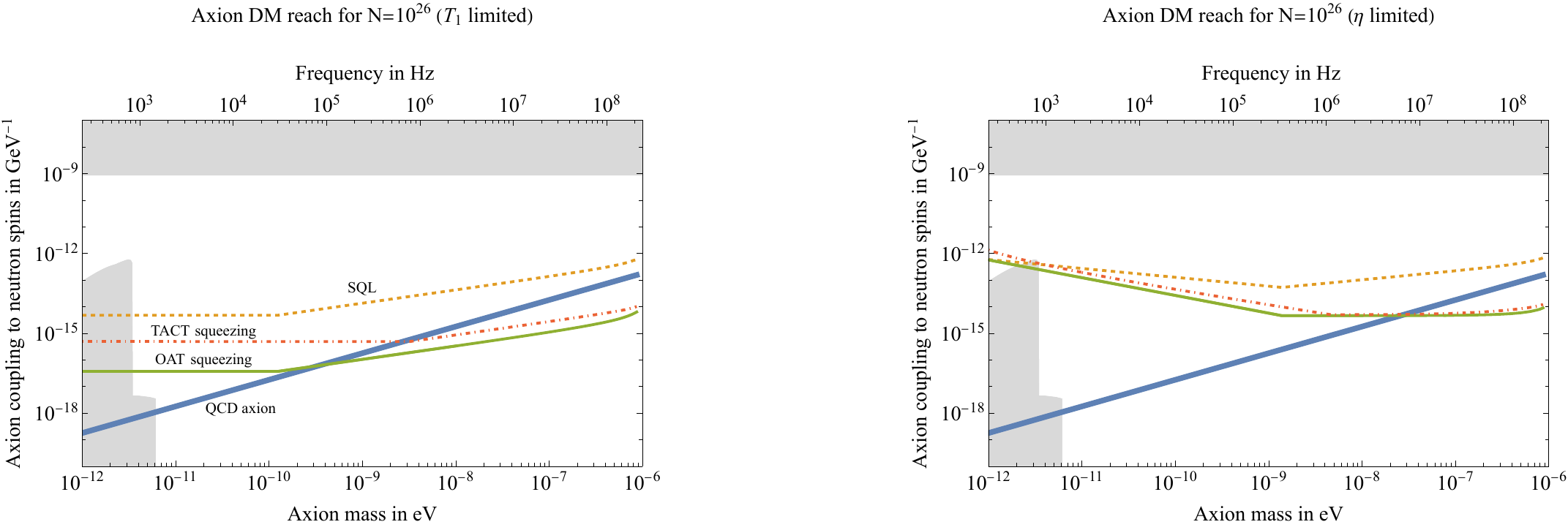}
    \caption{Sensitivity of a $N=10^{26}$ $^3$He spin sample of density $0.3\times 10^{20}~\text{cm}^{-3}$, to the axion-nucleon spin coupling as a function of the axion mass. Temperature is taken to be $T=1$~K. On the left panel, the spin decay through the circuit is dominated by capacitive effects and the sensitivity is limited by $T_1$. On the right panel, the spin decay through the circuit is dominated by inductive effects, and the sensitivity is limited by $\eta$. The curves are plotted on each panel for three different cases: {\it{(Dashed curve)}} The Standard quantum limit. {\it{(Dot-Dashed curve)}} The two-axis counter-twisting protocol parameter space.  {\it{(Solid curve)}} The one-axis twisting protocol parameter space.  { On the left panel, the OAT curve will take many years to be scanned, while the TACT curve can be covered within months.} {The TACT line should be taken as a potential benchmark
based on the heuristic derivations of App.~\ref{app:tact}. A more detailed treatment is required for a rigorous forecast, as
described in Sec.~\ref{sec:vacsq}}. On the right panel, times scales are fast enough that scanning the full frequency range takes much less than that. For both panels, gray shaded regions are excluded by astrophysics~\cite{SNTurner,SNRaffelt,SNOlive,SNRecent,BHSR-self-interactions} and the thick blue line shows the prediction for the QCD axion. { This projected reach assumes at least an SQL-limited readout, after steps (a)-(d) of Figs 2 and 3, and is not sensitive to phase noise of the driving Rabi pulses if step (c) can be performed in a different manner (such as the ones described in Sec.~V).}}
    \label{fig:axion}
\end{figure*}

 \begin{figure}[t!]
   \centering
    \includegraphics[width=0.46\textwidth]{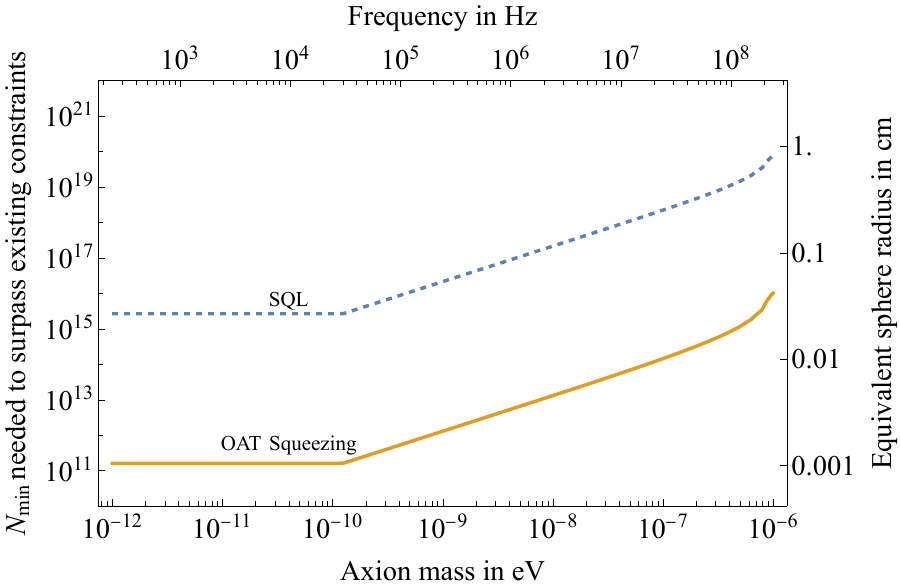}
    \caption{Minimum number of $^3$He spins for $0.3 \times 10^{20}~\text{cm}^{-3}$ density sample to surpass existing constraints~\cite{SNTurner,SNRaffelt,SNOlive,SNRecent} on the axion-nulceon derivative coupling as a function of the axion mass. {\it{(Dashed curve)}} The Standard quantum limit. {\it{(Solid curve)}} The one-axis twisting protocol. Both curves assume the decay of the spins to the circuit is dominated by capacitive effects. The OAT curve assumes that the effective spin decay is dominated by the capacitive circuit linewidth.}
    \label{fig:axionNmin}
\end{figure}

Axions can have a derivative coupling to fermions that makes axion DM manifest as an anomalous magnetic field that only couples to spin~\cite{krauss1985spin,Wilczek-spin-forces}, i.e $\mathcal{H}_\text{axion-int}=\frac{\vec{\nabla} a\cdot \vec{\sigma}}{f_a}$, where $a$ is the axion field. Such an interaction means that axions can be absorbed or emitted from a spin that is excited or de-excited. The sum of the stimulated axion absorption and interaction rates on a spin sample is given by:

\begin{eqnarray}
    \Gamma_++\Gamma_-\approx N^2\frac{\rho_\text{DM}}{f_a^2 m_a},
\end{eqnarray} in the Markovian regime inducing a change in $\text{var}(J_z')$. In the non-Markovian regime, the $\langle J_z \rangle$ rate of change is given by:
\begin{eqnarray}
    \langle \dot{J}_z \rangle\approx \sin \phi \frac{N}{2}\frac{2\rho_\text{DM} v_0^2}{f_a^2} t,
\end{eqnarray}
where $\phi$ is an (unknown) axion phase.

Unfortunately, there is no prediction for the absolute axion mass, leaving us with the need to scan. 
The scan needs to be optimized on an experiment by experiment basis, which goes beyond the scope of our paper. So we naively take the integration time per frequency bin, $\tau_\text{scan}$, to coincide with the squeezing time; if the entire squeezing and magnification protocol is applied, the $\tau_\text{scan}\approx 3 \tau_\text{sq}$, but maximum SNR is achieved only during $\tau_\text{sq}$. We will consider separately capacitive and inductive dominated circuit decay, that in turn result to the decay of the spins. We fix in both cases, the temperature to 1~K, the quality factor of the circuit to $Q=10^9$, and the density of the sample to be $0.3 \times 10^{20}~\text{cm}^{-3}$. $\eta$ being dominated by capacitive losses gives $\tau_\text{sq}\approx8.6~\text{sec}$ for the OAT protocol, while $\tau_\text{sq}\approx 0.33~\text{sec}$ for the TACT protocol. When inductive losses dominate $\eta$, $\tau_\text{sq}$ is $1.5~\text{sec}\pare{\frac{10^{-8}~\text{eV}}{\omega_0}}^{1/3}$ for the OAT protocol, while it ranges between 0.03-0.1~sec for the TACT protocol. Based on the above assumptions, we can derive a projected sensitivity of a nuclear-spin-based detector to the axion-nucleon spin coupling:

\begin{eqnarray}
    {g_{aNN}}_\text{min}= &\left[\sqrt{2} N \xi^2 ~\rho_\text{DM} \upsilon_0^2 \tau_\text{scan}\right]^{-1/2}\times \nonumber \\ 
    &\left[ \min \left(\tau_\text{scan} ,\frac{1}{m
   \upsilon_0 ^2}\right)\right]^{-1/2}
\end{eqnarray}

The sensitivity, assuming different squeezing protocols and different limiting cases for $\eta$ is shown in Fig.~\ref{fig:axion}. We have to point out that the amount of detuning has been numerically calculated in each case to maximize the reach at a given frequency. The OAT protocol not only greatly enhances the axion parameter space that can be explored, but may finally allow to probe the QCD axion derivative coupling to nucleons in a previously inaccessible regime. At the same time, the TACT protocol with a shorter squeezing time, demonstrates that squeezing can have another advantage beyond just increasing the maximum reach, as it can accelerate searches in a way that  what would take years, can now be scanned within mere months of running time.

Another way to illustrate the power of a $48$~dB squeezing and amplification protocol is shown in Fig.~\ref{fig:axionNmin}. This figure shows the minimum number of atoms needed to surpass existing constraints on the axion-nucleon coupling as a function of the axion mass. With the OAT protocol, a sample as small as $5\times 10^{15}$ is enough to overcome current constraints over 6 orders of magnitude in mass. 

To conclude, squeezing protocols not only improve axion dark matter search reach beyond the SQL, but they can also accelerate searches or go beyond what we already know without even resorting to macroscopic detector samples.

%%%%%%%%%%%%%%%%%%%%%%%%%%
\subsubsection{Dark photon DM searches}
\label{sec:dp}

\begin{figure*}[t!]
    \centering
    \includegraphics[width=1\textwidth]{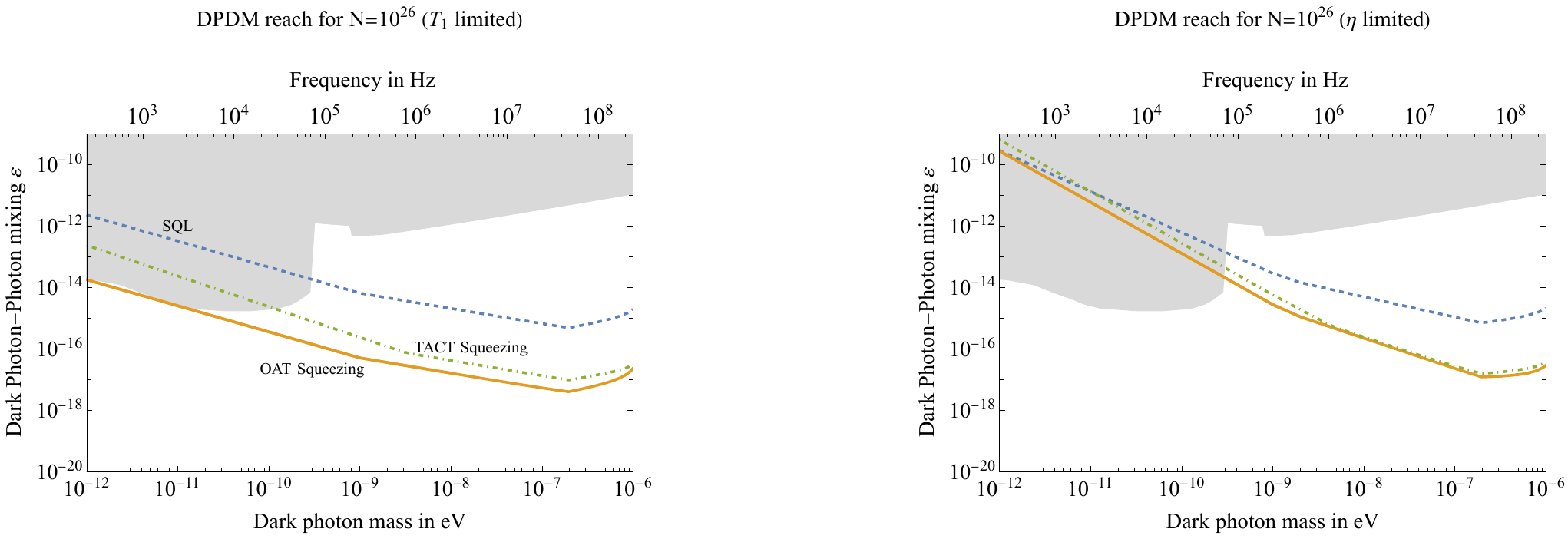}
    \caption{Sensitivity of a $N=10^{26}$ $^3$He spin sample of density $0.3 \times 10^{20}~\text{cm}^{-3}$,  to dark photon DM kinetically mixed with the ordinary photon as a function of the dark photon mass. Temperature is taken to be $T=1$~K. On the left panel, the spin decay through the circuit is dominated by capacitive effects and the sensitivity is limited by $T_1$. On the right panel, the spin decay through the circuit is dominated by inductive effects, and thus dominates the dynamics. The curves are plotted on each panel for three different cases: {\it{(Dashed curve)}} The Standard quantum limit. {\it{(Dot-Dashed curve)}} The two-axis counter-twisting protocol parameter space. {\it{(Solid curve)}} The one-axis twisting protocol parameter space. { On the left panel, the OAT curve will take many years to be scanned, while the TACT curve can be covered within months.} {The TACT line should be taken as a potential benchmark
based on the heuristic derivations of App.~\ref{app:tact}. A more detailed treatment is required for a rigorous forecast, as
described in Sec.~\ref{sec:vacsq}}. On the right panel, times scales are fast enough that scanning the full frequency range takes much less than that. Gray shaded regions are excluded by astrophysics~\cite{Caputo:2021eaa}. { This projected reach assumes at least an SQL-limited readout, after steps (a)-(d) of Figs 2 and 3, and is not sensitive to phase noise of the driving Rabi pulses if step (c) can be performed in a different manner (such as the ones described in Sec.~V).}}
    \label{fig:darkphoton}
\end{figure*}

 \begin{figure}
   \centering
    \includegraphics[width=0.46\textwidth]{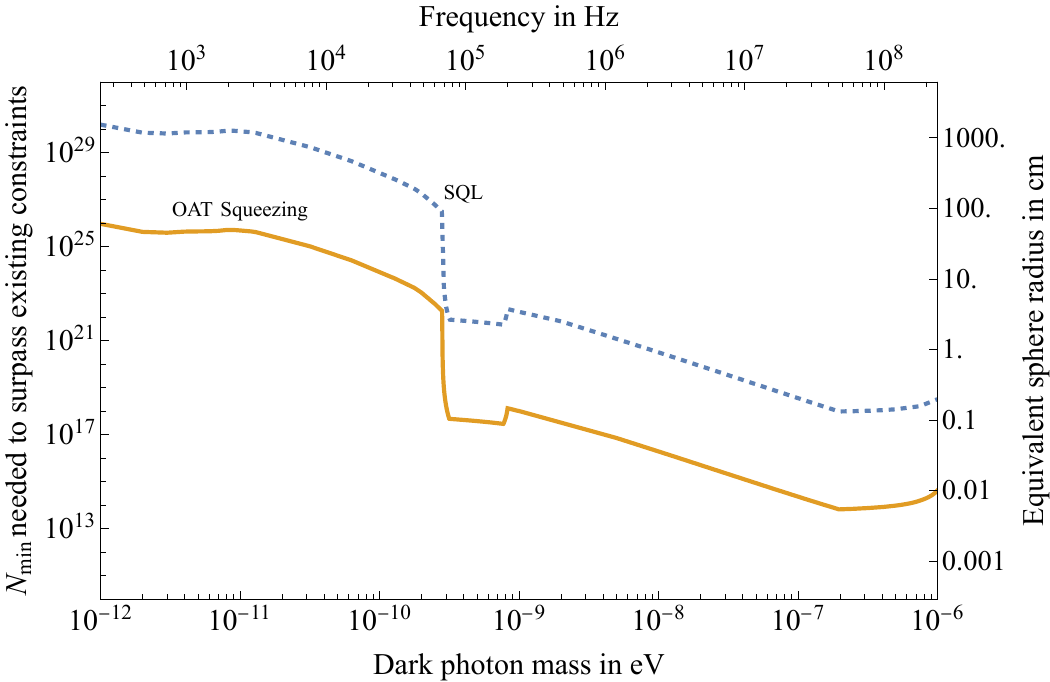}
    \caption{Minimum number of $^3$He spins for $0.3 \times 10^{20}~\text{cm}^{-3}$ density sample to surpass existing constraints on the dark photon dark matter kinetic mixing with the ordinary photon~\cite{Caputo:2021eaa} as a function of the dark photon mass. {\it{(Dashed curve)}} The Standard quantum limit. {\it{(Solid curve)}} The one-axis twisting protocol. The OAT curve assumes that the effective spin decay is dominated by the capacitive circuit linewidth.}
    \label{fig:darkphotonNmin}
\end{figure}

Dark Photons can have a marginal coupling to the SM through kinetic mixing with ordinary photons, $\mathcal{H}_\text{dp-inter}=\epsilon F_\text{dp}F_\text{EM}$, where $F_\text{EM}$ and $F_\text{dp}$ are the field strengths of the photon and dark photon, respectively. This kinetic mixing means that dark photon DM (DPDM) manifests as a background electric or magnetic field. The DPDM induced magnetic field is what can cause spin excitation and de-excitation. Taking into account the finite coherence time of DPDM similarly to the axion case, we find that the minimum kinetic mixing that can be probed is given by:
\begin{eqnarray}
    &\epsilon_\text{min}=\left[\sqrt{2}N \xi^2 ~g_f^2~ \rho_\text{DM} ~\tau_\text{scan} \right]^{-1/2} \times \\ \nonumber& \left[\min (1,m R)^2 ~\min
   \left(\tau_\text{scan} ,\frac{1}{m \upsilon_0^2}\right)\right]^{-1/2}.
\end{eqnarray}
The factor $\min (1,m R)$ accounts for the effects of the shield to the dark photon profile~\cite{DMradiooriginal}. We set the size of the shield to be 1~m. The sensitivity in $\epsilon$ achieved for the different squeezing possibilities previously described for the axion, is shown in Fig.~\ref{fig:darkphoton}. While there is no clear target in this part of the parameter space for DPDM, as there is for axions, the reach is greatly improved for a search that can run paracitically on an axion search. In addition, Fig.~\ref{fig:darkphotonNmin} shows the minimum number of atoms needed in a sample to overcome existing constraints on the dark photon-photon kinetic mixing. In this case, possible improvements due to squeezing are made abundantly obvious in the mass range between $10^{-12}~\text{eV}$, and $3\times 10^{-10}~\text{eV}$; while one would need a extraordinarily large sample of atoms to overcome astrophysical constraints without resorting to squeezing, the OAT protocol brings improvements beyond what we already know with minimum sample sizes of order $10^{25}$ spins.

\subsection{The cosmic neutrino background}

What makes abundantly clear how superradiant interactions  are different from other types of coherent effects, are the interactions of the C$\nu$B with an extended ensemble of spins prepared in an ECSS~\cite{Arvanitaki:2024taq}. Relic Majorana neutrinos from the Big Bang will excite or de-excite the spins confined in a sphere of radius R with a rate:

\begin{eqnarray}
&\Gamma_\text{total}=0.2~\text{Hz}\pare{\frac{m_\nu}{0.1~\text{eV}}}\pare{\frac{n_\text{S}}{3\times 10^{22}~\text{cm}^{-3}}}^2 \pare{\frac{R}{10~\text{cm}}}^4 \\
&\text{and} \nonumber \\
&\Gamma_\text{net}=10^{-3}~\text{Hz}\pare{\frac{m_\nu}{0.1~\text{eV}}}\pare{\frac{n_\text{S}}{3\times 10^{22}~\text{cm}^{-3}}}^2 \pare{\frac{R}{10~\text{cm}}}^3
\end{eqnarray}
where $\Gamma_\text{total}$ and $\Gamma_\text{net}$ are the sum and the difference of the excitation and de-excitation rates, respectively. Dirac neutrinos interact with half these rates. Unless $\xi^2$ is extremely large, the observable that provides the highest SNR is $\Gamma_\text{net}$, which corresponds to a net change in $\langle J_z \rangle$. Assuming a total running time for the experiment of 3 years, keeping the total number of spins in the sample fixed to $N=1.26\times10^{26}$, and assuming the OAT protocol, we find that SNR $\sim 1$ would correspond to a local C$\nu$B overdensity of:

\begin{eqnarray}
C_\nu\approx 10^{10} \pare{\frac{m_\nu}{0.1~\text{eV}}}^{-1}\pare{\frac{N}{1.26\times 10^{26}}}^{-1}\pare{\frac{R}{10\text{ cm}}}^{3/2},
\end{eqnarray}
where $R\geq 10$~cm.

{ Several comments about this sensitivity are in order. First, it would be an improvement of the KATRIN constraint~\cite{KATRIN:2022kkv} by up to an order of magnitude, which is currently the best direct detection constraint. Second, while Pauli blocking renders these over-densities completely unphysical even if one ignores cosmology, this constraint highlights the potential of these table-top size detectors, as they can compare favorably with much larger scale experiments. Third, and most importantly, it highlights the capabilities of NMR detectors to probe superradiant interactions of relics beyond wave-like particles, such as axions and dark photons, therefore establishing a long-term program for the improvement of this technology towards the ultimate detection of the C$\nu$B.}

%%%%%%%%%%%%%%%%%%%%%%%%%
\section{Technical noise}
\label{sec:tolerances}

While detailed technical noise considerations are going to be specific to a concrete experimental setup, below we comment on some universal sources of noise that can readily be expected and on their mitigation strategies. Detailed computations of most effects summarized here can be found in the Appendices. First we discuss technical noise sources that degrade squeezing, and then we comment on potential systematics that may emulate a cosmic relic signal.

\begin{table*}[t]
    \centering
    \begin{tabularx}{\textwidth}{X||c|c|X}
      Source of noise or systematic  &  Tolerance & Discussed in & Brief explanation\\\hline\hline
      Polarization &$\xi'=p\xi$ &App.~\ref{app:polarization} & Reduces total dipole $J_p=pJ=pN/2$\\
        Static Inhomogeneity &$ \delta g/g\lesssim \xi^{-1/2}$& App.~\ref{app:inh_static} & $\mathcal{O}(1)$ effective $N$ and polarization change as long as $\delta g/g$ is less than the amount stated.\\
        Time-dependent inhomogeneity: Brownian motion & $\xi^2\gtrsim 10^9 \pare{\frac{1/\text{m}}{g^{-1}\nabla g}}^2 $&App.~\ref{app:g_brownian} & Only for gaseous or liquid samples; $g^{-1}\nabla g$ is the vacuum mode spatial gradient.\\
        \hline
        Duration of Rabi pulses & $t_\text{R}\lesssim \min\{T_1,T_2\}$ & Sec.~\ref{sec:tolerances} & Upper limit on the time of the initialization Rabi pulse so as not to introduce noise in excess of shot noise. \\
        Phase stability of magnification protocol Rabi pulses& $N^{-1/2}$ & Sec.~\ref{sec:tolerances} & Prevents spurious signals during the magnification protocol implementation.\\
        Systematic (mean) drifts due to in-circuit decays, relaxation, etc. & Subtracted separately&Sec.~\ref{sec:tolerances}& \\
        Initialization & Insensitive to initial $\avg{J_z}$ & App.~\ref{app:rabipicture} & Initial displacement sets the origin; only relative shifts matter for magnification.\\
        \hline
      Decoherence of circuit state & $2\pare{\frac{gN}{2 \Delta}}^2\sqrt{\epsilon_\text{rms}^2+f_\text{rms}^2}\lesssim\sqrt{N\xi^{-2}}$ &App.~\ref{app:morenoise} & Level of control needed for $f_\text{rms}$ and $\epsilon_\text{rms}$ (see text for definitions).\\
        Mechanical fluctuations induced changes in the coupling and frequency & $ \frac{\delta x_\text{rms}}{V_\text{L}^{1/3}}\lesssim \frac{1}{3}\pare{\frac{g \sqrt{N}}{\Delta}}^{-2}\frac{1}{ \xi\sqrt{N}}$ &App.~\ref{app:mechfl}& Coil size fluctuations; $\delta x_\text{rms}$ is the rms coil wall displacement. $\delta x_\text{rms}\sim 0.1$~nm has already been demonstrated in~\cite{DarkSRF}, requiring $N\lesssim 4\times 10^{26}$.\\
        Circuit heating due to in-circuit decays (capacitive losses dominated)& $P_\text{decay}\approx 2 \times 10^{-12}~\text{W}\pare{\frac{N}{10^{26}}}^2\frac{10^9}{Q}\frac{1~\text{m}^3}{V}\pare{\frac{\omega_0}{10^{-8}~\text{eV}}}^3$ & Sec.~\ref{sec:tolerances} & Within the capabilities of dilution refrigerators at 10mK.\\Circuit heating due to in-circuit decays (inductive losses dominated)& $P_\text{decay}\approx 4 \times 10^{-8}~\text{W}\pare{\frac{N}{10^{26}}}^2\frac{10^9}{Q}\frac{1~\text{m}^3}{V}\pare{\frac{\omega_0}{10^{-8}~\text{eV}}}$ & Sec.~\ref{sec:tolerances} & Within the capabilities of dilution refrigerators at 10mK.\\
        
    \end{tabularx}
    \caption{Summary of several technical noise sources, as discussed in Sec.~\ref{sec:tolerances} and the corresponding appendices. This table does not include the fundamental limits to squeezing arising from spin decays, dephasing and relaxation, which are discussed in detail in Sec.~\ref{sec:magsqueeze}.}
    \label{tab:tolerances}
\end{table*}

So far we have assumed that the nuclear spins are perfectly polarized. If the polarization fraction of the spin sample is $p<1$, this corresponds to an effective change in $\langle J_x\rangle$ so that $\langle J_x\rangle=\frac{p N}{2}$ and the maximum squeezing that can be achieved is $\xi_\text{eff}^2=p^2 \xi^2$, with $\xi^2$ is given by Eq.~\eqref{eq:xi2}. This result is derived in App.~\ref{app:polarization}. While in solid state systems, large polarization fractions can only be achieved so far with large magnetic fields and cryogenic temperatures, hyperpolarization techniques for noble gases have demonstrated $p\sim \mathcal{O}(1)$ \cite{PhysRev.132.2561,PhysRevA.29.3092}, hence our choosing to focus on $^3\text{He}$.

Inhomogeneities in the vacuum mode, discussed in Apps.~\ref{app:inh_static} and~\ref{app:g_brownian} lead to spatially varying couplings and imperfect initial state preparation. For static spins, such as those in a solid, spatially varying couplings can be shown to change the effective number of spins being squeezed \cite{hu2015,wu2020}, and, as long as the fractional spatial variation $\delta g/g$ is less than $\xi^{-1/2}$, this effective number is $N_\text{eff}\approx N$. This statement is true as long as squeezing and readout is done with the spins in the same positions and with the same apparatus, which we prove rigorously in App.~\ref{app:inh_static}. For spins in a gas or liquid sample, diffusion due to Brownian motion makes the sample sensitive to large spatial gradients in $g$. In App.~\ref{app:g_brownian} we show that such effects are only relevant when $\xi^2\geq 10^9 \pare{\frac{1/\text{meter}}{g^{-1} \nabla g}}^2$. We emphasize that this effect is different from the inhomogeneous broadening arising due to external field gradients of $B_\text{ext}$, which can potentially shorten $T_\text{deph}$. The effects and requirements of $T_\text{deph}$ were discussed in Sec.~\ref{sec:magsqueeze}.

Next, the Rabi pulse needed to initialize the ECSS needs to be applied on time scales $t_\text{R}\lesssim \min\{ T_1,T_2\}$, so as to not induce excess noise (i.e. more than shot noise) during the initial preparation that will subsequently be magnified. Independent of Rabi pulse systematics, there are systematic drifts in $\langle J_z \rangle$ due to in-circuit 
decays and effects of spin relaxation. These drifts can be calculated, measured and subtracted.

The next question that needs to be addressed is to what degree Rabi pulse sequences need to be controlled. The initial Rabi pulse defines our effective axis origin. Any initial displacement in $\langle J_z\rangle$ away from the equator, after the pulse sequence described in Fig.~\ref{fig:protocolnet}, does not result in a shift in $\langle J_y \rangle$, and does not appear as a signal. Put differently, the protocol is insensitive to errors, stemming from pulse area noise, in preparing an exact initial $\langle J_z\rangle=0$ state. The pulse sequence is only sensitive to \emph{changes} in $\langle J_z \rangle$. In App.~\ref{app:rabipicture}, we show a pictorial representation of the above statements that may perhaps be more intuitive.  What introduces systematic noise and shifts is the phase stability of the RF Rabi pulses, since the protocol essentially maps a magnified version of changes in $J_z$ to $J_y$---i.e., to the spin precession phase. We find that this phase needs to be stable at the level of $1/\sqrt{N}$.  While this seems not to be a problem at the lower end of spin numbers considered in this work for achieving 96~dB of magnification ($N= 10^{7}$), it requires further considerations at the higher end---$N=10^{26}$, needed for new physics searches---requiring a phase stability of $10^{-13}$~rad between Rabi pulses. Today, the lowest-noise microwave and RF signals are generated by frequency dividing optical atomic clock signals through optical frequency combs \cite{diddams2020}. Keeping in mind that phase noise is linearly proportional to the final frequency, extrapolating from achieved performances at 10 GHz frequencies \cite{xie2017}, for frequencies below 100 MHz considered in this work, thermal-noise-limited room-temperature operation starting from 1-Hz noise frequencies can be taken as a benchmark. Further filtering with high-Q circuits of Hz-level linewidth assumed in this work in a 1~K cryogenic environment, another order of magnitude lower noise could be achieved (-220 dBc/Hz @ 20dBm instantaneous power). Assuming a reference protocol duration of 1 second and Rabi pulse durations of 10 ms---effectively implying a noise spectrum integration from 1 Hz to 100 Hz \cite{noise_calculator}---independent of the carrier frequency, this thermal noise limited phase noise would give $10^{-10}$~rad phase noise. 

While this \emph{absolute} phase sensitivity is not enough for more than $\sim 10^{20}$ spins, it is not necessarily the right figure of merit.
Instead, it is the \emph{relative} stability between the phase of the RF drive and the spin precession phase that needs to be stable. Therefore, it is equally critical to take into account the stability of the magnetic field bias that dictates this frequency. The best option for generating the bias fields is likely the utilization of persistent currents in superconducting coils that can be extremely low noise \cite{6612716, Takeda_2022, Brouwer_2022}. A potential solution, then, to further improving the relative phase stability could be phase locking a readily low-noise RF source to the in-circuit mean-field leaking out of the circuit that is generated by the spins themselves---e.g. through mixing the RF and the leakage inside of the cryogenic environment. A detailed analysis of such a scheme necessarily depends on the concrete experimental setup and is thus left to future work. 

Furthermore, we emphasize that the stability of the RF pulse is relevant only for the pulses associated with the magnification part of our protocol. For example, one could forego magnification, and be able to reach similar sensitivities to those of Figs.~\ref{fig:axion} and~\ref{fig:darkphoton}, at least in parts of the parameter space. It would amount to devising a way of flipping the sign of $\Delta_-$---and thus of $\chi$---once maximal squeezing is achieved, by swapping $\os\leftrightarrow\oc$. This would be equivalent to steps {\it{(b)}}$\to${\it{(d)}} in Figs.~\ref{fig:protocolnet} and~\ref{fig:protocoltot}. Then the signal can be read out as long as there is at least shot noise sensitivity, which has been recently achieved in macroscopic NMR samples~\cite{sushkovtalk}. We leave details of such workarounds to future work as they will entail their own experimental challenges.

Another potential source of noise coming from the coherent state $\ket{\alpha}$ produced in the circuit during the initialization of the ECSS, $|\alpha|=g N/{2\Delta}$.  Any decoherence of this state will back-react as noise onto the spins. This is a byproduct of the large densities needed to achieve sufficient squeezing within reasonable times. To our knowledge, there has not been a need to explore its effects in atom-cavity coupled systems because of the large couplings $g$ associated with such systems. Given this, we present what we think is a first step to an appropriate treatment of any decoherence effects due to the presence of $\ket{\alpha}$. We summarize our results below, but provide detailed computations in App.~\ref{app:morenoise}.

The decay of the spins through the circuit, parametrized by $\eta$ above, does not introduce any additional back-reaction on the system. However, fluctuations in the size of the coupling $g$ or the detuning $\Delta$ introduce a random walk of the mode population that decoheres the coherent circuit state $\ket{\alpha}$, which then backreacts as an additional decoherence channel onto the spins. Assuming that the fractional coupling and fractional detuning fluctuations have values $\epsilon_\text{rms}$ and $f_\text{rms}$, respectively, we estimate a bound on these fluctuations according to the inequality $2 \pare{\frac{\bar{g} N}{2 \Delta}}^2\sqrt{f_\text{rms}^2+\epsilon_\text{rms}^2}< \sqrt{N \xi^{-2}}$. This simple estimate is derived early on in App.~\ref{app:morenoise}, but a more rigorous---and complicated---expression for the excess variance in $J_z$ is also provided in the same appendix (Eq.~\eqref{eq:excessvarz}). For the parameters discussed in sec.~\ref{sec:sm_bsm}, this translates to a bound $f_\text{rms},\epsilon_\text{rms}\lesssim 10^{-9}-10^{-10}$. 

As a concrete example, we work out in detail the effects of white noise mechanical fluctuations in App.~\ref{app:mechfl}, as a source of coupling fluctuations contributing to both $\epsilon_\text{rms}$ and $f_\text{rms}$. Assuming a fractional wall jitter stability $\delta\equiv 10^{-10}$, as demonstrated by DarkSRF~\cite{DarkSRF,AsherSC} for a 1~m cavity, we show that the excess noise is $144 (gN/(2\Delta))^4 \delta^2$. This translates into a requirement $N\lesssim 4\times 10^{26}$, which independently sets the upper limit used in Table~\ref{tab:parameters}. We note that this is a qualitatively different requirement from other sources of decoherence (cf. Sec.~\ref{sec:vacsq}), which depend only on the (effective or physical) spin density. This is due to the fact that the coherent state $\ket{\alpha}$ is produced, and thus also backreacts, collectively.
Finally it is worth noting that the active decay of the spins would produce heat in the circuit. We find that for large detunings the power emitted by the spins is of order $P_\text{decay}= 2 N^2\frac{\mu^2 \omega_0^3}{ Q V \oc^2}=2.4 \times10^{-12}~\text{W}\pare{\frac{N}{10^{26}}}^2\frac{10^9}{Q}\frac{ 1~\text{m}^3}{V}\pare{\frac{\omega_0}{10^{-8}~\text{eV}}}^3$, in the case of capacitive dominated circuit decay. For an inductive decay on the other hand, $P_\text{decay}= 2 N^2\frac{\mu^2 \omega_0}{ Q V}= 4\times10^{-8}~\text{W}\pare{\frac{N}{10^{26}}}^2\frac{10^9}{Q}\frac{1~\text{m}^3}{V}\pare{\frac{\omega_0}{10^{-8}~\text{eV}}}$. Both of these scenarios appear to be within the capabilities of dilution refrigerators down to 10 mK temperatures~\cite{UHLIG1997279}.

%%%%%%%%%%%%%%%%%%%%%%%%%
\section{Discussion}
\label{sec:discussion}

{ In this paper, we propose the SIREN protocol and explore the possibility that macroscopic spin samples prepared in the ECSS can be manipulated to achieve unprecedented levels of quantum enhancement—beyond what has previously been explored in the literature~ \cite{serafin2021,boyers2025,boyers2022}.} Many properties of the coupled NMR-circuit system parallel those of the atom-cavity systems, while exhibiting some pronounced distinctions. One difference stems from the macroscopic number of spins required to achieve significant squeezing within a realistic amount of time. This large $N$ regime is nevertheless the natural setting for achieving high sensitivities regardless of squeezing. In this regime, a large coherent state $\ket{\alpha}$ (see Apps.~\ref{app:squeezingΗ} and~\ref{app:morenoise}) is present inside the circuit, an effect first presented here for quantum spin-circuit systems, and one that requires further study. Another distinction from atom-cavity systems lies in the origins of constraints on squeezing. Individual atomic spontaneous emission events in atom-cavity systems typically disturb the collective nature of the state and serve as the origin of ultimate limitation. However, such a loss channel does not exist for NMR systems, given that the sample size is well within the radiation wavelength and is unable to reveal which-atom information. In the case of the circuit decay being dominated by capacitive losses, this shifts the source of limitation eventually to the single spin relaxation events---related to the $T_1$ or $T_\text{deph}$ time.

There are several compelling reasons why the direction proposed in this work could and should be pursued:
\begin{itemize}
    \item The state preparation phase, during which the new physics signal is imprinted, can be separated from the readout phase which may have very different operational requirements;
    \item High-$Q$ LC circuits are under active development, and in combination with large detunings, radiation damping is no longer a limiting factor;
    \item As long as squeezing happens sufficiently fast, squeezing protocols such as one-axis twisting (OAT) are not sensitive to dephasing effects, which has traditionally been a major limitation in NMR setups. The large number of spins involved could further protect such protocols from dephasing effects, as recent experimental results in atomic systems suggest \cite{norcia2018,niu2025};
    \item The state preparation phase can include a magnification stage, which further facilitates the subsequent readout;
    \item The maximum amount of squeezing that can be achieved depends on whether the circuit decay is dominated by capacitive or inductive losses. This is relavant away from the RWA regime and needs to be studied in the context of a concrete setup;
    \item Besides improving the intrinsic SNR of these systems, squeezing can also improve the frequency scanning speed for new physics searches, such as those for axions;
    \item Finally—and perhaps most importantly—an ECSS is intrinsically sensitive to superradiant interactions with a broad class of cosmic relics beyond axions.
\end{itemize}

To conclude, we consider this work an initial step toward realizing quantum metrology protocols in NMR systems that surpass the performances demonstrated in atomic systems operating at much higher energies. While we focused on vacuum squeezing, we are confident that other protocols can be devised to further enhance the achievable squeezing. Taken together with the potential of superradiant interactions, NMR systems may evolve into a transformative new class of quantum-enhanced, ultra-low-threshold detectors.

\vspace{0.5cm}

\begin{acknowledgments}
We are grateful to Kent Irwin for illuminating discussions, and Monika Schleier-Smith, Mark Kasevich, Jason Hogan, Nicholas Rapidis, Michael Fedderke, Andy Geraci, Giorgio Gratta, David Schuster, Dima Budker, and Alex Sushkov for very valuable conversations.

OH acknowledges support by the Institute of Science and Technology Austria (ISTA). AA, SD and MG acknowledge the hospitality provided by ISTA, where part of this work was being completed.

AA, MG and OH acknowledge the hospitality provided by the Leinweber Institute for Theoretical Physics at Stanford, where part of this work was being completed.

SD acknowledges support
by NSF Grant PHY-2310429 and the Gordon and Betty
Moore Foundation Grant GBMF7946, and Perimeter Institute and  its hospitality. 

Research at Perimeter Institute is supported in part by the Government of Canada through the Department of Innovation, Science and Economic Development and by the Province of Ontario through the Ministry of Colleges and Universities. AA is grateful for the support of the Stavros Niarchos Foundation.

\end{acknowledgments}

\bibliography{biblio}

\begin{thebibliography}{64}%
\makeatletter
\providecommand \@ifxundefined [1]{%
 \@ifx{#1\undefined}
}%
\providecommand \@ifnum [1]{%
 \ifnum #1\expandafter \@firstoftwo
 \else \expandafter \@secondoftwo
 \fi
}%
\providecommand \@ifx [1]{%
 \ifx #1\expandafter \@firstoftwo
 \else \expandafter \@secondoftwo
 \fi
}%
\providecommand \natexlab [1]{#1}%
\providecommand \enquote  [1]{``#1''}%
\providecommand \bibnamefont  [1]{#1}%
\providecommand \bibfnamefont [1]{#1}%
\providecommand \citenamefont [1]{#1}%
\providecommand \href@noop [0]{\@secondoftwo}%
\providecommand \href [0]{\begingroup \@sanitize@url \@href}%
\providecommand \@href[1]{\@@startlink{#1}\@@href}%
\providecommand \@@href[1]{\endgroup#1\@@endlink}%
\providecommand \@sanitize@url [0]{\catcode `\\12\catcode `\$12\catcode
  `\&12\catcode `\#12\catcode `\^12\catcode `\_12\catcode `\%12\relax}%
\providecommand \@@startlink[1]{}%
\providecommand \@@endlink[0]{}%
\providecommand \url  [0]{\begingroup\@sanitize@url \@url }%
\providecommand \@url [1]{\endgroup\@href {#1}{\urlprefix }}%
\providecommand \urlprefix  [0]{URL }%
\providecommand \Eprint [0]{\href }%
\providecommand \doibase [0]{https://doi.org/}%
\providecommand \selectlanguage [0]{\@gobble}%
\providecommand \bibinfo  [0]{\@secondoftwo}%
\providecommand \bibfield  [0]{\@secondoftwo}%
\providecommand \translation [1]{[#1]}%
\providecommand \BibitemOpen [0]{}%
\providecommand \bibitemStop [0]{}%
\providecommand \bibitemNoStop [0]{.\EOS\space}%
\providecommand \EOS [0]{\spacefactor3000\relax}%
\providecommand \BibitemShut  [1]{\csname bibitem#1\endcsname}%
\let\auto@bib@innerbib\@empty
%</preamble>
\bibitem [{\citenamefont {Arvanitaki}\ \emph {et~al.}(2025)\citenamefont
  {Arvanitaki}, \citenamefont {Dimopoulos},\ and\ \citenamefont
  {Galanis}}]{Arvanitaki:2024taq}%
  \BibitemOpen
  \bibfield  {author} {\bibinfo {author} {\bibfnamefont {A.}~\bibnamefont
  {Arvanitaki}}, \bibinfo {author} {\bibfnamefont {S.}~\bibnamefont
  {Dimopoulos}},\ and\ \bibinfo {author} {\bibfnamefont {M.}~\bibnamefont
  {Galanis}},\ }\bibfield  {title} {\bibinfo {title} {{Superradiant
  interactions of the cosmic neutrino background, axions, dark matter, and
  reactor neutrinos}},\ }\href {https://doi.org/10.1103/PhysRevD.111.055015}
  {\bibfield  {journal} {\bibinfo  {journal} {Phys. Rev. D}\ }\textbf {\bibinfo
  {volume} {111}},\ \bibinfo {pages} {055015} (\bibinfo {year} {2025})},\
  \Eprint {https://arxiv.org/abs/2408.04021} {arXiv:2408.04021 [hep-ph]}
  \BibitemShut {NoStop}%
\bibitem [{\citenamefont {Hosten}\ \emph
  {et~al.}(2016{\natexlab{a}})\citenamefont {Hosten}, \citenamefont {Engelsen},
  \citenamefont {Krishnakumar},\ and\ \citenamefont {Kasevich}}]{Hosten2016}%
  \BibitemOpen
  \bibfield  {author} {\bibinfo {author} {\bibfnamefont {O.}~\bibnamefont
  {Hosten}}, \bibinfo {author} {\bibfnamefont {N.~J.}\ \bibnamefont
  {Engelsen}}, \bibinfo {author} {\bibfnamefont {R.}~\bibnamefont
  {Krishnakumar}},\ and\ \bibinfo {author} {\bibfnamefont {M.~A.}\ \bibnamefont
  {Kasevich}},\ }\bibfield  {title} {\bibinfo {title} {Measurement noise 100
  times lower than the quantum-projection limit using entangled atoms},\ }\href
  {https://doi.org/10.1038/nature16176} {\bibfield  {journal} {\bibinfo
  {journal} {Nature}\ }\textbf {\bibinfo {volume} {529}},\ \bibinfo {pages}
  {505} (\bibinfo {year} {2016}{\natexlab{a}})}\BibitemShut {NoStop}%
\bibitem [{\citenamefont {Cox}\ \emph {et~al.}(2016)\citenamefont {Cox},
  \citenamefont {Greve}, \citenamefont {Weiner},\ and\ \citenamefont
  {Thompson}}]{Cox2016}%
  \BibitemOpen
  \bibfield  {author} {\bibinfo {author} {\bibfnamefont {K.~C.}\ \bibnamefont
  {Cox}}, \bibinfo {author} {\bibfnamefont {G.~P.}\ \bibnamefont {Greve}},
  \bibinfo {author} {\bibfnamefont {J.~M.}\ \bibnamefont {Weiner}},\ and\
  \bibinfo {author} {\bibfnamefont {J.~K.}\ \bibnamefont {Thompson}},\
  }\bibfield  {title} {\bibinfo {title} {Deterministic squeezed states with
  collective measurements and feedback},\ }\href
  {https://doi.org/10.1103/PhysRevLett.116.093602} {\bibfield  {journal}
  {\bibinfo  {journal} {Phys. Rev. Lett.}\ }\textbf {\bibinfo {volume} {116}},\
  \bibinfo {pages} {093602} (\bibinfo {year} {2016})}\BibitemShut {NoStop}%
\bibitem [{\citenamefont {Pezz\`e}\ \emph {et~al.}(2018)\citenamefont
  {Pezz\`e}, \citenamefont {Smerzi}, \citenamefont {Oberthaler}, \citenamefont
  {Schmied},\ and\ \citenamefont {Treutlein}}]{Pezze18}%
  \BibitemOpen
  \bibfield  {author} {\bibinfo {author} {\bibfnamefont {L.}~\bibnamefont
  {Pezz\`e}}, \bibinfo {author} {\bibfnamefont {A.}~\bibnamefont {Smerzi}},
  \bibinfo {author} {\bibfnamefont {M.~K.}\ \bibnamefont {Oberthaler}},
  \bibinfo {author} {\bibfnamefont {R.}~\bibnamefont {Schmied}},\ and\ \bibinfo
  {author} {\bibfnamefont {P.}~\bibnamefont {Treutlein}},\ }\bibfield  {title}
  {\bibinfo {title} {Quantum metrology with nonclassical states of atomic
  ensembles},\ }\href {https://doi.org/10.1103/RevModPhys.90.035005} {\bibfield
   {journal} {\bibinfo  {journal} {Rev. Mod. Phys.}\ }\textbf {\bibinfo
  {volume} {90}},\ \bibinfo {pages} {035005} (\bibinfo {year}
  {2018})}\BibitemShut {NoStop}%
\bibitem [{\citenamefont {Szigeti}\ \emph {et~al.}(2021)\citenamefont
  {Szigeti}, \citenamefont {Hosten},\ and\ \citenamefont
  {Haine}}]{Szigeti2021}%
  \BibitemOpen
  \bibfield  {author} {\bibinfo {author} {\bibfnamefont {S.~S.}\ \bibnamefont
  {Szigeti}}, \bibinfo {author} {\bibfnamefont {O.}~\bibnamefont {Hosten}},\
  and\ \bibinfo {author} {\bibfnamefont {S.~A.}\ \bibnamefont {Haine}},\
  }\bibfield  {title} {\bibinfo {title} {Improving cold-atom sensors with
  quantum entanglement: Prospects and challenges},\ }\href@noop {} {\bibfield
  {journal} {\bibinfo  {journal} {Applied Physics Letters}\ }\textbf {\bibinfo
  {volume} {118}} (\bibinfo {year} {2021})}\BibitemShut {NoStop}%
\bibitem [{\citenamefont {Budker}\ \emph {et~al.}(2014)\citenamefont {Budker},
  \citenamefont {Graham}, \citenamefont {Ledbetter}, \citenamefont
  {Rajendran},\ and\ \citenamefont {Sushkov}}]{Budker:2013hfa}%
  \BibitemOpen
  \bibfield  {author} {\bibinfo {author} {\bibfnamefont {D.}~\bibnamefont
  {Budker}}, \bibinfo {author} {\bibfnamefont {P.~W.}\ \bibnamefont {Graham}},
  \bibinfo {author} {\bibfnamefont {M.}~\bibnamefont {Ledbetter}}, \bibinfo
  {author} {\bibfnamefont {S.}~\bibnamefont {Rajendran}},\ and\ \bibinfo
  {author} {\bibfnamefont {A.}~\bibnamefont {Sushkov}},\ }\bibfield  {title}
  {\bibinfo {title} {{Proposal for a Cosmic Axion Spin Precession Experiment
  (CASPEr)}},\ }\href {https://doi.org/10.1103/PhysRevX.4.021030} {\bibfield
  {journal} {\bibinfo  {journal} {Phys. Rev. X}\ }\textbf {\bibinfo {volume}
  {4}},\ \bibinfo {pages} {021030} (\bibinfo {year} {2014})},\ \Eprint
  {https://arxiv.org/abs/1306.6089} {arXiv:1306.6089 [hep-ph]} \BibitemShut
  {NoStop}%
\bibitem [{\citenamefont {Arvanitaki}\ and\ \citenamefont
  {Geraci}(2014)}]{Arvanitaki:2014dfa}%
  \BibitemOpen
  \bibfield  {author} {\bibinfo {author} {\bibfnamefont {A.}~\bibnamefont
  {Arvanitaki}}\ and\ \bibinfo {author} {\bibfnamefont {A.~A.}\ \bibnamefont
  {Geraci}},\ }\bibfield  {title} {\bibinfo {title} {{Resonantly Detecting
  Axion-Mediated Forces with Nuclear Magnetic Resonance}},\ }\href
  {https://doi.org/10.1103/PhysRevLett.113.161801} {\bibfield  {journal}
  {\bibinfo  {journal} {Phys. Rev. Lett.}\ }\textbf {\bibinfo {volume} {113}},\
  \bibinfo {pages} {161801} (\bibinfo {year} {2014})},\ \Eprint
  {https://arxiv.org/abs/1403.1290} {arXiv:1403.1290 [hep-ph]} \BibitemShut
  {NoStop}%
\bibitem [{\citenamefont {Rollano}\ \emph {et~al.}(2022)\citenamefont
  {Rollano}, \citenamefont {de~Ory}, \citenamefont {Buch}, \citenamefont
  {Rub{\'\i}n-Osanz}, \citenamefont {Zueco}, \citenamefont
  {S{\'a}nchez-Azqueta}, \citenamefont {Chiesa}, \citenamefont {Granados},
  \citenamefont {Carretta}, \citenamefont {Gomez} \emph
  {et~al.}}]{rollano2022}%
  \BibitemOpen
  \bibfield  {author} {\bibinfo {author} {\bibfnamefont {V.}~\bibnamefont
  {Rollano}}, \bibinfo {author} {\bibfnamefont {M.~C.}\ \bibnamefont {de~Ory}},
  \bibinfo {author} {\bibfnamefont {C.~D.}\ \bibnamefont {Buch}}, \bibinfo
  {author} {\bibfnamefont {M.}~\bibnamefont {Rub{\'\i}n-Osanz}}, \bibinfo
  {author} {\bibfnamefont {D.}~\bibnamefont {Zueco}}, \bibinfo {author}
  {\bibfnamefont {C.}~\bibnamefont {S{\'a}nchez-Azqueta}}, \bibinfo {author}
  {\bibfnamefont {A.}~\bibnamefont {Chiesa}}, \bibinfo {author} {\bibfnamefont
  {D.}~\bibnamefont {Granados}}, \bibinfo {author} {\bibfnamefont
  {S.}~\bibnamefont {Carretta}}, \bibinfo {author} {\bibfnamefont
  {A.}~\bibnamefont {Gomez}}, \emph {et~al.},\ }\bibfield  {title} {\bibinfo
  {title} {High cooperativity coupling to nuclear spins on a circuit quantum
  electrodynamics architecture},\ }\href@noop {} {\bibfield  {journal}
  {\bibinfo  {journal} {Communications Physics}\ }\textbf {\bibinfo {volume}
  {5}},\ \bibinfo {pages} {246} (\bibinfo {year} {2022})}\BibitemShut {NoStop}%
\bibitem [{\citenamefont {Kitagawa}\ and\ \citenamefont
  {Ueda}(1993)}]{Kitagawa93}%
  \BibitemOpen
  \bibfield  {author} {\bibinfo {author} {\bibfnamefont {M.}~\bibnamefont
  {Kitagawa}}\ and\ \bibinfo {author} {\bibfnamefont {M.}~\bibnamefont
  {Ueda}},\ }\bibfield  {title} {\bibinfo {title} {Squeezed spin states},\
  }\href {https://doi.org/10.1103/PhysRevA.47.5138} {\bibfield  {journal}
  {\bibinfo  {journal} {Phys. Rev. A}\ }\textbf {\bibinfo {volume} {47}},\
  \bibinfo {pages} {5138} (\bibinfo {year} {1993})}\BibitemShut {NoStop}%
\bibitem [{\citenamefont {Norcia}\ \emph {et~al.}(2018)\citenamefont {Norcia},
  \citenamefont {Lewis-Swan}, \citenamefont {Cline}, \citenamefont {Zhu},
  \citenamefont {Rey},\ and\ \citenamefont {Thompson}}]{norcia2018}%
  \BibitemOpen
  \bibfield  {author} {\bibinfo {author} {\bibfnamefont {M.~A.}\ \bibnamefont
  {Norcia}}, \bibinfo {author} {\bibfnamefont {R.~J.}\ \bibnamefont
  {Lewis-Swan}}, \bibinfo {author} {\bibfnamefont {J.~R.}\ \bibnamefont
  {Cline}}, \bibinfo {author} {\bibfnamefont {B.}~\bibnamefont {Zhu}}, \bibinfo
  {author} {\bibfnamefont {A.~M.}\ \bibnamefont {Rey}},\ and\ \bibinfo {author}
  {\bibfnamefont {J.~K.}\ \bibnamefont {Thompson}},\ }\bibfield  {title}
  {\bibinfo {title} {Cavity-mediated collective spin-exchange interactions in a
  strontium superradiant laser},\ }\href@noop {} {\bibfield  {journal}
  {\bibinfo  {journal} {Science}\ }\textbf {\bibinfo {volume} {361}},\ \bibinfo
  {pages} {259} (\bibinfo {year} {2018})}\BibitemShut {NoStop}%
\bibitem [{\citenamefont {Leroux}\ \emph {et~al.}(2010)\citenamefont {Leroux},
  \citenamefont {Schleier-Smith},\ and\ \citenamefont
  {Vuleti{\'c}}}]{leroux2010}%
  \BibitemOpen
  \bibfield  {author} {\bibinfo {author} {\bibfnamefont {I.~D.}\ \bibnamefont
  {Leroux}}, \bibinfo {author} {\bibfnamefont {M.~H.}\ \bibnamefont
  {Schleier-Smith}},\ and\ \bibinfo {author} {\bibfnamefont {V.}~\bibnamefont
  {Vuleti{\'c}}},\ }\bibfield  {title} {\bibinfo {title} {Implementation of
  cavity squeezing of a collective atomic spin},\ }\href@noop {} {\bibfield
  {journal} {\bibinfo  {journal} {Physical Review Letters}\ }\textbf {\bibinfo
  {volume} {104}},\ \bibinfo {pages} {073602} (\bibinfo {year}
  {2010})}\BibitemShut {NoStop}%
\bibitem [{\citenamefont {Hosten}\ \emph
  {et~al.}(2016{\natexlab{b}})\citenamefont {Hosten}, \citenamefont
  {Krishnakumar}, \citenamefont {Engelsen},\ and\ \citenamefont
  {Kasevich}}]{Hosten2016Science}%
  \BibitemOpen
  \bibfield  {author} {\bibinfo {author} {\bibfnamefont {O.}~\bibnamefont
  {Hosten}}, \bibinfo {author} {\bibfnamefont {R.}~\bibnamefont
  {Krishnakumar}}, \bibinfo {author} {\bibfnamefont {N.~J.}\ \bibnamefont
  {Engelsen}},\ and\ \bibinfo {author} {\bibfnamefont {M.~A.}\ \bibnamefont
  {Kasevich}},\ }\bibfield  {title} {\bibinfo {title} {Quantum phase
  magnification},\ }\href {https://doi.org/10.1126/science.aaf3397} {\bibfield
  {journal} {\bibinfo  {journal} {Science}\ }\textbf {\bibinfo {volume}
  {352}},\ \bibinfo {pages} {1552} (\bibinfo {year}
  {2016}{\natexlab{b}})}\BibitemShut {NoStop}%
\bibitem [{\citenamefont {Li}\ \emph {et~al.}(2023)\citenamefont {Li},
  \citenamefont {Colombo}, \citenamefont {Shu}, \citenamefont {Velez},
  \citenamefont {Pilatowsky-Cameo}, \citenamefont {Schmied}, \citenamefont
  {Choi}, \citenamefont {Lukin}, \citenamefont {Pedrozo-Pe{\~n}afiel},\ and\
  \citenamefont {Vuleti{\'c}}}]{li2023}%
  \BibitemOpen
  \bibfield  {author} {\bibinfo {author} {\bibfnamefont {Z.}~\bibnamefont
  {Li}}, \bibinfo {author} {\bibfnamefont {S.}~\bibnamefont {Colombo}},
  \bibinfo {author} {\bibfnamefont {C.}~\bibnamefont {Shu}}, \bibinfo {author}
  {\bibfnamefont {G.}~\bibnamefont {Velez}}, \bibinfo {author} {\bibfnamefont
  {S.}~\bibnamefont {Pilatowsky-Cameo}}, \bibinfo {author} {\bibfnamefont
  {R.}~\bibnamefont {Schmied}}, \bibinfo {author} {\bibfnamefont
  {S.}~\bibnamefont {Choi}}, \bibinfo {author} {\bibfnamefont {M.}~\bibnamefont
  {Lukin}}, \bibinfo {author} {\bibfnamefont {E.}~\bibnamefont
  {Pedrozo-Pe{\~n}afiel}},\ and\ \bibinfo {author} {\bibfnamefont
  {V.}~\bibnamefont {Vuleti{\'c}}},\ }\bibfield  {title} {\bibinfo {title}
  {Improving metrology with quantum scrambling},\ }\href@noop {} {\bibfield
  {journal} {\bibinfo  {journal} {Science}\ }\textbf {\bibinfo {volume}
  {380}},\ \bibinfo {pages} {1381} (\bibinfo {year} {2023})}\BibitemShut
  {NoStop}%
\bibitem [{\citenamefont {Strobel}\ \emph {et~al.}(2014)\citenamefont
  {Strobel}, \citenamefont {Muessel}, \citenamefont {Linnemann}, \citenamefont
  {Zibold}, \citenamefont {Hume}, \citenamefont {Pezz{\`e}}, \citenamefont
  {Smerzi},\ and\ \citenamefont {Oberthaler}}]{strobel2014}%
  \BibitemOpen
  \bibfield  {author} {\bibinfo {author} {\bibfnamefont {H.}~\bibnamefont
  {Strobel}}, \bibinfo {author} {\bibfnamefont {W.}~\bibnamefont {Muessel}},
  \bibinfo {author} {\bibfnamefont {D.}~\bibnamefont {Linnemann}}, \bibinfo
  {author} {\bibfnamefont {T.}~\bibnamefont {Zibold}}, \bibinfo {author}
  {\bibfnamefont {D.~B.}\ \bibnamefont {Hume}}, \bibinfo {author}
  {\bibfnamefont {L.}~\bibnamefont {Pezz{\`e}}}, \bibinfo {author}
  {\bibfnamefont {A.}~\bibnamefont {Smerzi}},\ and\ \bibinfo {author}
  {\bibfnamefont {M.~K.}\ \bibnamefont {Oberthaler}},\ }\bibfield  {title}
  {\bibinfo {title} {Fisher information and entanglement of non-gaussian spin
  states},\ }\href@noop {} {\bibfield  {journal} {\bibinfo  {journal}
  {Science}\ }\textbf {\bibinfo {volume} {345}},\ \bibinfo {pages} {424}
  (\bibinfo {year} {2014})}\BibitemShut {NoStop}%
\bibitem [{\citenamefont {Berrada}\ \emph {et~al.}(2013)\citenamefont
  {Berrada}, \citenamefont {Van~Frank}, \citenamefont {B{\"u}cker},
  \citenamefont {Schumm}, \citenamefont {Schaff},\ and\ \citenamefont
  {Schmiedmayer}}]{berrada2013}%
  \BibitemOpen
  \bibfield  {author} {\bibinfo {author} {\bibfnamefont {T.}~\bibnamefont
  {Berrada}}, \bibinfo {author} {\bibfnamefont {S.}~\bibnamefont {Van~Frank}},
  \bibinfo {author} {\bibfnamefont {R.}~\bibnamefont {B{\"u}cker}}, \bibinfo
  {author} {\bibfnamefont {T.}~\bibnamefont {Schumm}}, \bibinfo {author}
  {\bibfnamefont {J.-F.}\ \bibnamefont {Schaff}},\ and\ \bibinfo {author}
  {\bibfnamefont {J.}~\bibnamefont {Schmiedmayer}},\ }\bibfield  {title}
  {\bibinfo {title} {Integrated mach--zehnder interferometer for bose--einstein
  condensates},\ }\href@noop {} {\bibfield  {journal} {\bibinfo  {journal}
  {Nature communications}\ }\textbf {\bibinfo {volume} {4}},\ \bibinfo {pages}
  {2077} (\bibinfo {year} {2013})}\BibitemShut {NoStop}%
\bibitem [{\citenamefont {Davis}\ \emph {et~al.}(2016)\citenamefont {Davis},
  \citenamefont {Bentsen},\ and\ \citenamefont {Schleier-Smith}}]{Davis16}%
  \BibitemOpen
  \bibfield  {author} {\bibinfo {author} {\bibfnamefont {E.}~\bibnamefont
  {Davis}}, \bibinfo {author} {\bibfnamefont {G.}~\bibnamefont {Bentsen}},\
  and\ \bibinfo {author} {\bibfnamefont {M.}~\bibnamefont {Schleier-Smith}},\
  }\bibfield  {title} {\bibinfo {title} {Approaching the {H}eisenberg limit
  without single-particle detection},\ }\href
  {https://doi.org/10.1103/PhysRevLett.116.053601} {\bibfield  {journal}
  {\bibinfo  {journal} {Phys. Rev. Lett.}\ }\textbf {\bibinfo {volume} {116}},\
  \bibinfo {pages} {053601} (\bibinfo {year} {2016})}\BibitemShut {NoStop}%
\bibitem [{\citenamefont {Backes}\ \emph {et~al.}(2021)\citenamefont {Backes},
  \citenamefont {Palken}, \citenamefont {Kenany}, \citenamefont {Brubaker},
  \citenamefont {Cahn}, \citenamefont {Droster}, \citenamefont {Hilton},
  \citenamefont {Ghosh}, \citenamefont {Jackson}, \citenamefont {Lamoreaux},
  \citenamefont {Leder}, \citenamefont {Lehnert}, \citenamefont {Lewis},
  \citenamefont {Malnou}, \citenamefont {Maruyama}, \citenamefont {Rapidis},
  \citenamefont {Simanovskaia}, \citenamefont {Singh}, \citenamefont {Speller},
  \citenamefont {Urdinaran}, \citenamefont {Vale}, \citenamefont {van
  Assendelft}, \citenamefont {van Bibber},\ and\ \citenamefont
  {Wang}}]{haystacsqueeze}%
  \BibitemOpen
  \bibfield  {author} {\bibinfo {author} {\bibfnamefont {K.~M.}\ \bibnamefont
  {Backes}}, \bibinfo {author} {\bibfnamefont {D.~A.}\ \bibnamefont {Palken}},
  \bibinfo {author} {\bibfnamefont {S.~A.}\ \bibnamefont {Kenany}}, \bibinfo
  {author} {\bibfnamefont {B.~M.}\ \bibnamefont {Brubaker}}, \bibinfo {author}
  {\bibfnamefont {S.~B.}\ \bibnamefont {Cahn}}, \bibinfo {author}
  {\bibfnamefont {A.}~\bibnamefont {Droster}}, \bibinfo {author} {\bibfnamefont
  {G.~C.}\ \bibnamefont {Hilton}}, \bibinfo {author} {\bibfnamefont
  {S.}~\bibnamefont {Ghosh}}, \bibinfo {author} {\bibfnamefont
  {H.}~\bibnamefont {Jackson}}, \bibinfo {author} {\bibfnamefont {S.~K.}\
  \bibnamefont {Lamoreaux}}, \bibinfo {author} {\bibfnamefont {A.~F.}\
  \bibnamefont {Leder}}, \bibinfo {author} {\bibfnamefont {K.~W.}\ \bibnamefont
  {Lehnert}}, \bibinfo {author} {\bibfnamefont {S.~M.}\ \bibnamefont {Lewis}},
  \bibinfo {author} {\bibfnamefont {M.}~\bibnamefont {Malnou}}, \bibinfo
  {author} {\bibfnamefont {R.~H.}\ \bibnamefont {Maruyama}}, \bibinfo {author}
  {\bibfnamefont {N.~M.}\ \bibnamefont {Rapidis}}, \bibinfo {author}
  {\bibfnamefont {M.}~\bibnamefont {Simanovskaia}}, \bibinfo {author}
  {\bibfnamefont {S.}~\bibnamefont {Singh}}, \bibinfo {author} {\bibfnamefont
  {D.~H.}\ \bibnamefont {Speller}}, \bibinfo {author} {\bibfnamefont
  {I.}~\bibnamefont {Urdinaran}}, \bibinfo {author} {\bibfnamefont {L.~R.}\
  \bibnamefont {Vale}}, \bibinfo {author} {\bibfnamefont {E.~C.}\ \bibnamefont
  {van Assendelft}}, \bibinfo {author} {\bibfnamefont {K.}~\bibnamefont {van
  Bibber}},\ and\ \bibinfo {author} {\bibfnamefont {H.}~\bibnamefont {Wang}},\
  }\bibfield  {title} {\bibinfo {title} {A quantum enhanced search for dark
  matter axions},\ }\href {https://doi.org/10.1038/s41586-021-03226-7}
  {\bibfield  {journal} {\bibinfo  {journal} {Nature}\ }\textbf {\bibinfo
  {volume} {590}},\ \bibinfo {pages} {238} (\bibinfo {year}
  {2021})}\BibitemShut {NoStop}%
\bibitem [{\citenamefont {Aker}\ \emph {et~al.}(2022)\citenamefont {Aker} \emph
  {et~al.}}]{KATRIN:2022kkv}%
  \BibitemOpen
  \bibfield  {author} {\bibinfo {author} {\bibfnamefont {M.}~\bibnamefont
  {Aker}} \emph {et~al.} (\bibinfo {collaboration} {KATRIN}),\ }\bibfield
  {title} {\bibinfo {title} {{New Constraint on the Local Relic Neutrino
  Background Overdensity with the First KATRIN Data Runs}},\ }\href
  {https://doi.org/10.1103/PhysRevLett.129.011806} {\bibfield  {journal}
  {\bibinfo  {journal} {Phys. Rev. Lett.}\ }\textbf {\bibinfo {volume} {129}},\
  \bibinfo {pages} {011806} (\bibinfo {year} {2022})},\ \Eprint
  {https://arxiv.org/abs/2202.04587} {arXiv:2202.04587 [nucl-ex]} \BibitemShut
  {NoStop}%
\bibitem [{\citenamefont {Nagaoka}(1909)}]{1390853649533379968}%
  \BibitemOpen
  \bibfield  {author} {\bibinfo {author} {\bibfnamefont {H.}~\bibnamefont
  {Nagaoka}},\ }\bibfield  {title} {\bibinfo {title} {The inductance
  coefficients of solenoids},\ }\href {https://doi.org/10.15083/00037799}
  {\bibfield  {journal} {\bibinfo  {journal} {The journal of the College of
  Science, Imperial University of Tokyo, Japan}\ }\textbf {\bibinfo {volume}
  {27}},\ \bibinfo {pages} {1} (\bibinfo {year} {1909})}\BibitemShut {NoStop}%
\bibitem [{\citenamefont {Hu}\ \emph {et~al.}(2017)\citenamefont {Hu},
  \citenamefont {Chen}, \citenamefont {Vendeiro}, \citenamefont {Urvoy},
  \citenamefont {Braverman},\ and\ \citenamefont {Vuleti{\'c}}}]{hu2017}%
  \BibitemOpen
  \bibfield  {author} {\bibinfo {author} {\bibfnamefont {J.}~\bibnamefont
  {Hu}}, \bibinfo {author} {\bibfnamefont {W.}~\bibnamefont {Chen}}, \bibinfo
  {author} {\bibfnamefont {Z.}~\bibnamefont {Vendeiro}}, \bibinfo {author}
  {\bibfnamefont {A.}~\bibnamefont {Urvoy}}, \bibinfo {author} {\bibfnamefont
  {B.}~\bibnamefont {Braverman}},\ and\ \bibinfo {author} {\bibfnamefont
  {V.}~\bibnamefont {Vuleti{\'c}}},\ }\bibfield  {title} {\bibinfo {title}
  {Vacuum spin squeezing},\ }\href@noop {} {\bibfield  {journal} {\bibinfo
  {journal} {Physical Review A}\ }\textbf {\bibinfo {volume} {96}},\ \bibinfo
  {pages} {050301} (\bibinfo {year} {2017})}\BibitemShut {NoStop}%
\bibitem [{\citenamefont {Colombo}\ \emph {et~al.}(2022)\citenamefont
  {Colombo}, \citenamefont {Pedrozo-Pe{\~n}afiel}, \citenamefont {Adiyatullin},
  \citenamefont {Li}, \citenamefont {Mendez}, \citenamefont {Shu},\ and\
  \citenamefont {Vuleti{\'c}}}]{colombo2022}%
  \BibitemOpen
  \bibfield  {author} {\bibinfo {author} {\bibfnamefont {S.}~\bibnamefont
  {Colombo}}, \bibinfo {author} {\bibfnamefont {E.}~\bibnamefont
  {Pedrozo-Pe{\~n}afiel}}, \bibinfo {author} {\bibfnamefont {A.~F.}\
  \bibnamefont {Adiyatullin}}, \bibinfo {author} {\bibfnamefont
  {Z.}~\bibnamefont {Li}}, \bibinfo {author} {\bibfnamefont {E.}~\bibnamefont
  {Mendez}}, \bibinfo {author} {\bibfnamefont {C.}~\bibnamefont {Shu}},\ and\
  \bibinfo {author} {\bibfnamefont {V.}~\bibnamefont {Vuleti{\'c}}},\
  }\bibfield  {title} {\bibinfo {title} {Time-reversal-based quantum metrology
  with many-body entangled states},\ }\href@noop {} {\bibfield  {journal}
  {\bibinfo  {journal} {Nature Physics}\ }\textbf {\bibinfo {volume} {18}},\
  \bibinfo {pages} {925} (\bibinfo {year} {2022})}\BibitemShut {NoStop}%
\bibitem [{\citenamefont {Arvanitaki}\ \emph {et~al.}(2026)\citenamefont
  {Arvanitaki}, \citenamefont {Dimopoulos}, \citenamefont {Galanis},\ and\
  \citenamefont {Rapidis}}]{upcomingRapidis}%
  \BibitemOpen
  \bibfield  {author} {\bibinfo {author} {\bibfnamefont {A.}~\bibnamefont
  {Arvanitaki}}, \bibinfo {author} {\bibfnamefont {S.}~\bibnamefont
  {Dimopoulos}}, \bibinfo {author} {\bibfnamefont {M.}~\bibnamefont
  {Galanis}},\ and\ \bibinfo {author} {\bibfnamefont {N.}~\bibnamefont
  {Rapidis}},\ }\bibfield  {title} {\bibinfo {title} {Title forthcoming}}
  (\bibinfo {year} {2026}),\ \bibinfo {note} {in preparation}\BibitemShut
  {NoStop}%
\bibitem [{\citenamefont {Newbury}\ \emph {et~al.}(1993)\citenamefont
  {Newbury}, \citenamefont {Barton}, \citenamefont {Cates}, \citenamefont
  {Happer},\ and\ \citenamefont {Middleton}}]{PhysRevA.48.4411}%
  \BibitemOpen
  \bibfield  {author} {\bibinfo {author} {\bibfnamefont {N.~R.}\ \bibnamefont
  {Newbury}}, \bibinfo {author} {\bibfnamefont {A.~S.}\ \bibnamefont {Barton}},
  \bibinfo {author} {\bibfnamefont {G.~D.}\ \bibnamefont {Cates}}, \bibinfo
  {author} {\bibfnamefont {W.}~\bibnamefont {Happer}},\ and\ \bibinfo {author}
  {\bibfnamefont {H.}~\bibnamefont {Middleton}},\ }\bibfield  {title} {\bibinfo
  {title} {Gaseous $^{3}\text{He}$--$^{3} \text{He}$ magnetic dipolar spin
  relaxation},\ }\href {https://doi.org/10.1103/PhysRevA.48.4411} {\bibfield
  {journal} {\bibinfo  {journal} {Phys. Rev. A}\ }\textbf {\bibinfo {volume}
  {48}},\ \bibinfo {pages} {4411} (\bibinfo {year} {1993})}\BibitemShut
  {NoStop}%
\bibitem [{\citenamefont {Kolevatov}\ \emph {et~al.}(2026)\citenamefont
  {Kolevatov}, \citenamefont {Chaudhuri},\ and\ \citenamefont {Page}}]{highq}%
  \BibitemOpen
  \bibfield  {author} {\bibinfo {author} {\bibfnamefont {R.}~\bibnamefont
  {Kolevatov}}, \bibinfo {author} {\bibfnamefont {S.}~\bibnamefont
  {Chaudhuri}},\ and\ \bibinfo {author} {\bibfnamefont {L.}~\bibnamefont
  {Page}},\ }\bibfield  {title} {\bibinfo {title} {High-q superconducting
  lumped-element resonators for low-mass axion searches},\ }\href
  {https://doi.org/10.1063/5.0311286} {\bibfield  {journal} {\bibinfo
  {journal} {Review of Scientific Instruments}\ }\textbf {\bibinfo {volume}
  {97}},\ \bibinfo {pages} {024503} (\bibinfo {year} {2026})}\BibitemShut
  {NoStop}%
\bibitem [{\citenamefont {Irwin}(2026)}]{kentprivate}%
  \BibitemOpen
  \bibfield  {author} {\bibinfo {author} {\bibfnamefont {K.}~\bibnamefont
  {Irwin}},\ }\bibfield  {title} {\bibinfo {title} {Private communication}}
  (\bibinfo {year} {2026}),\ \bibinfo {note} {personal
  communication}\BibitemShut {NoStop}%
\bibitem [{\citenamefont {Muessel}\ \emph {et~al.}(2015)\citenamefont
  {Muessel}, \citenamefont {Strobel}, \citenamefont {Linnemann}, \citenamefont
  {Zibold}, \citenamefont {Juli{\'a}-D{\'\i}az},\ and\ \citenamefont
  {Oberthaler}}]{muessel2015}%
  \BibitemOpen
  \bibfield  {author} {\bibinfo {author} {\bibfnamefont {W.}~\bibnamefont
  {Muessel}}, \bibinfo {author} {\bibfnamefont {H.}~\bibnamefont {Strobel}},
  \bibinfo {author} {\bibfnamefont {D.}~\bibnamefont {Linnemann}}, \bibinfo
  {author} {\bibfnamefont {T.}~\bibnamefont {Zibold}}, \bibinfo {author}
  {\bibfnamefont {B.}~\bibnamefont {Juli{\'a}-D{\'\i}az}},\ and\ \bibinfo
  {author} {\bibfnamefont {M.}~\bibnamefont {Oberthaler}},\ }\bibfield  {title}
  {\bibinfo {title} {Twist-and-turn spin squeezing in bose-einstein
  condensates},\ }\href@noop {} {\bibfield  {journal} {\bibinfo  {journal}
  {Physical Review A}\ }\textbf {\bibinfo {volume} {92}},\ \bibinfo {pages}
  {023603} (\bibinfo {year} {2015})}\BibitemShut {NoStop}%
\bibitem [{\citenamefont {Niu}\ \emph {et~al.}(2025)\citenamefont {Niu},
  \citenamefont {Sch{\"a}fer}, \citenamefont {Zhang}, \citenamefont {Wagner},
  \citenamefont {Taylor}, \citenamefont {Young}, \citenamefont {Song},
  \citenamefont {Chu}, \citenamefont {Rey},\ and\ \citenamefont
  {Thompson}}]{niu2025}%
  \BibitemOpen
  \bibfield  {author} {\bibinfo {author} {\bibfnamefont {Z.}~\bibnamefont
  {Niu}}, \bibinfo {author} {\bibfnamefont {V.~M.}\ \bibnamefont
  {Sch{\"a}fer}}, \bibinfo {author} {\bibfnamefont {H.}~\bibnamefont {Zhang}},
  \bibinfo {author} {\bibfnamefont {C.}~\bibnamefont {Wagner}}, \bibinfo
  {author} {\bibfnamefont {N.~R.}\ \bibnamefont {Taylor}}, \bibinfo {author}
  {\bibfnamefont {D.~J.}\ \bibnamefont {Young}}, \bibinfo {author}
  {\bibfnamefont {E.~Y.}\ \bibnamefont {Song}}, \bibinfo {author}
  {\bibfnamefont {A.}~\bibnamefont {Chu}}, \bibinfo {author} {\bibfnamefont
  {A.~M.}\ \bibnamefont {Rey}},\ and\ \bibinfo {author} {\bibfnamefont {J.~K.}\
  \bibnamefont {Thompson}},\ }\bibfield  {title} {\bibinfo {title} {Many-body
  gap protection against motional dephasing of an optical clock transition},\
  }\href@noop {} {\bibfield  {journal} {\bibinfo  {journal} {Physical Review
  Letters}\ }\textbf {\bibinfo {volume} {134}},\ \bibinfo {pages} {113403}
  (\bibinfo {year} {2025})}\BibitemShut {NoStop}%
\bibitem [{\citenamefont {Sushkov}(2025)}]{sushkovtalk}%
  \BibitemOpen
  \bibfield  {author} {\bibinfo {author} {\bibfnamefont {A.}~\bibnamefont
  {Sushkov}},\ }\href@noop {} {\bibinfo {title} {Quantum-limited magnetic
  resonance detection for the casper axion dark matter search}},\ \bibinfo
  {howpublished} {Talk presented at the workshop ``Enabling the discovery of
  QCD axion Dark Matter at the GUT scale''} (\bibinfo {year} {2025}),\ \bibinfo
  {note} {{Lawrence Berkeley National Laboratory (LBNL)}, May 7--9,
  2025}\BibitemShut {NoStop}%
\bibitem [{\citenamefont {AlShirawi}\ \emph {et~al.}(2023)\citenamefont
  {AlShirawi} \emph {et~al.}}]{DMRadio:2023igr}%
  \BibitemOpen
  \bibfield  {author} {\bibinfo {author} {\bibfnamefont {A.}~\bibnamefont
  {AlShirawi}} \emph {et~al.} (\bibinfo {collaboration} {DMRadio}),\ }\bibfield
   {title} {\bibinfo {title} {{Electromagnetic modeling and science reach of
  DMRadio-m$^3$}},\ }\href@noop {} {\bibfield  {journal} {\bibinfo  {journal}
  {arXiv}\ } (\bibinfo {year} {2023})},\ \Eprint
  {https://arxiv.org/abs/2302.14084} {arXiv:2302.14084 [hep-ex]} \BibitemShut
  {NoStop}%
\bibitem [{\citenamefont {Weinberg}(1978)}]{Weinberg:1977ma}%
  \BibitemOpen
  \bibfield  {author} {\bibinfo {author} {\bibfnamefont {S.}~\bibnamefont
  {Weinberg}},\ }\bibfield  {title} {\bibinfo {title} {{A New Light Boson?}},\
  }\href {https://doi.org/10.1103/PhysRevLett.40.223} {\bibfield  {journal}
  {\bibinfo  {journal} {Phys. Rev. Lett.}\ }\textbf {\bibinfo {volume} {40}},\
  \bibinfo {pages} {223} (\bibinfo {year} {1978})}\BibitemShut {NoStop}%
\bibitem [{\citenamefont {Wilczek}(1978)}]{Wilczek:1977pj}%
  \BibitemOpen
  \bibfield  {author} {\bibinfo {author} {\bibfnamefont {F.}~\bibnamefont
  {Wilczek}},\ }\bibfield  {title} {\bibinfo {title} {{Problem of Strong $P$
  and $T$ Invariance in the Presence of Instantons}},\ }\href
  {https://doi.org/10.1103/PhysRevLett.40.279} {\bibfield  {journal} {\bibinfo
  {journal} {Phys. Rev. Lett.}\ }\textbf {\bibinfo {volume} {40}},\ \bibinfo
  {pages} {279} (\bibinfo {year} {1978})}\BibitemShut {NoStop}%
\bibitem [{\citenamefont {Dine}\ and\ \citenamefont
  {Fischler}(1983)}]{Dine:1982ah}%
  \BibitemOpen
  \bibfield  {author} {\bibinfo {author} {\bibfnamefont {M.}~\bibnamefont
  {Dine}}\ and\ \bibinfo {author} {\bibfnamefont {W.}~\bibnamefont
  {Fischler}},\ }\bibfield  {title} {\bibinfo {title} {{The Not So Harmless
  Axion}},\ }\href {https://doi.org/10.1016/0370-2693(83)90639-1} {\bibfield
  {journal} {\bibinfo  {journal} {Phys. Lett. B}\ }\textbf {\bibinfo {volume}
  {120}},\ \bibinfo {pages} {137} (\bibinfo {year} {1983})}\BibitemShut
  {NoStop}%
\bibitem [{\citenamefont {Preskill}\ \emph {et~al.}(1983)\citenamefont
  {Preskill}, \citenamefont {Wise},\ and\ \citenamefont
  {Wilczek}}]{Preskill:1982cy}%
  \BibitemOpen
  \bibfield  {author} {\bibinfo {author} {\bibfnamefont {J.}~\bibnamefont
  {Preskill}}, \bibinfo {author} {\bibfnamefont {M.~B.}\ \bibnamefont {Wise}},\
  and\ \bibinfo {author} {\bibfnamefont {F.}~\bibnamefont {Wilczek}},\
  }\bibfield  {title} {\bibinfo {title} {{Cosmology of the Invisible Axion}},\
  }\href {https://doi.org/10.1016/0370-2693(83)90637-8} {\bibfield  {journal}
  {\bibinfo  {journal} {Phys. Lett. B}\ }\textbf {\bibinfo {volume} {120}},\
  \bibinfo {pages} {127} (\bibinfo {year} {1983})}\BibitemShut {NoStop}%
\bibitem [{\citenamefont {Turner}(1988)}]{SNTurner}%
  \BibitemOpen
  \bibfield  {author} {\bibinfo {author} {\bibfnamefont {M.~S.}\ \bibnamefont
  {Turner}},\ }\bibfield  {title} {\bibinfo {title} {Axions from sn1987a},\
  }\href {https://doi.org/10.1103/PhysRevLett.60.1797} {\bibfield  {journal}
  {\bibinfo  {journal} {Phys. Rev. Lett.}\ }\textbf {\bibinfo {volume} {60}},\
  \bibinfo {pages} {1797} (\bibinfo {year} {1988})}\BibitemShut {NoStop}%
\bibitem [{\citenamefont {Raffelt}\ and\ \citenamefont
  {Seckel}(1988)}]{SNRaffelt}%
  \BibitemOpen
  \bibfield  {author} {\bibinfo {author} {\bibfnamefont {G.}~\bibnamefont
  {Raffelt}}\ and\ \bibinfo {author} {\bibfnamefont {D.}~\bibnamefont
  {Seckel}},\ }\bibfield  {title} {\bibinfo {title} {Bounds on exotic-particle
  interactions from sn1987a},\ }\href
  {https://doi.org/10.1103/PhysRevLett.60.1793} {\bibfield  {journal} {\bibinfo
   {journal} {Phys. Rev. Lett.}\ }\textbf {\bibinfo {volume} {60}},\ \bibinfo
  {pages} {1793} (\bibinfo {year} {1988})}\BibitemShut {NoStop}%
\bibitem [{\citenamefont {Mayle}\ \emph {et~al.}(1988)\citenamefont {Mayle},
  \citenamefont {Wilson}, \citenamefont {Ellis}, \citenamefont {Olive},
  \citenamefont {Schramm},\ and\ \citenamefont {Steigman}}]{SNOlive}%
  \BibitemOpen
  \bibfield  {author} {\bibinfo {author} {\bibfnamefont {R.}~\bibnamefont
  {Mayle}}, \bibinfo {author} {\bibfnamefont {J.~R.}\ \bibnamefont {Wilson}},
  \bibinfo {author} {\bibfnamefont {J.}~\bibnamefont {Ellis}}, \bibinfo
  {author} {\bibfnamefont {K.}~\bibnamefont {Olive}}, \bibinfo {author}
  {\bibfnamefont {D.~N.}\ \bibnamefont {Schramm}},\ and\ \bibinfo {author}
  {\bibfnamefont {G.}~\bibnamefont {Steigman}},\ }\bibfield  {title} {\bibinfo
  {title} {Constraints on axions from sn 1987a},\ }\href
  {https://doi.org/https://doi.org/10.1016/0370-2693(88)91595-X} {\bibfield
  {journal} {\bibinfo  {journal} {Physics Letters B}\ }\textbf {\bibinfo
  {volume} {203}},\ \bibinfo {pages} {188} (\bibinfo {year}
  {1988})}\BibitemShut {NoStop}%
\bibitem [{\citenamefont {Lella}\ \emph {et~al.}(2024)\citenamefont {Lella},
  \citenamefont {Carenza}, \citenamefont {Co'}, \citenamefont {Lucente},
  \citenamefont {Giannotti}, \citenamefont {Mirizzi},\ and\ \citenamefont
  {Rauscher}}]{SNRecent}%
  \BibitemOpen
  \bibfield  {author} {\bibinfo {author} {\bibfnamefont {A.}~\bibnamefont
  {Lella}}, \bibinfo {author} {\bibfnamefont {P.}~\bibnamefont {Carenza}},
  \bibinfo {author} {\bibfnamefont {G.}~\bibnamefont {Co'}}, \bibinfo {author}
  {\bibfnamefont {G.}~\bibnamefont {Lucente}}, \bibinfo {author} {\bibfnamefont
  {M.}~\bibnamefont {Giannotti}}, \bibinfo {author} {\bibfnamefont
  {A.}~\bibnamefont {Mirizzi}},\ and\ \bibinfo {author} {\bibfnamefont
  {T.}~\bibnamefont {Rauscher}},\ }\bibfield  {title} {\bibinfo {title}
  {Getting the most on supernova axions},\ }\href
  {https://doi.org/10.1103/PhysRevD.109.023001} {\bibfield  {journal} {\bibinfo
   {journal} {Phys. Rev. D}\ }\textbf {\bibinfo {volume} {109}},\ \bibinfo
  {pages} {023001} (\bibinfo {year} {2024})}\BibitemShut {NoStop}%
\bibitem [{\citenamefont {Baryakhtar}\ \emph {et~al.}(2021)\citenamefont
  {Baryakhtar}, \citenamefont {Galanis}, \citenamefont {Lasenby},\ and\
  \citenamefont {Simon}}]{BHSR-self-interactions}%
  \BibitemOpen
  \bibfield  {author} {\bibinfo {author} {\bibfnamefont {M.}~\bibnamefont
  {Baryakhtar}}, \bibinfo {author} {\bibfnamefont {M.}~\bibnamefont {Galanis}},
  \bibinfo {author} {\bibfnamefont {R.}~\bibnamefont {Lasenby}},\ and\ \bibinfo
  {author} {\bibfnamefont {O.}~\bibnamefont {Simon}},\ }\bibfield  {title}
  {\bibinfo {title} {{Black hole superradiance of self-interacting scalar
  fields}},\ }\href@noop {} {\bibfield  {journal} {\bibinfo  {journal} {Phys.
  Rev. D}\ }\textbf {\bibinfo {volume} {103}},\ \bibinfo {pages} {095019}
  (\bibinfo {year} {2021})},\ \Eprint {https://arxiv.org/abs/2011.11646}
  {arXiv:2011.11646 [hep-ph]} \BibitemShut {NoStop}%
\bibitem [{\citenamefont {Krauss}\ \emph {et~al.}(1985)\citenamefont {Krauss},
  \citenamefont {Moody}, \citenamefont {Wilczek},\ and\ \citenamefont
  {Morris}}]{krauss1985spin}%
  \BibitemOpen
  \bibfield  {author} {\bibinfo {author} {\bibfnamefont {L.}~\bibnamefont
  {Krauss}}, \bibinfo {author} {\bibfnamefont {J.}~\bibnamefont {Moody}},
  \bibinfo {author} {\bibfnamefont {F.}~\bibnamefont {Wilczek}},\ and\ \bibinfo
  {author} {\bibfnamefont {D.}~\bibnamefont {Morris}},\ }\bibfield  {title}
  {\bibinfo {title} {Spin coupled axion detection},\ }\href@noop {} {\bibfield
  {journal} {\bibinfo  {journal} {Preprint HUTP-85 A}\ }\textbf {\bibinfo
  {volume} {6}},\ \bibinfo {pages} {1985} (\bibinfo {year} {1985})}\BibitemShut
  {NoStop}%
\bibitem [{\citenamefont {Moody}\ and\ \citenamefont
  {Wilczek}(1984)}]{Wilczek-spin-forces}%
  \BibitemOpen
  \bibfield  {author} {\bibinfo {author} {\bibfnamefont {J.~E.}\ \bibnamefont
  {Moody}}\ and\ \bibinfo {author} {\bibfnamefont {F.}~\bibnamefont
  {Wilczek}},\ }\bibfield  {title} {\bibinfo {title} {New macroscopic
  forces?},\ }\href@noop {} {\bibfield  {journal} {\bibinfo  {journal} {Phys.
  Rev. D}\ }\textbf {\bibinfo {volume} {30}},\ \bibinfo {pages} {130} (\bibinfo
  {year} {1984})}\BibitemShut {NoStop}%
\bibitem [{\citenamefont {Caputo}\ \emph {et~al.}(2021)\citenamefont {Caputo},
  \citenamefont {O'Hare}, \citenamefont {Millar},\ and\ \citenamefont
  {Vitagliano}}]{Caputo:2021eaa}%
  \BibitemOpen
  \bibfield  {author} {\bibinfo {author} {\bibfnamefont {A.}~\bibnamefont
  {Caputo}}, \bibinfo {author} {\bibfnamefont {C.~A.~J.}\ \bibnamefont
  {O'Hare}}, \bibinfo {author} {\bibfnamefont {A.~J.}\ \bibnamefont {Millar}},\
  and\ \bibinfo {author} {\bibfnamefont {E.}~\bibnamefont {Vitagliano}},\
  }\bibfield  {title} {\bibinfo {title} {{Dark photon limits: a cookbook}},\
  }\href@noop {} {\bibfield  {journal} {\bibinfo  {journal} {arXiv}\ }
  (\bibinfo {year} {2021})},\ \Eprint {https://arxiv.org/abs/2105.04565}
  {arXiv:2105.04565 [hep-ph]} \BibitemShut {NoStop}%
\bibitem [{\citenamefont {Chaudhuri}\ \emph {et~al.}(2015)\citenamefont
  {Chaudhuri}, \citenamefont {Graham}, \citenamefont {Irwin}, \citenamefont
  {Mardon}, \citenamefont {Rajendran},\ and\ \citenamefont
  {Zhao}}]{DMradiooriginal}%
  \BibitemOpen
  \bibfield  {author} {\bibinfo {author} {\bibfnamefont {S.}~\bibnamefont
  {Chaudhuri}}, \bibinfo {author} {\bibfnamefont {P.~W.}\ \bibnamefont
  {Graham}}, \bibinfo {author} {\bibfnamefont {K.}~\bibnamefont {Irwin}},
  \bibinfo {author} {\bibfnamefont {J.}~\bibnamefont {Mardon}}, \bibinfo
  {author} {\bibfnamefont {S.}~\bibnamefont {Rajendran}},\ and\ \bibinfo
  {author} {\bibfnamefont {Y.}~\bibnamefont {Zhao}},\ }\bibfield  {title}
  {\bibinfo {title} {{Radio for hidden-photon dark matter detection}},\ }\href
  {https://doi.org/10.1103/PhysRevD.92.075012} {\bibfield  {journal} {\bibinfo
  {journal} {Phys. Rev. D}\ }\textbf {\bibinfo {volume} {92}},\ \bibinfo
  {pages} {075012} (\bibinfo {year} {2015})},\ \Eprint
  {https://arxiv.org/abs/1411.7382} {arXiv:1411.7382 [hep-ph]} \BibitemShut
  {NoStop}%
\bibitem [{\citenamefont {Grassellino}\ \emph {et~al.}(2020)\citenamefont
  {Grassellino}, \citenamefont {Harnik}, \citenamefont {Liu},\ and\
  \citenamefont {Romanenko}}]{DarkSRF}%
  \BibitemOpen
  \bibfield  {author} {\bibinfo {author} {\bibfnamefont {A.}~\bibnamefont
  {Grassellino}}, \bibinfo {author} {\bibfnamefont {R.}~\bibnamefont {Harnik}},
  \bibinfo {author} {\bibfnamefont {Z.}~\bibnamefont {Liu}},\ and\ \bibinfo
  {author} {\bibfnamefont {A.}~\bibnamefont {Romanenko}},\ }\href@noop {}
  {\bibinfo {title} {{First results of Dark SRF: a dark photon search with SRF
  cavities}}},\ \bibinfo {howpublished} {Presentation at Aspen Center for
  Physics (Aspen 2)} (\bibinfo {year} {2020}),\ \bibinfo {note}
  {\url{https://indico.physics.lbl.gov/event/939/contributions/4371/attachments/2162/2915/DarkSRF-Aspen-2.pdf}}\BibitemShut
  {NoStop}%
\bibitem [{\citenamefont {Colegrove}\ \emph {et~al.}(1963)\citenamefont
  {Colegrove}, \citenamefont {Schearer},\ and\ \citenamefont
  {Walters}}]{PhysRev.132.2561}%
  \BibitemOpen
  \bibfield  {author} {\bibinfo {author} {\bibfnamefont {F.~D.}\ \bibnamefont
  {Colegrove}}, \bibinfo {author} {\bibfnamefont {L.~D.}\ \bibnamefont
  {Schearer}},\ and\ \bibinfo {author} {\bibfnamefont {G.~K.}\ \bibnamefont
  {Walters}},\ }\bibfield  {title} {\bibinfo {title} {Polarization of
  ${\mathrm{he}}^{3}$ gas by optical pumping},\ }\href
  {https://doi.org/10.1103/PhysRev.132.2561} {\bibfield  {journal} {\bibinfo
  {journal} {Phys. Rev.}\ }\textbf {\bibinfo {volume} {132}},\ \bibinfo {pages}
  {2561} (\bibinfo {year} {1963})}\BibitemShut {NoStop}%
\bibitem [{\citenamefont {Happer}\ \emph {et~al.}(1984)\citenamefont {Happer},
  \citenamefont {Miron}, \citenamefont {Schaefer}, \citenamefont {Schreiber},
  \citenamefont {van Wijngaarden},\ and\ \citenamefont
  {Zeng}}]{PhysRevA.29.3092}%
  \BibitemOpen
  \bibfield  {author} {\bibinfo {author} {\bibfnamefont {W.}~\bibnamefont
  {Happer}}, \bibinfo {author} {\bibfnamefont {E.}~\bibnamefont {Miron}},
  \bibinfo {author} {\bibfnamefont {S.}~\bibnamefont {Schaefer}}, \bibinfo
  {author} {\bibfnamefont {D.}~\bibnamefont {Schreiber}}, \bibinfo {author}
  {\bibfnamefont {W.~A.}\ \bibnamefont {van Wijngaarden}},\ and\ \bibinfo
  {author} {\bibfnamefont {X.}~\bibnamefont {Zeng}},\ }\bibfield  {title}
  {\bibinfo {title} {Polarization of the nuclear spins of noble-gas atoms by
  spin exchange with optically pumped alkali-metal atoms},\ }\href
  {https://doi.org/10.1103/PhysRevA.29.3092} {\bibfield  {journal} {\bibinfo
  {journal} {Phys. Rev. A}\ }\textbf {\bibinfo {volume} {29}},\ \bibinfo
  {pages} {3092} (\bibinfo {year} {1984})}\BibitemShut {NoStop}%
\bibitem [{\citenamefont {Hu}\ \emph {et~al.}(2015)\citenamefont {Hu},
  \citenamefont {Chen}, \citenamefont {Vendeiro}, \citenamefont {Zhang},\ and\
  \citenamefont {Vuleti{\'c}}}]{hu2015}%
  \BibitemOpen
  \bibfield  {author} {\bibinfo {author} {\bibfnamefont {J.}~\bibnamefont
  {Hu}}, \bibinfo {author} {\bibfnamefont {W.}~\bibnamefont {Chen}}, \bibinfo
  {author} {\bibfnamefont {Z.}~\bibnamefont {Vendeiro}}, \bibinfo {author}
  {\bibfnamefont {H.}~\bibnamefont {Zhang}},\ and\ \bibinfo {author}
  {\bibfnamefont {V.}~\bibnamefont {Vuleti{\'c}}},\ }\bibfield  {title}
  {\bibinfo {title} {Entangled collective-spin states of atomic ensembles under
  nonuniform atom-light interaction},\ }\href@noop {} {\bibfield  {journal}
  {\bibinfo  {journal} {Physical Review A}\ }\textbf {\bibinfo {volume} {92}},\
  \bibinfo {pages} {063816} (\bibinfo {year} {2015})}\BibitemShut {NoStop}%
\bibitem [{\citenamefont {Wu}\ \emph {et~al.}(2020)\citenamefont {Wu},
  \citenamefont {Krishnakumar}, \citenamefont {Mart{\'\i}nez-Rinc{\'o}n},
  \citenamefont {Malia}, \citenamefont {Hosten},\ and\ \citenamefont
  {Kasevich}}]{wu2020}%
  \BibitemOpen
  \bibfield  {author} {\bibinfo {author} {\bibfnamefont {Y.}~\bibnamefont
  {Wu}}, \bibinfo {author} {\bibfnamefont {R.}~\bibnamefont {Krishnakumar}},
  \bibinfo {author} {\bibfnamefont {J.}~\bibnamefont
  {Mart{\'\i}nez-Rinc{\'o}n}}, \bibinfo {author} {\bibfnamefont {B.~K.}\
  \bibnamefont {Malia}}, \bibinfo {author} {\bibfnamefont {O.}~\bibnamefont
  {Hosten}},\ and\ \bibinfo {author} {\bibfnamefont {M.~A.}\ \bibnamefont
  {Kasevich}},\ }\bibfield  {title} {\bibinfo {title} {Retrieval of
  cavity-generated atomic spin squeezing after free-space release},\
  }\href@noop {} {\bibfield  {journal} {\bibinfo  {journal} {Physical Review
  A}\ }\textbf {\bibinfo {volume} {102}},\ \bibinfo {pages} {012224} (\bibinfo
  {year} {2020})}\BibitemShut {NoStop}%
\bibitem [{\citenamefont {Diddams}\ \emph {et~al.}(2020)\citenamefont
  {Diddams}, \citenamefont {Vahala},\ and\ \citenamefont {Udem}}]{diddams2020}%
  \BibitemOpen
  \bibfield  {author} {\bibinfo {author} {\bibfnamefont {S.~A.}\ \bibnamefont
  {Diddams}}, \bibinfo {author} {\bibfnamefont {K.}~\bibnamefont {Vahala}},\
  and\ \bibinfo {author} {\bibfnamefont {T.}~\bibnamefont {Udem}},\ }\bibfield
  {title} {\bibinfo {title} {Optical frequency combs: Coherently uniting the
  electromagnetic spectrum},\ }\href@noop {} {\bibfield  {journal} {\bibinfo
  {journal} {Science}\ }\textbf {\bibinfo {volume} {369}},\ \bibinfo {pages}
  {eaay3676} (\bibinfo {year} {2020})}\BibitemShut {NoStop}%
\bibitem [{\citenamefont {Xie}\ \emph {et~al.}(2017)\citenamefont {Xie},
  \citenamefont {Bouchand}, \citenamefont {Nicolodi}, \citenamefont {Giunta},
  \citenamefont {H{\"a}nsel}, \citenamefont {Lezius}, \citenamefont {Joshi},
  \citenamefont {Datta}, \citenamefont {Alexandre}, \citenamefont {Lours} \emph
  {et~al.}}]{xie2017}%
  \BibitemOpen
  \bibfield  {author} {\bibinfo {author} {\bibfnamefont {X.}~\bibnamefont
  {Xie}}, \bibinfo {author} {\bibfnamefont {R.}~\bibnamefont {Bouchand}},
  \bibinfo {author} {\bibfnamefont {D.}~\bibnamefont {Nicolodi}}, \bibinfo
  {author} {\bibfnamefont {M.}~\bibnamefont {Giunta}}, \bibinfo {author}
  {\bibfnamefont {W.}~\bibnamefont {H{\"a}nsel}}, \bibinfo {author}
  {\bibfnamefont {M.}~\bibnamefont {Lezius}}, \bibinfo {author} {\bibfnamefont
  {A.}~\bibnamefont {Joshi}}, \bibinfo {author} {\bibfnamefont
  {S.}~\bibnamefont {Datta}}, \bibinfo {author} {\bibfnamefont
  {C.}~\bibnamefont {Alexandre}}, \bibinfo {author} {\bibfnamefont
  {M.}~\bibnamefont {Lours}}, \emph {et~al.},\ }\bibfield  {title} {\bibinfo
  {title} {Photonic microwave signals with zeptosecond-level absolute timing
  noise},\ }\href@noop {} {\bibfield  {journal} {\bibinfo  {journal} {nature
  photonics}\ }\textbf {\bibinfo {volume} {11}},\ \bibinfo {pages} {44}
  (\bibinfo {year} {2017})}\BibitemShut {NoStop}%
\bibitem [{noi(2009)}]{noise_calculator}%
  \BibitemOpen
  \href@noop {} {\bibinfo {title} {Marki microwave phase noise calculator}},\
  \bibinfo {howpublished}
  {\url{https://markimicrowave.com/technical-resources/tools/phase-noise-jitter-calculator/}}
  (\bibinfo {year} {2009})\BibitemShut {NoStop}%
\bibitem [{\citenamefont {Yamamoto}\ \emph {et~al.}(2014)\citenamefont
  {Yamamoto}, \citenamefont {Konii}, \citenamefont {Tanabe}, \citenamefont
  {Yokoyama}, \citenamefont {Matsuda},\ and\ \citenamefont {Yamada}}]{6612716}%
  \BibitemOpen
  \bibfield  {author} {\bibinfo {author} {\bibfnamefont {S.}~\bibnamefont
  {Yamamoto}}, \bibinfo {author} {\bibfnamefont {K.}~\bibnamefont {Konii}},
  \bibinfo {author} {\bibfnamefont {H.}~\bibnamefont {Tanabe}}, \bibinfo
  {author} {\bibfnamefont {S.}~\bibnamefont {Yokoyama}}, \bibinfo {author}
  {\bibfnamefont {T.}~\bibnamefont {Matsuda}},\ and\ \bibinfo {author}
  {\bibfnamefont {T.}~\bibnamefont {Yamada}},\ }\bibfield  {title} {\bibinfo
  {title} {Super-stable superconducting mri magnet operating for 25 years},\
  }\href {https://doi.org/10.1109/TASC.2013.2283228} {\bibfield  {journal}
  {\bibinfo  {journal} {IEEE Transactions on Applied Superconductivity}\
  }\textbf {\bibinfo {volume} {24}},\ \bibinfo {pages} {1} (\bibinfo {year}
  {2014})}\BibitemShut {NoStop}%
\bibitem [{\citenamefont {Takeda}\ \emph {et~al.}(2022)\citenamefont {Takeda},
  \citenamefont {Maeda}, \citenamefont {Ohki},\ and\ \citenamefont
  {Yanagisawa}}]{Takeda_2022}%
  \BibitemOpen
  \bibfield  {author} {\bibinfo {author} {\bibfnamefont {Y.}~\bibnamefont
  {Takeda}}, \bibinfo {author} {\bibfnamefont {H.}~\bibnamefont {Maeda}},
  \bibinfo {author} {\bibfnamefont {K.}~\bibnamefont {Ohki}},\ and\ \bibinfo
  {author} {\bibfnamefont {Y.}~\bibnamefont {Yanagisawa}},\ }\bibfield  {title}
  {\bibinfo {title} {Review of the temporal stability of the magnetic field for
  ultra-high field superconducting magnets with a particular focus on
  superconducting joints between hts conductors},\ }\href
  {https://doi.org/10.1088/1361-6668/ac5645} {\bibfield  {journal} {\bibinfo
  {journal} {Superconductor Science and Technology}\ }\textbf {\bibinfo
  {volume} {35}},\ \bibinfo {pages} {043002} (\bibinfo {year}
  {2022})}\BibitemShut {NoStop}%
\bibitem [{\citenamefont {Brouwer}\ \emph {et~al.}(2022)\citenamefont
  {Brouwer}, \citenamefont {Shen}, \citenamefont {Norris}, \citenamefont
  {Hafalia}, \citenamefont {Schlueter}, \citenamefont {Wang}, \citenamefont
  {Ciston}, \citenamefont {Ercius}, \citenamefont {Ji}, \citenamefont {Mankos},
  \citenamefont {Ophus}, \citenamefont {Stibor}, \citenamefont {Schmid},
  \citenamefont {Minor},\ and\ \citenamefont {Denes}}]{Brouwer_2022}%
  \BibitemOpen
  \bibfield  {author} {\bibinfo {author} {\bibfnamefont {L.}~\bibnamefont
  {Brouwer}}, \bibinfo {author} {\bibfnamefont {T.}~\bibnamefont {Shen}},
  \bibinfo {author} {\bibfnamefont {R.}~\bibnamefont {Norris}}, \bibinfo
  {author} {\bibfnamefont {A.}~\bibnamefont {Hafalia}}, \bibinfo {author}
  {\bibfnamefont {R.}~\bibnamefont {Schlueter}}, \bibinfo {author}
  {\bibfnamefont {L.}~\bibnamefont {Wang}}, \bibinfo {author} {\bibfnamefont
  {J.}~\bibnamefont {Ciston}}, \bibinfo {author} {\bibfnamefont
  {P.}~\bibnamefont {Ercius}}, \bibinfo {author} {\bibfnamefont
  {Q.}~\bibnamefont {Ji}}, \bibinfo {author} {\bibfnamefont {M.}~\bibnamefont
  {Mankos}}, \bibinfo {author} {\bibfnamefont {C.}~\bibnamefont {Ophus}},
  \bibinfo {author} {\bibfnamefont {A.}~\bibnamefont {Stibor}}, \bibinfo
  {author} {\bibfnamefont {A.}~\bibnamefont {Schmid}}, \bibinfo {author}
  {\bibfnamefont {A.~M.}\ \bibnamefont {Minor}},\ and\ \bibinfo {author}
  {\bibfnamefont {P.}~\bibnamefont {Denes}},\ }\bibfield  {title} {\bibinfo
  {title} {Stabilization and control of persistent current magnets using
  variable inductance},\ }\href {https://doi.org/10.1088/1361-6668/ac549b}
  {\bibfield  {journal} {\bibinfo  {journal} {Superconductor Science and
  Technology}\ }\textbf {\bibinfo {volume} {35}},\ \bibinfo {pages} {045011}
  (\bibinfo {year} {2022})}\BibitemShut {NoStop}%
\bibitem [{\citenamefont {Berlin}\ \emph {et~al.}(2020)\citenamefont {Berlin},
  \citenamefont {D'Agnolo}, \citenamefont {Ellis}, \citenamefont {Nantista},
  \citenamefont {Neilson}, \citenamefont {Schuster}, \citenamefont {Tantawi},
  \citenamefont {Toro},\ and\ \citenamefont {Zhou}}]{AsherSC}%
  \BibitemOpen
  \bibfield  {author} {\bibinfo {author} {\bibfnamefont {A.}~\bibnamefont
  {Berlin}}, \bibinfo {author} {\bibfnamefont {R.~T.}\ \bibnamefont
  {D'Agnolo}}, \bibinfo {author} {\bibfnamefont {S.~A.~R.}\ \bibnamefont
  {Ellis}}, \bibinfo {author} {\bibfnamefont {C.}~\bibnamefont {Nantista}},
  \bibinfo {author} {\bibfnamefont {J.}~\bibnamefont {Neilson}}, \bibinfo
  {author} {\bibfnamefont {P.}~\bibnamefont {Schuster}}, \bibinfo {author}
  {\bibfnamefont {S.}~\bibnamefont {Tantawi}}, \bibinfo {author} {\bibfnamefont
  {N.}~\bibnamefont {Toro}},\ and\ \bibinfo {author} {\bibfnamefont
  {K.}~\bibnamefont {Zhou}},\ }\bibfield  {title} {\bibinfo {title} {{Axion
  Dark Matter Detection by Superconducting Resonant Frequency Conversion}},\
  }\href {https://doi.org/10.1007/JHEP07(2020)088} {\bibfield  {journal}
  {\bibinfo  {journal} {JHEP}\ }\textbf {\bibinfo {volume} {07}}\bibfield
  {number} {\bibinfo  {number} { (07)},\ \bibinfo {pages} {088}},\ }\Eprint
  {https://arxiv.org/abs/1912.11048} {arXiv:1912.11048 [hep-ph]} \BibitemShut
  {NoStop}%
\bibitem [{\citenamefont {Uhlig}\ and\ \citenamefont
  {Hehn}(1997)}]{UHLIG1997279}%
  \BibitemOpen
  \bibfield  {author} {\bibinfo {author} {\bibfnamefont {K.}~\bibnamefont
  {Uhlig}}\ and\ \bibinfo {author} {\bibfnamefont {W.}~\bibnamefont {Hehn}},\
  }\bibfield  {title} {\bibinfo {title} {3he4he dilution refrigerator precooled
  by gifford-mcmahon refrigerator},\ }\href
  {https://doi.org/https://doi.org/10.1016/S0011-2275(97)00026-X} {\bibfield
  {journal} {\bibinfo  {journal} {Cryogenics}\ }\textbf {\bibinfo {volume}
  {37}},\ \bibinfo {pages} {279} (\bibinfo {year} {1997})}\BibitemShut
  {NoStop}%
\bibitem [{\citenamefont {Serafin}\ \emph {et~al.}(2021)\citenamefont
  {Serafin}, \citenamefont {Fadel}, \citenamefont {Treutlein},\ and\
  \citenamefont {Sinatra}}]{serafin2021}%
  \BibitemOpen
  \bibfield  {author} {\bibinfo {author} {\bibfnamefont {A.}~\bibnamefont
  {Serafin}}, \bibinfo {author} {\bibfnamefont {M.}~\bibnamefont {Fadel}},
  \bibinfo {author} {\bibfnamefont {P.}~\bibnamefont {Treutlein}},\ and\
  \bibinfo {author} {\bibfnamefont {A.}~\bibnamefont {Sinatra}},\ }\bibfield
  {title} {\bibinfo {title} {Nuclear spin squeezing in helium-3 by continuous
  quantum nondemolition measurement},\ }\href@noop {} {\bibfield  {journal}
  {\bibinfo  {journal} {Physical review letters}\ }\textbf {\bibinfo {volume}
  {127}},\ \bibinfo {pages} {013601} (\bibinfo {year} {2021})}\BibitemShut
  {NoStop}%
\bibitem [{\citenamefont {Boyers}\ \emph {et~al.}(2025)\citenamefont {Boyers},
  \citenamefont {Goldstein},\ and\ \citenamefont {Sushkov}}]{boyers2025}%
  \BibitemOpen
  \bibfield  {author} {\bibinfo {author} {\bibfnamefont {E.}~\bibnamefont
  {Boyers}}, \bibinfo {author} {\bibfnamefont {G.}~\bibnamefont {Goldstein}},\
  and\ \bibinfo {author} {\bibfnamefont {A.~O.}\ \bibnamefont {Sushkov}},\
  }\bibfield  {title} {\bibinfo {title} {Spin squeezing of macroscopic nuclear
  spin ensembles},\ }\href@noop {} {\bibfield  {journal} {\bibinfo  {journal}
  {Physical Review D}\ }\textbf {\bibinfo {volume} {111}},\ \bibinfo {pages}
  {052004} (\bibinfo {year} {2025})}\BibitemShut {NoStop}%
\bibitem [{\citenamefont {Boyers}(2022)}]{boyers2022}%
  \BibitemOpen
  \bibfield  {author} {\bibinfo {author} {\bibfnamefont {E.}~\bibnamefont
  {Boyers}},\ }\emph {\bibinfo {title} {{Prospects for spin squeezing in
  nuclear magnetic resonance dark matter searches}}},\ \href@noop {} {Ph.D.
  thesis},\ \bibinfo  {school} {Boston U.} (\bibinfo {year} {2022})\BibitemShut
  {NoStop}%
\bibitem [{\citenamefont {Yurke}\ and\ \citenamefont {Denker}(1984)}]{qnt}%
  \BibitemOpen
  \bibfield  {author} {\bibinfo {author} {\bibfnamefont {B.}~\bibnamefont
  {Yurke}}\ and\ \bibinfo {author} {\bibfnamefont {J.~S.}\ \bibnamefont
  {Denker}},\ }\bibfield  {title} {\bibinfo {title} {Quantum network theory},\
  }\href {https://doi.org/10.1103/PhysRevA.29.1419} {\bibfield  {journal}
  {\bibinfo  {journal} {Phys. Rev. A}\ }\textbf {\bibinfo {volume} {29}},\
  \bibinfo {pages} {1419} (\bibinfo {year} {1984})}\BibitemShut {NoStop}%
\bibitem [{\citenamefont {Blais}\ \emph {et~al.}(2021)\citenamefont {Blais},
  \citenamefont {Grimsmo}, \citenamefont {Girvin},\ and\ \citenamefont
  {Wallraff}}]{cqed}%
  \BibitemOpen
  \bibfield  {author} {\bibinfo {author} {\bibfnamefont {A.}~\bibnamefont
  {Blais}}, \bibinfo {author} {\bibfnamefont {A.~L.}\ \bibnamefont {Grimsmo}},
  \bibinfo {author} {\bibfnamefont {S.}~\bibnamefont {Girvin}},\ and\ \bibinfo
  {author} {\bibfnamefont {A.}~\bibnamefont {Wallraff}},\ }\bibfield  {title}
  {\bibinfo {title} {Circuit quantum electrodynamics},\ }\bibfield  {journal}
  {\bibinfo  {journal} {Reviews of Modern Physics}\ }\textbf {\bibinfo {volume}
  {93}},\ \href {https://doi.org/10.1103/revmodphys.93.025005}
  {10.1103/revmodphys.93.025005} (\bibinfo {year} {2021})\BibitemShut {NoStop}%
\bibitem [{\citenamefont {Schleier‑Smith}(2011)}]{Monikathesis}%
  \BibitemOpen
  \bibfield  {author} {\bibinfo {author} {\bibfnamefont {M.~H.}\ \bibnamefont
  {Schleier‑Smith}},\ }\href@noop {} {\bibinfo {title} {Cavity‑enabled spin
  squeezing for a quantum‑enhanced atomic clock}} (\bibinfo {year} {2011}),\
  \bibinfo {note} {{PhD Thesis}}\BibitemShut {NoStop}%
\bibitem [{\citenamefont {Schleier-Smith}\ \emph {et~al.}(2010)\citenamefont
  {Schleier-Smith}, \citenamefont {Leroux},\ and\ \citenamefont
  {Vuleti\ifmmode~\acute{c}\else \'{c}\fi{}}}]{inh_Monika}%
  \BibitemOpen
  \bibfield  {author} {\bibinfo {author} {\bibfnamefont {M.~H.}\ \bibnamefont
  {Schleier-Smith}}, \bibinfo {author} {\bibfnamefont {I.~D.}\ \bibnamefont
  {Leroux}},\ and\ \bibinfo {author} {\bibfnamefont {V.}~\bibnamefont
  {Vuleti\ifmmode~\acute{c}\else \'{c}\fi{}}},\ }\bibfield  {title} {\bibinfo
  {title} {States of an ensemble of two-level atoms with reduced quantum
  uncertainty},\ }\href {https://doi.org/10.1103/PhysRevLett.104.073604}
  {\bibfield  {journal} {\bibinfo  {journal} {Phys. Rev. Lett.}\ }\textbf
  {\bibinfo {volume} {104}},\ \bibinfo {pages} {073604} (\bibinfo {year}
  {2010})}\BibitemShut {NoStop}%
\bibitem [{\citenamefont {Mandel}\ and\ \citenamefont
  {Wolf}(1995)}]{Mandel_Wolf_1995}%
  \BibitemOpen
  \bibfield  {author} {\bibinfo {author} {\bibfnamefont {L.}~\bibnamefont
  {Mandel}}\ and\ \bibinfo {author} {\bibfnamefont {E.}~\bibnamefont {Wolf}},\
  }\href@noop {} {\bibinfo {title} {Optical coherence and quantum optics}}
  (\bibinfo {year} {1995})\BibitemShut {NoStop}%
\bibitem [{\citenamefont {Dicke}(1953)}]{Dicke_narrowing}%
  \BibitemOpen
  \bibfield  {author} {\bibinfo {author} {\bibfnamefont {R.~H.}\ \bibnamefont
  {Dicke}},\ }\bibfield  {title} {\bibinfo {title} {The effect of collisions
  upon the doppler width of spectral lines},\ }\href
  {https://doi.org/10.1103/PhysRev.89.472} {\bibfield  {journal} {\bibinfo
  {journal} {Phys. Rev.}\ }\textbf {\bibinfo {volume} {89}},\ \bibinfo {pages}
  {472} (\bibinfo {year} {1953})}\BibitemShut {NoStop}%
\end{thebibliography}%

\clearpage
\onecolumngrid
\setcounter{section}{0}
\appendix

\section{Unitary Evolution}
\label{app:unitary}

In this appendix we derive the Dicke Hamiltonian describing our system, App.~\ref{app:TCH}, and from that we derive the OAT Hamiltonian, App.~\ref{app:squeezingΗ}, in the RWA, for simplicity. Finally, we derive exact expressions of the squeezing Hamiltonian in the general, non-RWA case, App.~\ref{app:nonRWA}. We have set everywhere $\hbar=c=1$, unless otherwise noted.

%%%%%%%%%%%%%%%%%%%%%
\subsection{The Dicke Hamiltonian}
\label{app:TCH} 

Here we derive the spin-circuit Hamiltonian Eq.~\eqref{eq:htc}. The classical Hamiltonian of an LC circuit is $H_\text{LC}=\frac{LI^2}{2}+\frac{q^2}{2C}$, where $L$ is the inductance, $C$ the capacitance, $I$ the current and $q$ the charge in the circuit. The resonance frequency is $\omega_\text{LC}=1/\sqrt{LC}$. Standard quantization of the charge and the flux, such that $\phi=LI\to i\sqrt{\frac{\omega_\text{LC}L}{2}}(a^\dagger-a)$, leads to a quantized magnetic field $B=\frac{i}{N_\text{c}A}\sqrt{\frac{\omega_\text{LC} L}{2}}(a^\dagger-a)=i\sqrt{\frac{\omega_\text{LC}}{2V_\text{L}}}(a^\dagger-a)$, where $a^\dagger$ and $a$ are the creation and annihilation operators of quanta in the circuit so that $H_\text{LC}=\omega_\text{LC}(a^\dagger a+1/2)$, $V_\text{L}=A\ell$ is the physical volume of a coil of length $\ell$ and cross-section $A$, and $N_\text{c}$ the number of turns of the coil.

The fundamental interaction of a spin with the magnetic field of the coil is $H_\text{int}=-\gamma\mathbf{s}\cdot\mathbf{B}\approx -\gamma s_yB_y$, where $\gamma$ is the gyromagnetic ratio of the spin, assuming the solenoid axis is along the $y$ direction, in a coordinate system where $B_\text{ext}$ is along the $z$-direction. Then $H_\text{int}^{(1)}=-\frac{\gamma}{2}\sqrt{\frac{\omega_\text{LC}}{2V_\text{L}}}(s_+-s_-)(a^\dagger-a)$. In the case of $N$ spins and the limit where the wavelength $\lambda_\text{LC}=2\pi/\omega_\text{LC}\gg V_\text{L}$, this becomes the Dicke Hamiltonian

\begin{equation}
    H_\text{int}=g(a - a^\dagger) (J_+-J_-),
\end{equation}
where $g\equiv\mu[\omega_\text{LC}/(2V_\text{L})]^{1/2}$ and $\mu=\gamma/2$ for a spin-1/2 particle.

In the Rotating Wave Approximation (RWA) one spin obeys the Janes-Cummings Hamiltonian $H_\text{JC}\approx g(s_+ a+s_-a^\dagger)$, while $N$ spins obey the Tavis-Cummings Hamiltonian

\begin{equation}
    H_\text{TC}=g(aJ_++a^\dagger J_-).
\end{equation}

%%%%%%%%%%%%%%%%%%%%%
\subsection{Squeezing Hamiltonian}
\label{app:squeezingΗ} 

Here we derive the squeezing Hamiltonian, Eq.~\eqref{eq:hsqueeze} in the RWA, which makes the computation more tractable. The full non-RWA result is presented in the next section. First, we polarize the spins in their ground state, $\prod \ket{g}$ and apply a slow Rabi oscillation of frequency $\Omega\ll\Delta$ to bring them to the $\ket{\css}$, where $\Delta\equiv \Delta_-$ in the RWA. The Hamiltonian of the combined circuit-spin system is

\begin{equation}
    H_\text{Rabi}=\omega_0 J_z+\oc a^\dagger a+g (aJ_++a^\dagger J_-)+\frac{1}{2}\pare{\Omega e^{-i\omega_\text{d} t}J_++\Omega^*e^{+i\omega_\text{d} t} J_-},
\end{equation}
where the driving of angular frequency $\omega_\text{d}$ can be directly on the spins, or through an off-resonant driving of the cavity, of the form $\parea{\frac{\Omega(\omega_\text{d}-\oc)}{2g} a e^{i\omega_\text{d} t}+\text{h.c.}}$. Mathematically, the two are equivalent, but the circuit driving requires $\sim|\Omega|^2/g^2$ circuit quanta.

The Heisenberg equations of motion for the coupled system in the limit of no losses is

\begin{align}
    \dot{\bar{a}}&=-\frac{ig}{\Delta}\bar{J}_- e^{-i\tau}\label{eq:app_a}\\
    \dot{\bar{J}}_-&=\frac{2ig}{\Delta}\bar{a} J_z e^{i\tau}+\frac{i\Omega}{\Delta}J_z \label{eq:app_jm}\\
    \dot{J}_z&=-\frac{ig}{\Delta}(\bar{a}\bar{J}_+ e^{i\tau}-\bar{a}^\dagger \bar{J}_-e^{-i\tau})-\frac{i}{2\Delta} (\Omega \bar{J}_+-\Omega^* \bar{J}_- ),\label{eq:app_jz}
\end{align}
where dots denote derivatives with respect to $\tau\equiv\Delta t$, we've set the drive to be on resonance with the spins, and we have gone to the interaction picture, $\bar{a}\equiv a e^{i\oc t}$ and $\bar{J}_-\equiv J_-e^{i\os t}$. As long as $|\Omega|\ll\Delta$ and the back-reaction of the circuit to the spins is negligible, $g\avg{\bar{a}}\ll \Omega$ , to leading order the spins will perform a Rabi oscillation and the circuit operator will adiabatically follow the spins, $\bar{a}\approx \bar{a}(0)+\frac{g}{\Delta}[\bar{J}_-e^{-i\tau}-\bar{J}_-(0)]$.

At $\tau=\tau_0=\pi\Delta/(2|\Omega|)$ we stop the driving, reaching $\Omega=0$. Bringing the state to the equator must be done adiabatically, namely slower than $\Delta^{-1}$. In the Schr\"{o}dinger picture, since we worked in the limit of no circuit back-reaction to the spins, the spins are in the $\ket{\css}$. The photon operator in the Heisenberg picture can be written as  $\bar{a}(\tau_0)=\bar{a}(0)-\frac{g}{\Delta}\bar{J}_-(0)+g\bar{J}_-(\tau_0)e^{-i\tau_0}/\Delta=\bar{a}(0)-\frac{g}{\Delta}\bar{J}_-(0)+g(\bar{J}_-(\tau_0)-N/2)e^{-i\tau_0}/\Delta+gNe^{-i\tau_0}/(2\Delta)$, where the third term is small in the $\ket{\css}$. Neglecting both this term and the second term, which are of the same order, the photon operator has the form of a displaced operator by an amplitude $\alpha(\tau_0)=\frac{gN}{2\Delta}e^{-i
\tau_0}$, meaning that, in the Schr\"{o}dinger picture, the state of the photon field at $\tau=\tau_0$ is the coherent state $\ket{\alpha(\tau_0)}$, and so there is a mean photon occupation number $\avg{a^\dagger a}\approx \frac{g^2N^2}{4\Delta^2}$ in the circuit. Despite this being the leading order effect, in general the cavity operator will exhibit suppressed oscillations around this mean value with amplitudes at most linear in $|\Omega|/\Delta$. These oscillations can, in principle, be canceled by an appropriate initialization of the circuit in a coherent state, with an occupation number parametrically smaller than the final equilibrium mean value. In the presence of cavity decays, these oscillations decay exponentially with a timescale $\kappa^{-1}$. Adiabaticity and suppressed circuit back-reaction impose the conditions $(g \sqrt{N}/\Delta)^2\ll \Omega/\Delta\ll 1$.

The system can thus be thought of as initialized in $\ket{\css}\otimes\ket{\alpha(\tau_0)}$  at $\tau=\tau_0$. The coherent state can be removed by a similarity transformation on the circuit operators, so that we can describe the effects of the circuit on the spins \emph{on top of} the coherent circuit state. In this transformed system, the circuit is in the vaccum state $D^\dagger(\alpha)\ket{\alpha}=\ket{0}$, and the Hamiltonian in the interaction picture is transformed to be

\begin{equation}
    H'=\frac{Ng^2}{\Delta}\pare{J_x-\frac{N}{2}}+g\parea{\pare{J_--\frac{N}{2}}a^\dagger e^{-i\tau}+\pare{J_+-\frac{N}{2}}a e^{i\tau}}.
    \label{eq:hprime}
\end{equation}

One can derive the squeezing Hamiltonian either by perturbative diagonalization or by manipulating the Heisenberg equations of motion. Here we will do the former as it is the more straightforward. We set $V_0\equiv \frac{Ng^2}{\Delta}\pare{J_x-\frac{N}{2}}$ and $V_1\equiv \frac{g}{\Delta}\parea{\pare{J_--\frac{N}{2}}a^\dagger e^{-i\tau}+\pare{J_+-\frac{N}{2}}a e^{i\tau}}$. Diagonalizing the Hamiltonian is equivalent to removing $V_1$. We apply a similarity transformation $e^{iS}$ to the system, such that the new Hamiltonian is $H_\text{eff}=V_0+V_\text{eff}$, with $V_\text{eff}=[iS,V_0]+\frac{1}{2}[iS,V_1]$ and $iS$ obeying $i(i\dot{S})=-V_1$, to leading order in perturbation theory. This gives $iS=-\frac{g}{\Delta}(J_--N/2)a^\dagger e^{-i\tau}-\text{h.c.}$. Perturbation theory is thus manifestly under control as long as $g\sqrt{N}/\Delta\ll1$, since the operators $(J_\pm-N/2)$ have zero mean and variance of order $N$ in the $\ket{\css}$.

Dropping constant terms, the effective Hamiltonian is now

\begin{equation}
    H_\text{eff}=-\frac{g^2}{\Delta}J_z^{2}+\frac{2g^2}{\Delta}J_z\pare{a^\dagger a+\frac{1}{2}}
    \label{eq:heff}
\end{equation}
where, to be precise, all operators here have received small admixtures, i.e. $\mathcal{O}'\approx \mathcal{O}+[iS,\mathcal{O}]$, for the operator $\mathcal{O}$. The dominant term is the first one, the OAT Hamiltonian. Therefore, we can think of our system as starting from the state $\ket{\css}\otimes\ket{0}$, and evolving under $H_\text{eff}$, but keeping in mind that, physically, the circuit has a mean photon number $g^2N^2/(4\Delta^2)$, that is acting as a spectator to the subsequent dynamics. In the realistic case of coupling and frequency fluctuations, decoherence of $\ket{\alpha}$ can be a source of noise, which we discuss later in App.~\ref{app:morenoise}.

%%%%%%%%%%%%%%%%%%%%%
\subsection{Beyond the rotating wave approximation}
\label{app:nonRWA}

In the protocol described in the main paper, the large squeezing factor at low frequencies depends on large-detunings where the system does not obey the RWA: $\os\ll\oc$. Here we briefly derive the OAT Hamiltonian without the RWA, omitting the details that can be easily adapted from the formalism developed in the two previous sections.

The exact Hamiltonian in the interaction picture is

\begin{eqnarray}
    H_\text{non-RWA} =g \pare{ae^{-i\oc t}-a^\dagger e^{i\oc t}}\pare{J_+e^{i\os t}-J_-e^{-i\os t}}.
    \label{eq:fullDicke}
\end{eqnarray}

We may remove the large circuit quanta coherent state as in the previous section, by noting now that $\alpha(\tau_0)\approx\frac{gN}{2\Delta_-}e^{-i\tau_0}\parea{1+\frac{\Delta_-}{\Delta_+} e^{i(\Delta_+/\Delta_-+1)\tau_0}}$, where $\Delta_\pm\equiv \os\pm\oc$. Perturbative diagonalization then proceeds as above, but now $iS=-\frac{g}{\Delta_-}(J_--N/2) a^\dagger e^{-i\Delta_- t} - \frac{g}{\Delta_+}(J_+-N/2) a^\dagger e^{i\Delta_+ t}-\text{h.c.}$. The effective Hamiltonian is then

\begin{eqnarray}
    H_\text{eff, non-RWA} \approx -\frac{2g^2\oc}{\os^2-\oc^2}J_z^2 + \frac{4g^2\os}{\os^2-\oc^2}J_z\pare{a^\dagger a +\frac{1}{2}},
    \label{eq:HnonRWA}
\end{eqnarray}
so $\chi$ in the full Dicke model is $\chi\equiv\frac{2g^2\oc}{\os^2-\oc^2}$, which should be compared to $\chi$ within the RWA, $\chi_\text{RWA}\equiv \frac{g^2}{\Delta_-}=\chi \frac{\Delta_+}{2\oc}$. In the limit $\Delta_+\approx\oc$, which is where we are operating for most of this work, $\chi\approx2\chi_\text{RWA}$. We note that the full Dicke model, when perturbatively diagonalized, also includes terms oscillating at frequencies $2\os$ and $2\oc$ in Eq.~\eqref{eq:HnonRWA}. In a setup where every interrogation time, including the squeezing timescale and any Rabi pulse timescale, is much slower than $\os$, these terms average to zero and are thus neglected here.

We also note that the operators after perturbative diagonalization receive corrections, such that, for example, $a'\approx a +\frac{g}{\Delta_-}\pare{J_--\frac{N}{2}}e^{-i\Delta_- t} + \frac{g}{\Delta_+}\pare{J_+-\frac{N}{2}}e^{i\Delta_+ t}$.

%%%%%%%%%%%%%%%%%%%%%
\section{Fundamental limits to squeezing}
\label{app:limits}

In this appendix we compute the amount of achievable squeezing with our protocol, in the idealized quantum-limited case, App.~\ref{app:QL}, and in the presence of decoherence, App.~\ref{app:decoherence}. 

%%%%%%%%%%%%%%%%%%%%%
\subsection{Bloch sphere curvature}
\label{app:QL}

The squeezing Hamiltonian $H_\text{eff}=-\chi J_z^2$ leads to the following dynamics of the collective spin operators:

\begin{align}
    \dot{J}_z&=0\\
    \dot{J}_- &= -2i\chi J_z J_- - i\chi J_-,\label{eq:jm_H}
\end{align}
whose formal solution is $J_z(t)=J_z(0)$ and $J_-(t)=e^{2i\chi t (J_z(0)+1/2)}J_-(0)$. For a system initially in the $\ket{\css}$, the evolution of operator expectation values is well-known~\cite{Kitagawa93}:

\begin{align}
    \avg{J_z}&=\avg{J_y}=0\\
    \avg{J_z^2} &= \frac{N}{4}\\
    \avg{J_y^2} &= \frac{N}{4} +\frac{N}{8}\pare{N-1}\parea{1-\cos^{N-2}(2\chi t)}\\
    \avg{J_yJ_z+J_zJ_y} &= \frac{N}{2}\pare{N-1}\sin(\chi t)\cos^{N-2}(\chi t).
\end{align}

We define the rotated observable $J_z'[\theta]\equiv \cos\theta J_z+\sin\theta J_y$. The angle $\theta_c$ that minimizes $\text{var}(J_z')$ defines the squeezing parameter $\xi^{-2}=\text{var}(J_z')[\theta_c]/(N/4)$. In the limit $N\chi t\gg 1$ and $\sqrt{N}\chi t\ll 1$, it can be approximated as

\begin{equation}
    \xi^{-2}\approx \frac{1}{(N\chi t)^2}+\frac{1}{6}\pare{N^{1/2}\chi t}^4,
    \label{eq:xi2_qnf}
\end{equation}
at $\theta_c\approx -(N\chi t)^{-1}$. The second term limits squeezing due to the curvature of the Bloch sphere, giving a quantum floor of

\begin{equation}\label{eq:qnf}
    \text{var}(J_z')_\text{quantum floor}\approx \frac{3^{2/3}}{8}N^{1/3}
\end{equation}
This corresponds to approximately $133~\text{dB} 
\frac{\log_{10}N}{20}$ of squeezing, which corresponds to 133~dB for $N= 10^{20}$ spins or 173~dB for $N=10^{26}$ spins, which is thus the ultimate quantum limit for this protocol. We note that this quantum noise floor is 48~dB, the maximum squeezing achieved in our protocol, for $N_\text{min}\approx 2\times 10^{7}$. Decohering processes will affect squeezing much before curvature effects become important for $N\gg N_\text{min}$, so in what follows we will neglect the second term in Eq.~\eqref{eq:xi2_qnf}.

Since we aim to describe the effects of decoherence on the squeezing process, it is more convenient to treat the time-evolution perturbatively. Then, the Heisenberg equations of motion yield

\begin{align}
    J_y^2(t)&\approx \tz{J_y^2} + \chi t\tz{\pare{J_xJ_zJ_y+J_zJ_xJ_y +J_y J_x J_z + J_y J_z J_x}} + \chi^2 t^2 \tz{\pare{J_xJ_z + J_zJ_x}^2}\label{eq:jy2_H}\\
    \pare{J_y J_z + J_zJ_y}(t)&\approx \tz{\pare{J_y J_z + J_z J_y}} + \chi t\tz{\pare{J_xJ_z^2+2J_zJ_xJ_z + J_z^2J_x}}\label{eq:jyjz_H},
\end{align}
where we kept only the terms whose expectation values give the leading order effects in $N$ and have assumed only $\sqrt{N}\chi t\ll 1$. The $|_0$ notation means that the operators are computed at the initial time $t=0$, where the state of the system is the $\ket{\css}$. In the limit of $N\chi t\gg 1$, we recover $\xi^2\approx (N\chi t)^{2}$, without the curvature term. We will use these equations to compute the effects of decoherence on the squeezing process.

%%%%%%%%%%%%%%%%%%%%%
\subsection{Decoherence}
\label{app:decoherence}

Squeezing is limited by any interaction of the spins or the circuit with environmental degrees of freedom. Some of these are fundamental, in that they occur because of the nature of nuclear spins and LC circuits. These include decay to thermal circuit modes, App.~\ref{app:spin_decays}, which is dominated by collective effects for nuclear spins, and relaxation and dephasing processes, App.~\ref{app:T1T2}, that include all other non-radiative interactions of the spins. The induced decoherence then limits squeezing, App.~\ref{app:lindblad_time}. Other decohering processes that can be suppressed by appropriate control of experimental parameters are studied in detail in App.~\ref{app:tech_noise} and in Sec.~\ref{sec:tolerances}.

%%%%%%%%%%%%%%%%%%
\subsubsection{Spin decays}
\label{app:spin_decays}

Since the spin-circuit system is inside a large superconducting shield with a very large quality factor ($Q_\text{shield}\gtrsim 10^{10})$ and linear size shorter than the wavelength of radiation $\lambda_0=2\pi/\omega_0$, the spins do not decay to free space significantly. Instead, through their mixing with the circuit operators, whose linewidth is $\kappa$, the dominant decay is into the thermal modes of the circuit, through the circuit. Assuming the circuit is at a temperature $T$, powered by a fridge strong enough to remove any instantaneous heat produced by this decay (see Sec.~\ref{sec:tolerances}), these environmental thermal modes have thermal occupation numbers, $\avg{a_k^\dagger a_k}=(e^{\omega_k/T}-1)^{-1}\equiv \bar{n}(\omega_k)$, where $a_k$ is the environmental mode operator with energy $\omega_k$.

Dissipation of the circuit mode can be modeled to occur due to both inductive and capacitive couplings to the environment~\cite{qnt,cqed}. The relevant coupling strengths depend on geometry, parasitic capacitances and inductances and in general on the details of the circuit. Thus, a realistic phenomenological description is beyond the scope of this work. In the RWA, this difference does not matter and the cavity linewidth $\kappa=\omega_\text{LC}/Q$ can be used to describe either coupling. However, because the spins couple inductively to the circuit, their mixing through the circuit with inductive and capacitive couplings to environmental modes yields different spin decay rates in the full Dicke model. To illustrate this point, we couple the cavity operators both inductively and capacitively to two independent baths, $a_k^\text{(E)}$ and $a_k^\text{(B)}$, respectively. In an actual experiment there may be additional mixing by coupling the cavity inductively and capacitively through the \emph{same} environmental operator. In practice, one of these will dominate the decay rate. This is an optimization problem that will be considered in follow-up work with a concrete experimental design. 

The circuit-bath Hamiltonian is then

\begin{equation}
    H_\text{bc} = \sqrt{\frac{\kappa_\text{E}}{2L}} 
    \sum_k \pare{a_k^\text{(E)} + a_k^{\text{(E)}\dagger}}\pare{a + a^\dagger} + \sqrt{\frac{\kappa_\text{B}}{2L}} 
    \sum_k \pare{a_k^\text{(B)} - a_k^{\text{(B)}\dagger}}\pare{a - a^\dagger},
    \label{eq:Hbc}
\end{equation}
where the environmental operators are canonically normalized such that their free Hamiltonians are $H_{\text{bath, }k}=\omega_ka^\dagger_ka_k$. The relative phase between the two terms doesn't matter for our purposes, but if the same environmental operator is used for both couplings, then care must be taken to account for it. The cavity linewidth is $\kappa\equiv \kappa_\text{E} + \kappa_\text{B}$.

We add the Hamiltonian Eq.~\eqref{eq:Hbc}---in the interaction picture---to the Dicke model Hamiltonian Eq.~\eqref{eq:fullDicke}. Perturbative diagonalization mixes the spins and the environment, by adding a term $[iS,H_\text{bc}]$ in the effective Hamiltonian Eq.~\eqref{eq:HnonRWA}. Now it is appropriate to take the RWA, because the decay of the spins into the circuit must conserve energy for integration times $t\gg\omega_0^{-1}$. Then the spins are described by

\begin{equation}
\begin{split}
    H_\text{eff, non-RWA} \approx\, & -\frac{2g^2\oc}{\os^2-\oc^2}J_z^2 + \frac{4g^2\os}{\os^2-\oc^2}J_z\pare{a^\dagger a +\frac{1}{2}} \\
    &+ \sqrt{\frac{\kappa_\text{E}}{2L}}\frac{2g\os }{\os^2-\oc^2}\sum_k\pare{a_k^\text{(E)} J_+ e^{i(\os-\omega_k)t}+\text{h.c.}}- \sqrt{\frac{\kappa_\text{B}}{2L}}\frac{2g\oc }{\os^2-\oc^2}\sum_k\pare{a_k^\text{(B)} J_+ e^{i(\os-\omega_k)t}+\text{h.c.}},
    \label{eq:HnonRWAdiss}
\end{split}
\end{equation}
which holds for $\os\neq\oc$, but not within $|\Delta_-|\lesssim\kappa$, i.e. near resonance. Neglecting the second term in the first line, which is small since $a$ is in its vacuum state, spins and circuit are decoupled from one another.

In the limit of weak decays, it is straighforward to derive a Lindbladian for the interaction of the spins with the thermal modes of the circuit. The single-spin decay rate is then

\begin{eqnarray}
    \eta\equiv \frac{4g^2\pare{\kappa_\text{E}\os^2+\kappa_\text{B}\oc^2}}{(\os^2-\oc^2)^2}.
\end{eqnarray}

Clearly, in this model, optimizing the inductive and capacitive couplings such that $\kappa_\text{E}\sim\kappa\gg\kappa_\text{B}$ (or, more correctly, $\kappa_\text{E}
\os^2\gg\kappa_\text{B}\oc^2$) can allow for suppressed decay rates in the far-detuned limit $\oc\gg\os$. However, we emphasize again that care must be taken to derive the correct phenomenological Hamiltonian, Eq.~\eqref{eq:Hbc}.

The corresponding Lindbladian, after integrating out the bath operators in the equation for the reduced density matrix of the spin system $\rho_\text{S}$, is

\begin{equation}
    \mathcal{L}_\text{rad}[\rho_\text{S}]=\eta(\bar{n}(\os)+1)\pare{J_-\rho_\text{S} J_+-\frac{1}{2}\rho_\text{S} J_+J_--\frac{1}{2}J_+J_-\rho_\text{S}}+\eta\bar{n}(\os)\pare{J_+\rho_\text{S} J_--\frac{1}{2}\rho_\text{S} J_-J_+-\frac{1}{2}J_-J_+\rho_\text{S}}.
    \label{eq:lind_s_rad}
\end{equation}

We note that this Lindbladian preserves the total dipole $J^2$, as it describes \emph{collective} effects. New physics signals due to superradiant interactions have the same structure~\cite{Arvanitaki:2024taq}.

%%%%%%%%%%%%%%%%%
\subsubsection{Relaxation and dephasing}
\label{app:T1T2}

Relaxation and dephasing effects can be described through the terms

\begin{align}
    \mathcal{L}_\text{relax}[\rho_\text{S}]&=\frac{1}{4T_+}\sum_\alpha\pare{\sigma_+^{(\alpha)}\rho_\text{S}\sigma_-^{(\alpha)}-\frac{1}{2}\sigma_-^{(\alpha)}\sigma_+^{(\alpha)}\rho_\text{S}-\frac{1}{2}\rho_\text{S}\sigma_-^{(\alpha)}\sigma_+^{(\alpha)}}+\frac{1}{4T_-}\sum_\alpha\pare{\sigma_-^{(\alpha)}\rho_\text{S}\sigma_+^{(\alpha)}-\frac{1}{2}\sigma_+^{(\alpha)}\sigma_-^{(\alpha)}\rho_\text{S}-\frac{1}{2}\rho_\text{S}\sigma_+^{(\alpha)}\sigma_-^{(\alpha)}}\\
    \mathcal{L}_\text{deph}[\rho_\text{S}]&=\frac{1}{2T_\text{deph}}\sum_\alpha \pare{\sigma_z^{(a)}\rho_\text{S}\sigma_z^{(a)}-\rho_\text{S}},
\end{align}
where $\sigma^{(\alpha)}_i$ are the Pauli matrices of the spin $\alpha$. The timescales $T_\pm$ are defined such that $J_z$ reaches the equilibrium value $J_z^\text{eq}\equiv\frac{N}{2}\frac{T_--T_+}{T_-+T_+}$ in a characteristic timescale $T_1\equiv(T_+^{-1}+T_-^{-1})^{-1}$, i.e. $\avg{\dot{J}_z}= T_1^{-1}(J_z^\text{eq}-\avg{J_z})$. The timescale $T_\text{deph}$ is defined such that the transverse magnetization in a free spin system decays as $\avg{J_x}=\avg{J_x(0)}e^{-t/T_2}$, with $T_2\equiv (1/(2T_1)+1/T_\text{deph})^{-1}$. These terms do not conserve $J^2$.

%%%%%%%%%%%%%%
\subsubsection{Time evolution}
\label{app:lindblad_time}

The time evolution of the expectation value of the spin operator $\mathcal{O}$ will then be

\begin{equation}
    \mathcal{\avg{\dot{O}}}=i\avg{[H_\text{eff},\mathcal{O}]} + \text{Tr}_\text{S}\parea{\mathcal{O}(\mathcal{L}_\text{rad}[\rho_S]+\mathcal{L}_\text{deph}[\rho_S]+\mathcal{L}_\text{relax}[\rho_S])}.
\end{equation}

The first term is responsible for creating spin-spin correlations, and the second one for destroying them. Additional interacting particles that exhibit superradiant interactions can be trivially included in this formalism by adding Lindbladians of the form $\mathcal{L}_\text{rad}[\rho_\text{S}]$ and substituting $\eta(\bar{n}+1)\to\gamma_-$ and $\eta\bar{n}\to\gamma_+$ in the two terms, where $\gamma_\pm$ are the single-spin interaction rates~\cite{Arvanitaki:2024taq}. For particles that exhibit long-time coherence, such as axions, the Markovian approximation assumed here does not apply, but one can still use a similar equation to describe the evolution of the system~\cite{Arvanitaki:2024taq}. For the purposes of this work, a coherent axion field can be thought of as a unitary Rabi drive, but see~\cite{Arvanitaki:2024taq} for the subtleties between this approach and the Lindblad formalism.

Focusing on the Lindbladian contribution for the moment, the operators we are interested in obey the following equations

\begin{align}
    \avg{\dot{J}_x}=&\,i\avg{[H_\text{eff},J_x]}-\frac{1}{T_2}\avg{J_x}\label{eq:jx}\\
    \avg{\dot{J}_y}=&\,i\avg{[H_\text{eff},J_y]}-\frac{1}{T_2}\avg{J_y}\label{eq:jy}\\
    \avg{\dot{J}_z}=&\,\frac{1}{T_1}\parea{J_z^\text{eq}-\avg{J_z}}-\eta\avg{J_+J_-}\label{eq:jz}\\
    \avg{\dot{J}_x^2}=&\,i\avg{[H_\text{eff},J_x^2]}-\frac{2}{T_2}\pare{\avg{J_x^2}-\frac{N}{4}}-2\eta\bar{n}\pare{\avg{J_x^2}-\avg{J_z^2}}\nonumber\\
    &    -\frac{\eta}{2}\avg{J_+J_-+J_x^2}+\frac{\eta}{2}\avg{J_y^2 + J_-J_zJ_-+J_+J_zJ_+ + (J^2-J_z^2+J_z)J_z}\label{eq:jx2}\\
    \avg{\dot{J}_y^2}=&\,i\avg{[H_\text{eff},J_y^2]}-\frac{2}{T_2}\pare{\avg{J_y^2}-\frac{N}{4}}-2\eta\bar{n}\pare{\avg{J_y^2}-\avg{J_z^2}}\nonumber\\
    &+ \frac{\eta}{4}\avg{J_+^2 + J_-^2 - 2J_+J_--2 J_-J_zJ_--2 J_+J_zJ_+ + 2 (J^2-J_z^2+J_z)J_z}\label{eq:jy2}\\
    \avg{\dot{J}_z^2}=&\,-\frac{2}{T_1}\pare{\avg{J_z^2}-\frac{N}{4}}+\frac{2 J_z^\text{eq}-J_z^\text{eq}/(N/2)}{T_1}\avg{J_z}+2\eta\bar{n}\pare{\avg{J^2}-3\avg{J_z^2}}-\eta\avg{(2J_z-1)J_+J_-}\label{eq:jz2}\\
    \frac{\di}{\di t}\avg{J_yJ_z+J_zJ_y}=&\,i\avg{[H_\text{eff},J_yJ_z+J_zJ_y]}-\pare{\frac{3}{2T_1}+\frac{1}{T_\text{deph}}}\avg{J_yJ_z+J_zJ_y} +\frac{2(1-1/N)J_z^\text{eq}}{T_1}\avg{J_y}\label{eq:jyjz}
\end{align}

In the ECSS and large $N$ we may approximate $\avg{J_+J_-}\approx \avg{J^2}$. In addition, the last terms in both Eqs.~\eqref{eq:jx2} and~\eqref{eq:jy2} are $\mathcal{O}(N)$ in the ECSS. Thus, there is no term $\propto N^2\eta$ for $\avg{J_y^2}$, but there is one for $\avg{J_x^2}$ and $\avg{J_z^2}$. Nevertheless, the total dipole $\avg{J^2}$ is preserved under collective radiative decay, as expected. These approximations fail when either the system is close to its ground state, or after time $\approx \min\{T_1,T_2\}$, whichever happens first. In our system, the superradiant decay timescale $(N\eta/4)^{-1}$ and $T_1$  are much longer than any reasonable experimental runtime, but $T_2$ limits the time to run each shot of the experiment. We elaborate below.

After a straightforward---but tedious--- computation in perturbation theory using Eqs.~\eqref{eq:jy2_H} and~\eqref{eq:jyjz_H}, we find

\begin{align}
    \avg{J_y}&\approx 0\\
    \avg{J_y^2}&\approx \frac{N}{4} + \frac{1}{4}N^3\chi^2t^2\pare{1-\frac{2t}{T_\text{deph}} - \frac{4t}{3T_1}}\label{eq:jy2_sq}\\
    \avg{J_z}&\approx -\frac{J_z^\text{eq} t}{T_1}-\frac{1}{4}N^2\eta t \label{eq:avgjz}\\
    \text{var}(J_z)&\approx  \frac{N}{4} +\frac{N^2}{4}(2\bar{n}+1)\eta t - \frac{(J_z^\text{eq})^2}{N/2}\frac{t^2}{2T_1^2}\label{eq:varz}\\
    \avg{J_zJ_y+J_yJ_z}&\approx \frac{1}{2}N^2\chi t\pare{1-\frac{t}{T_\text{deph}} - \frac{t}{T_1}}\label{eq:cov_sq} ,
\end{align}
where the approximations hold as long as $t\ll T_1,\, T_\text{deph}, (N \eta/4)^{-1}$ and we have kept only the leading order terms in $N$. Zero-mean superradiant interactions in the Markovian regime contribute a term $\sim+N^2\gamma_\text{tot} t/4$ to the variance of $J_z$, Eq.~\eqref{eq:varz}, while non-zero mean ones (such as the C$\nu$B) also contribute a $\sim +N^2\gamma_\text{net} t/4$ term in Eq.~\eqref{eq:avgjz}~\cite{Arvanitaki:2024taq}. We see that both dephasing and relaxation decrease linearly the amount of correlations achieved with the squeezing Hamiltonian.

We define the observable $J_z'=\cos\theta J_z+\sin\theta J_y$, whose variance is minimized. This variance is minimum for $\theta_\text{c}\approx - (N\chi t)^{-1}$ to leading order. The squeezing parameter is again defined as $\xi^{-2}=\text{var}(J_z')[\theta_c]/(N/4)$, so that

\begin{equation}
    \xi^{-2}\approx N\eta(2\bar{n}+1)t + \frac{2 t}{3T_1}
    +\frac{1+t/T_\text{deph}}{(N\chi t)^2},
    \label{eq:sq_noise}
\end{equation}
neglecting the $\propto (J_z^\text{eq})^2$ term, which is small. Evidently, both $T_1$ and $T_\text{deph}$ limit squeezing linearly in time, but the effects of dephasing are suppressed by $\theta_\text{c}^2$. Thus, as long as $t_\text{shot}<T_\text{deph}$, squeezing is not affected by dephasing effects.

The different behavior of $T_1$ and $T_\text{deph}$ is not a numerical coincidence, but is directly related to the fact that dephasing does not affect $J_z$. A computationally simpler, but less rigorous, way to understand the effects of dephasing and relaxation is to start with a state that was being squeezed for a time $t_0$ in the \emph{absence} of any decoherence effects, and then evolve the state with the Lindblad master equations in the \emph{presence} of relaxation and dephasing. We thus start with a state whose correlations are those of Eqs.~\eqref{eq:jy2_sq} and~\eqref{eq:cov_sq}, with $t=t_0$ and $1/T_1=1/T_\text{deph}=0$, such that $N\chi t_0\gg 1$. Then it is clear from the Lindblad terms of Eqs.~\eqref{eq:jy2} and~\eqref{eq:jyjz} that both $\avg{J_yJ_z+J_zJ_y}$ and $\avg{J_y^2}-N/4$ simply decay exponentially with rates $1/T_\text{deph}+3/(2T_1)$ and $2/T_\text{deph} + 1/T_1$, respectively. The squeezing parameter then

\begin{equation}
\begin{split}
    \xi^2(t)&=\frac{1}{N/4}\pare{\text{var}[J_z]+\text{var}[J_y] - \sqrt{(\text{var}[J_z]-\text{var}[J_y])^2+\text{cov}[J_y,J_z]}}\\
    &\approx 1-e^{-2t/T_1} + \frac{e^{2t/T_\text{deph}-3t/T_1}}{(N\chi t_0)^2},
\end{split}
\end{equation}
in the limit $\xi^{-2}\ll 1$, and, thus, only $T_1$ appreciably degrades squeezing at short times. We defined the covariance of $J_y$ and $J_z$ as $\text{cov}[J_y,J_z]\equiv\avg{J_yJ_z+J_zJ_y}-2\avg{J_y}\avg{J_z}$.

Therefore, the optimal squeezing time and the corresponding squeezing factor $\xi^2_\text{opt}$ are

\begin{align}
    \tau_\text{sq}&\approx\frac{2^{1/3}}{N}\parea{\frac{1}{\chi^2[\eta(2\bar{n}+1)+2/(3NT_1)]}}^{1/3}\label{eq:OAT_tsq}\\
    \xi^2_\text{opt}&\approx \frac{2^{2/3}}{3}\parea{\frac{\chi}{\eta(2\bar{n}+1)+2/(3NT_1)}}^{2/3},
    \label{eq:OAT_xi}
\end{align}
so that $\text{var}(J_z')[\theta_c]\approx N/(4\xi^2_\text{opt})$, and $\theta_\text{c}\approx [(\eta(2\bar{n}+1)+2/(3NT_1))/(2\chi)]^{1/3}=1/(\sqrt{3}\xi_\text{opt})$.

%%%%%%%%%%%%%%%%%%%%%
\section{Two-axis counter-twisting}
\label{app:tact}

The Hamiltonian 

\begin{eqnarray}
    H_\text{TACT}=\frac{N\chi}{2}J_x + \chi J_z^2,
\end{eqnarray}
has been shown to be equivalent to two-axis countertwisting (TACT)~\cite{hu2017}. TACT offers two advantages by 1. being faster, as squeezing proceeds exponentially $\xi^2 \approx e^{N \chi t}$ along the diagonal in the $z-y$ plane~\cite{muessel2015,li2023}, and 2. not being limited by curvature effects as OAT is (cf. Eq.~\eqref{eq:qnf}), thus being able to reach the Heisenberg limit. The decoherence computation that follows is heuristic.

Since TACT proceeds along the diagonal, decoherence in phase is equally important as decoherence in amplitude. Therefore, we expect that $T_\text{deph}$ effects along the squeezing direction should not be suppressed, as is the case in OAT (cf. Eq.~\eqref{eq:sq_noise}). Thus, in analogy with Eq.~\eqref{eq:sq_noise}, we expect for this protocol

\begin{eqnarray}
    \xi^{-2}\sim N\eta(2\bar{n}+1)t + \frac{2t}{3T_1} + \frac{2t}{T_\text{deph}} + e^{-N\chi t}.
\end{eqnarray}

From this we find 

\begin{align}
    \tsq &\sim \frac{1}{N\chi} \log\pare{\frac{N\chi}{\frac{2}{3T_1}+\frac{2}{T_\text{deph}} + N\eta(2\bar{n}+1)}},\\
    \xi^{-2} &\sim \parea{\frac{(2\bar{n}+1)\eta}{\chi} + \frac{2}{3N\chi T_1}+ \frac{2}{N\chi T_\text{deph}}}\parea{1+\log\pare{\frac{N\chi}{\frac{2}{3T_1}+\frac{2}{T_\text{deph}} + N\eta(2\bar{n}+1)}}}.
\end{align}

These should be compared to Eqs.~\eqref{eq:OAT_tsq} and~\eqref{eq:OAT_xi}. For instance, in the limit where radiative decays are the dominant source of decoherence, here $\xi^{2}\propto Q$, instead of the $Q^{2/3}$ of the OAT protocol. While this would correspond to three orders of magnitude enhancement for $Q=10^9$, the relatively large logarithm reduces the total gain. For a $^3\text{He}$ system with $n_\text{eff}\approx n_\text{S}=10^{20}~\text{cm}^{-3}$, $\os=10^{-6}$~eV, $\oc=1.3\times 10^{-6}$~eV, $Q=10^9$, and $T_\text{deph}\approx 1000~\text{s}\frac{3\times 10^{22}~\text{cm}^{-3}}{n_\text{S}}$ cooled at 10~mK, TACT can achieve a $\sim 12.6$~dB boost to the maximum amount of squeezing, bringing it to a total of 60~dB. Since maximum squeezing is achieved for $\os\approx\oc$ close to the upper limit $10^{-6}$~eV, whether the circuit decay is capacitive or inductive is not important. For capacitive decays this amount of squeezing persists to lower spin frequencies.  In addition, $\tau_\text{sq}$ is reduced by a factor of $\sim45$ compared to the OAT protocol. This speed-up is especially important for searches requiring frequency scanning, such as the ones for axions or dark photons (see Sec.~\ref{sec:axion_and_dp}).

For systems where $T_\text{deph}$ is reduced compared to the expectation from spin-spin interactions (see Table~\ref{tab:parameters}) due to technical noise, whether TACT is preferable to OAT depends on the frequency, the quality factor of the circuit and its dominant decay mode (capacitive or inductive). Nevertheless, it has recently been shown that certain dephasing effects contributing to $T_\text{deph}$ can be alleviated by the large gap $g \sqrt{N}$ needed to change the total dipole $J^2$~\cite{norcia2018,niu2025}. 

Due to such rich dynamics, the lack of a straightforward analytic treatment for TACT observables---in contrast to OAT---and the potential experimental challenges associated with an additional rotation (see Sec.~\ref{sec:tolerances} and App.~\ref{app:tech_noise}), we restrict ourselves to heuristic arguments here and leave a detailed study of this protocol for future work.

%%%%%%%%%%%%%%%%%%%%%

\section{Magnification}
\label{app:magnification}

After the state has been squeezed, Eq.~\eqref{eq:sq_noise}, we apply a small rotation by an angle $2\theta_\text{c}$, as described in the main text, which essentially results in changing $\avg{J_zJ_y+J_yJ_z}\to-\avg{J_zJ_y+J_yJ_z}$. Then the unitary Hamiltonian un-squeezes the state. To see that, we ignore any decoherence arising from now on and take $J_x\sim N/2$. While there is an $\mathcal{O}(Nt/T_\text{deph},Nt/T_1)$ correction to $\avg{J_x}$ at the end of squeezing, this does not affect the argument that follows. We can use Eqs.~\eqref{eq:jy2_H} and~\eqref{eq:jyjz_H}, taking as the initial time the end of squeezing, assuming that the rotation is much faster than the squeezing timescale. Then after time $t$ we find

\begin{align}
    \avg{J_y^2}&\approx \frac{N}{4} + \frac{N^3\chi^2}{4}\parea{\tsq^2 \pare{1-\frac{2\tsq}{T_\text{deph}}-\frac{4\tsq}{3T_1}}-2 t\tsq \pare{1-\frac{\tsq}{T_\text{deph}}-\frac{\tsq}{T_1}} + t^2 \pare{1+N\eta (2\bar{n}+1)t + \frac{1}{N/4}S(\tsq)}},\\
    \avg{J_yJ_z+J_zJ_y}&\approx \frac{N^2\chi}{2}\parea{-\tsq\pare{1-\frac{\tsq}{T_\text{deph}}-\frac{\tsq}{T_1}}+t\pare{1+N\eta(2\bar{n}+1) \tsq+\frac{1}{N/4}S(\tsq)}}\label{eq:cov_mag}
\end{align}
where we added the accumulated signal from superradiant interactions as $S(\tsq)$.

For $t=\tsq$, we see that the signal has been mapped from an increase of the variance of $J_z$ to that of $J_y$, and is further magnified by $\xi^2$:

\begin{equation}
    \avg{J_y^2}=\frac{N}{4} + \frac{\xi^2}{4}\pare{\frac{2N\tsq}{3T_1} + N^2\eta(2\bar{n}+1) \tsq + S(\tsq)}.
    \label{eq:mag_step1}
\end{equation}

In addition, because of the non-zero covariance, Eq.~\eqref{eq:cov_mag}, the final state of this step is angled with respect to the $y$-axis. In the ideal case of $T_1,T_\text{deph}\to\infty$, this angle is simply $-\xi^{-1}$, as shown in panel \textit{(d)} of Fig.~\ref{fig:protocoltot}. When these timescales are present there are minor corrections, which we do not compute rigorously here.

While Eq.~\eqref{eq:mag_step1} has not increased the sensitivity of the state compared to that before this first magnification step, the effect can be read out as long as readout is at the shot-noise level. In other words, the magnification protocol has \emph{magnified} the fluctuations of the state so that they are easier to be read out.

We can go one step further, and magnify the fluctuations even more. A $(\pi/2-\xi^{-1})$-pulse along the $x$-axis now maps $J_y$ to $J_z$, and vice versa. Allowing the unitary Hamiltonian to act for another $\tsq$, we finally get

\begin{eqnarray}
    \avg{J_y^2} = \frac{N}{4} + \xi^2\parea{\frac{N}{4} + \xi^2\pare{\frac{2N\tsq}{3T_1} + N^2\eta(2\bar{n}+1) \tsq + S(\tsq)}},
\end{eqnarray}
so that now readout needs to be sensitive only down to $\xi^2 N/4$, namely, much above shot noise. The fluctuations due to the signal (and of all other decohering interactions that mimic the signal) have been enhanced by $\xi^4$, which can be as large as 9.6 orders of magnitude for the one-axis twisting protocol.

Similarly, if there is a mean displacement signal, so that at the end of squeezing $J_z=J_z(0)+S(\tsq)$, for instance $S(\tsq)=-|\Gamma_\text{net}| \tsq$ for the net rate of the C$\nu$B, a time $t=\tsq$ \emph{after} the initial squeezing and the first $2\theta_c$ rotation, we find $\avg{J_y}=-\xi \pare{S(\tsq) -N^2\eta \tsq/4 -J_z^\text{eq}\tsq/T_1}$. Mapping $J_y$ to $J_z$ with a fast $(\pi/2-\xi^{-1})$-pulse, and allowing for a unitary evolution for another $\tsq$, we finally get

\begin{eqnarray}
    \avg{J_y} \approx -\xi^2 \pare{S(\tsq) -N^2\eta \tsq/4 -J_z^\text{eq}\tsq/T_1},
\end{eqnarray}
and the variance is correspondingly enhanced as before, $\avg{J_y^2} \approx \xi^2\parea{\frac{N}{4} + \xi^2\pare{\frac{2N\tsq}{3T_1} + N^2\eta(2\bar{n}+1) \tsq}}$. As a result, the relative enhancement of the SNR for the mean displacement remains $\mathcal{O}(\xi)$, but, again, it is magnified orders of magnitude above shot noise.

%%%%%%%%%%%%%%%%%%%%%
\section{Technical noise}
\label{app:tech_noise}

In this appendix we compute several effects that arise through technical imperfections and may limit squeezing if not addressed by the experimentalist. In App.~\ref{app:polarization} we compute the effects of finite sample polarization, in Apps.~\ref{app:inh_static} and~\ref{app:g_brownian} the effects of static and time-dependent vacuum mode inhomogeneities across the spin sample, in App.~\ref{app:morenoise} the effects of time-dependent coupling and frequency fluctuations on the circuit coherent state $\ket{\alpha}$ and its backreaction on the spins, and in App.~\ref{app:rabipicture} we illustrate our protocol's insensitivity to initial mean values of $\avg{J_z}$.

%%%%%%%%%%%%%%%%%%%%%
\subsection{Finite polarization}
\label{app:polarization}

A finite sample polarization $p$ means that the initial state is not described by the pure state matrix $\prod_i\ket{g}_i$, but rather by the density matrix $2^{-N}\prod_i [(1+p)\ket{g}\bra{g} + (1-p)\ket{e}\bra{e}]_i$, which is a mixed state. Under a coherent Rabi $\pi/2$ pulse, this state evolves to 

\begin{equation}
    \rho_p=2^{-N}\prod_i\parea{(1+p)\ket{+}\bra{+}+(1-p)\ket{-}\bra{-}}_i,
\end{equation}
where $\ket{\pm}_i$ are the spin eigenstates of the single-spin operators $\sigma_x^{(i)}/2$. For $p=1$ this reduces to $\ket{\css}\bra{\css}$. For the vacuum squeezing protocol described in this work, the density matrix $\rho_p$ has a reduced total angular momentum $\avg{J^2}_p=\text{tr}(\rho_p J^2)\approx N^2p^2/4$, to leading order. Thus, the mean photon occupation number in the cavity after the initial slow Rabi pulse of $\Omega\ll \Delta$, is reduced by $p^2$, such that $\avg{a^\dagger a}=g^2 N^2p^2/(4\Delta^2)$.

The effects of finite polarization can then be included in the full solution for the squeezing coefficient $\xi^2$ of the one-axis twisting Hamiltonian by using the formal solution of Eq.~\eqref{eq:jm_H}, $J_-(t)=e^{2i\chi t (J_z(0)+1/2)}J_-(0)$. The result is

\begin{align}
    \avg{J_z}&=\avg{J_y}=0\\
    \avg{J_z^2} &= \frac{N}{4}\\
    \avg{J_y^2} &= \frac{N}{4} +\frac{N}{8}\pare{N-1}p^2\parea{1-\cos^{N-2}(2\chi t)}\\
    \avg{J_yJ_z+J_zJ_y} &= \frac{N}{2}\pare{N-1}p\sin(\chi t)\cos^{N-2}(\chi t), 
\end{align}
which translates to $\xi^2\sim (Np\chi t)^{2}$.

Including decoherence as before, by utilizing Eqs.~\eqref{eq:jy2_H} and~\eqref{eq:jyjz_H}, the squeezing coefficient of Eq.~\eqref{eq:sq_noise} becomes

\begin{equation}
    \xi^2\approx Np^2\eta(2\bar{n}+1)t + \frac{2 t}{3T_1} 
    +\frac{1+t/T_\text{deph}}{(N p\chi t)^2},
    \label{eq:sq_noise_pol}
\end{equation}
and the optimal angle is $\theta_c\approx-(Np\chi  t)^{-1}$. Similarly to the effect on the photons, namely the first term in the equation above, finite polarization also reduces the signal from any superradiant interactions by $p^2$.

%%%%%%%%%%%%%%%%%%%%%%%%%%
\subsection{Static inhomogeneity}
\label{app:inh_static}

In a realistic experimental setup, the cavity coupling would vary from spin to spin because of inhomogeneities in the vacuum mode. For solid samples, these are static, while the Brownian motion exhibited by the spins in a gaseous or liquid sample makes those time-dependent as well (addressed in the next section). The effects inhomogeneities on the sensitivity of the experiment depend both on their amount as well as on the experimental observable~\cite{Monikathesis,inh_Monika}. 

If the spin-circuit coupling is inhomogeneous, the TC Hamiltonian is modified to

\begin{equation}
    H_\text{inh} = \sum_i g_i\pare{a\sigma_+^{(i)}e^{i\Delta t}+a^\dagger\sigma_-^{(i)}e^{-i\Delta t}},
\end{equation}
in the interaction picture. A similar procedure for the elimination of the cavity mode can be carried out here as well, with the coherent state in the circuit now being $\alpha=-\sum_i g_ie^{-i\Delta t}/(2\Delta)$ at the end of the initial slow Rabi drive. The effective Hamiltonian is then, to leading order,

\begin{equation}
    H_\text{eff,inh}=\sum_{i,j} \frac{g_ig_j}{\Delta} s_z^{(i)}s_z^{(j)},
\end{equation}
where $s_i^{(\beta)}\equiv\sigma_i^{(\beta)}/2$, with $\sigma_i^{(\beta)}$ are the Pauli matrices for the spin $\beta$. We define then the \emph{weighted} collective spin operators $S_k\equiv\frac{1}{\bar{g}}\sum_i g_i s_k^{(i)}$, where $\bar{g}\equiv\sum_i g_i^2/(\sum_i g_i)$. As long as measurements are done through the circuit, the observables are the inhomogeneously-weighted spin operators $S_i$, rather than the homogeneously-weighted ones $J_i$~\cite{Monikathesis,inh_Monika}. 

The single spin operators $s_z^{(i)}$ still commute with $H_\text{eff,inh}$, so that $s_z^{(i)}(t)=s_z^{(i)}(0)$. The Heisenberg equation of motion for $s_-^{(i)}\equiv s_x^{(i)}-is_y^{(i)}$ under $H_\text{eff,inh}$ then has the formal solution

\begin{equation}
    s_-^{(i)} = \exp\parea{-i\bar{\chi}t\frac{g_i}{\bar{g}}\pare{\frac{g_i}{\bar{g}}+\sum_j \frac{g_j}{\bar{g}}s_z^{(j)}}}s_-^{(i)}(0),
    \label{eq:sminus_inh}
\end{equation}
where $\bar{\chi}\equiv \bar{g}^2/\Delta$.

First, we consider the simpler case where the initial state is $\ket{\css}$. It is then straightforward to show that 

\begin{align}
    \avg{S_z^2} &= \frac{N_\text{eff}}{4} \label{eq:inh_z2}\\
    \avg{S_y^2} &= \frac{N_\text{eff}}{4} +  \frac{1}{8}\sum_{i\neq j}\frac{g_i}{g}\frac{g_j}{g}
    \parea{\prod_{k\neq i,j} \cos\pare{ \frac{g_k}{g}\frac{g_i-g_j}{g}\bar{\chi} t}
    -\prod_{k\neq i,j} \cos\pare{ \frac{g_k}{g}\frac{g_i+g_j}{g}\bar{\chi} t}} \label{eq:inh_y2}\\
     &\approx \frac{N_\text{eff}}{4} + \frac{1}{4}N_\text{eff}^3 \bar{\chi}^2 t^2 \parea{1+ \mathcal{O}[(\sqrt{N}\bar{\chi}t)^2]}\nonumber\\
     \avg{S_yS_z+S_zS_y} &= \frac{1}{2}\sum_{i\neq j} \frac{g_i}{g}\frac{g_j}{g}\sin\pare{\frac{g_i}{g}\frac{g_j}{g}\bar{\chi}t}\prod_{k\neq i,j}\cos\pare{\frac{g_i}{g}\frac{g_j}{g}\bar{\chi}t} \label{eq:inh_yz}\\
     &=\frac{1}{2}N_\text{eff}^2\chi t\parea{1+ \mathcal{O}[(\sqrt{N}\bar{\chi}t)]},
    \nonumber
\end{align}
where we defined $N_\text{eff}\equiv (\sum_i g_i)^2/(\sum_i g_i^2)$. Clearly squeezing is not degraded by inhomogeneities, if the probe operators and the squeezing operators are the same. For instance, for a non-homogeneous geometry of the vaccum mode such that $B_\text{vac}=B(\mathbf{r})$, we would have $g\propto B$ and so 

\begin{equation}
    N_\text{eff}=\frac{\pare{\int n_\text{S} B(\mathbf{r})\di^3 r}^2}{\int n_\text{S} B^2(\mathbf{r})\di^3 r},
\end{equation}
where $n_\text{S}$ is the sample number density. For a uniform number density, assuming the sample fills a perfect, but finite solenoid, $N_\text{eff}$ has the minimum value $\sim 0.98 N$ at $R/l\sim 0.26$, where $R$ is the solenoid radius and $l$ its length. Therefore, the inhomogeneity of the coupling has a negligible effect on squeezing, despite the fact that the magnetic field varies $\mathcal{O}(1)$ between the center and the edges of the solenoid. 

However, if readout was done in a different basis~\cite{Monikathesis,inh_Monika}, then we would need to project the squeezing operator to the experimental observable. The excess noise would be the remaining component, i.e.

\begin{equation}
    \xi^{-2}_\text{excess} = 1-\frac{(\sum_i g_i)^2}{N\sum_i g_i^2}=1-\frac{N_\text{eff}}{N},
\end{equation}
so that, if the observable was $J_z'\equiv\cos\theta J_z + \sin\theta J_y$ instead of $S_z' \equiv \cos\theta S_z + \sin\theta S_y$, the \emph{maximum} excess noise would be $\xi^{-2}_\text{excess,max}\simeq0.02$. Nevertheless, even in this case, a ``flat'' solenoid of radius-to-length ratio $R/l\gtrsim 6$ would suppress this to $\xi_\text{excess}^{-2}\lesssim 10^{-5}$.

However, in a practical setup, the same coil responsible for squeezing may be used for initially polarizing the spin sample along the $x$-axis. In this case, the initial state will be

\begin{eqnarray}
    \ket{\psi}=\prod_{i=1}^N\parea{\sin\frac{g_i t_\text{R}}{2}\ket{g}_i+\cos\frac{g_i t_\text{R}}{2}\ket{e}_i},
\end{eqnarray}
where $t_\text{R}$ is the time for the initial $\pi/2$-pulse. A more involved computation now yields

\begin{align}
    \text{var}(S_z)=&\frac{1}{4}\sum_i\frac{g_i^2}{\bar{g}^2}(1-\cos^2\theta_i)\\
    \text{var}(S_y)=&\sum_i\frac{g_i^2}{\bar{g}^2}-\frac{1}{4}\sum_{i\neq j}\frac{\sin \theta_i}{2}\frac{\sin \theta_j}{2}\parea{\prod_{k\neq i,j} \pare{\cos\frac{g_k(g_i+g_j)t}{\Delta}+i\cos\theta_k\sin\frac{g_k(g_i+g_j)t}{\Delta}}+\text{h.c.}}\\
    &+\frac{1}{4}\sum_{i\neq j}\frac{\sin \theta_i}{2}\frac{\sin \theta_j}{2}\parea{\prod_{k\neq i,j} \pare{\cos\frac{g_k(g_i-g_j)t}{\Delta}+i\cos\theta_k\sin\frac{g_k(g_i-g_j)t}{\Delta}}+\text{h.c.}}\\
    &+\pareb{\sum_i\frac{g_i}{\bar{g}}\frac{\sin\theta_i}{2}\parea{\prod_{k\neq i}\pare{\cos\frac{g_ig_kt}{\Delta}+i\cos\theta_k\sin\frac{g_ig_kt}{\Delta}}-\text{h.c.}}}^2\\
\text{cov}(S_y,S_z) &= \frac{1}{2}\sum_{i\neq j}\frac{g_ig_j}{\bar{g}^2}\frac{\sin\theta_i}{2}\parea{\pare{\sin\frac{g_ig_jt}{\Delta}-i\cos\theta_j\cos\frac{g_ig_jt}{\Delta}}\prod_{k\neq i,j}\pare{\cos\frac{g_ig_jt}{\Delta}+i\cos\theta_k\sin\frac{g_ig_jt}{\Delta}}+\text{h.c.}}\\
&-\frac{1}{2i}\sum_{i,j}\frac{g_ig_j}{\bar{g}^2}\frac{\sin\theta_i}{2}\cos\theta_j\parea{\prod_{k\neq i}\pare{\cos\frac{g_ig_kt}{\Delta}+i\cos\theta_k\sin\frac{g_ig_kt}{\Delta}}-\text{h.c.}},
\end{align}
where we set $\theta_i\equiv g_i t_\text{R}$ for brevity.

We may choose $t_\text{R}$ such that $\avg{S_z}=-(1/2)\sum_i \frac{g_i}{\bar{g}}\cos \theta_i=0$ by the Intermediate Value Theorem. Then it can be shown that in the limit $N\chi t\gg 1\gg\sqrt{N}\chi t$, 

\begin{align}
    \text{var}(S_z) &= \frac{N_\text{eff}}{4}\\
    \text{var}(S_y) &= \frac{N_\text{eff}}{4}  + \frac{1}{4}N_\text{eff}^3 p_\text{eff}^2 \bar{\chi}^2 t^2 \\
     \text{cov}(S_y,S_z) &=\frac{1}{2}N_\text{eff}^2 p_\text{eff}\bar{\chi} t,
\end{align}
where we have defined an effective number of atoms $N_\text{eff}$ and an effective polarization $p_\text{eff}$, given by

\begin{equation}
    N_\text{eff}\equiv\sum_i\frac{g_i}{\bar{g}}\sin^2\theta_i,\quad\text{and}\quad p_\text{eff}\equiv \frac{1}{N_\text{eff}}\sum_i\frac{g_i}{\bar{g}}\sin\theta_i.
\end{equation}

For this perturbative expansion to hold, we also need to impose $\sum_{k\neq i,j} \cos\theta_k \frac{g_i g_j t}{\Delta}\ll 1$, namely roughly~$(1-g_i/\bar{g})^2 (N\bar{\chi}t)\ll1$, assuming $|\bar{g}-g_i|\ll \bar{g}$ and $t_\text{R}\approx \pi/(2\bar{g})$. Then the requirement on the homogeneity of the magnetic field is $|\nabla B/B|\propto |g_i-\bar{g}|/\bar{g}\ll \xi^{-1/2}$, which is on the order of $6\%$ for 48dB of squeezing.

The effect is the same as that of a polarization fraction $p_\text{eff}$ of a sample with $N_\text{eff}$ spins. Thus, the critical angle becomes $\theta_\text{c}=-1/(N_\text{eff}p_\text{eff}\bar{\chi} t)$ and the squeezing parameter, accordingly, is $\xi^{-2}=\text{var}(J_z'(\theta_\text{c}))/(N_\text{eff}/4)=(N_\text{eff}p_\text{eff}\bar{\chi}t)^{-2}$.

%%%%%%%%%%%%%%%%%%%%%

\subsection{Time-dependent inhomogeneity}
\label{app:g_brownian}

In a liquid or gaseous system, the spins are free to move, so that they explore different regions of the vacuum magnetic field during the duration of the experiment. 

In this case each spin couples with $g_i=\bar{g}_i+\delta g_i(t)$ to the circuit, where $\bar{g}_i$ is the coupling at the beginning of the experiment, which depends on the location of the spin in the volume, and is thus time-independent, while the correction $\delta g_i(t)\propto \delta y_i(t)$, i.e. the displacement of the spin, along the vacuum mode (we take the solenoid mode to be along the $y$ axis). 

The spins exhibit brownian motion with velocities obeying $\avg{v_i(t)}_\text{cl}=0$ and $\avg{v_i(t)v_i(t')}_\text{cl}=\bar{v}^2\tau_c\delta(t-t')$, where $\tau_c$ is the collision time, $\bar{v}=\sqrt{T/m}$, where $T$ is the temperature of the gas and $m$ the atomic mass, and we use the subscript ``cl'' to denote classical averaging with respect to the kinematic stochastic variables of the spins. The collision time is $\tau_c=\bar{v}\ell$, where $\ell\equiv (n_\text{S}\sigma_\text{sph})^{-1}$, where $\sigma_\text{sph}\approx 1$\AA~is the cross-section for hard-sphere atom-atom collisions, $n_\text{S}$ is the physical sample number density, and $\ell$ is the mean-free path. Taking into account the finite correlation time for the velocities, i.e. $\avg{v(t)v(t')}_\text{cl}=\bar{v}^2 e^{-|t-t'|/\tau_c}$ will not change the conclusions below, since $\tau_c$ is the shortest timescale in the problem.

The spin-cavity coupling is now

\begin{equation}
    H_\text{inh,t} = \sum_i g_i(t)\pare{a\sigma_+^{(i)}e^{i\Delta(t+\int_0^t \mathbf{y}\cdot\mathbf{v}_i(t')\di t')}+a^\dagger\sigma_-^{(i)}e^{-i\Delta (t+\int_0^t \mathbf{y}\cdot\mathbf{v}_i(t')\di t')}},
\end{equation}
where we also included the Doppler shift to the frequency. 

Since we wish to be in the limit where all the effects of inhomogeneities, Doppler broadenings, and stochastic coupling variations are small, it is more instructive to treat each effect separately in perturbation theory and derive tolerances for each.

We start by considering the time-dependence in $g_i(t)$ only from the Brownian motion of the spins, and ignore Doppler shifts.

The displacement of the cavity operator and the perturbative diagonalization has to be carried out from the beginning. Following the procedure of App.~\ref{app:squeezingΗ} we may absorb the average number of quanta in the circuit by displacing the photon operator using $D(\alpha)$ with 

\begin{equation}
\begin{split}
        \alpha &\equiv\sum_i\alpha_i
         = -\sum_i\parea{\frac{i}{2}\int_0^t g_i(t') e^{-i\Delta t'} - \frac{\bar{g}_i}{2\Delta}}
         = \sum_i\parea{\frac{\bar{g}_i e^{-i\Delta t}}{2\Delta} - \frac{i}{2}\int_0^t \delta g_i(t') e^{-i\Delta t'}\di t'}
         \equiv \sum_i \parea{\bar{\alpha}_i+\delta\alpha_i(t)},
\end{split}
\end{equation}
with the obvious definition in the last equality. 

Perturbative diagonalization can be done by choosing

\begin{eqnarray}
    iS = 2\sum_i \parea{\alpha_i^* a \pare{\sigma_+^{(i)}-\frac{1}{2}} - \alpha_i a^\dagger \pare{\sigma_-^{(i)}-\frac{1}{2}}}.
\end{eqnarray}

The effective Hamiltonian then becomes

\begin{equation}
    H_\text{eff} = \sum_{i,j}\text{Re}(\chi_{ij}) \parea{\sigma^{2(ij)} -\frac{1}{2}} -\sum_{i,j}\text{Re}(\chi_{ij})\sigma_z^{(i)}\sigma_z^{(j)} + \sum_i\Omega_x^{(i)}\sigma_x^{(i)}+ \sum_i\Omega_y^{(i)}\sigma_y^{(i)},
\end{equation}
where we defined the following quantities

\begin{align}
    \sigma^{2(ij)}&\equiv \sigma_x^{(i)}\sigma_x^{(j)} + \sigma_y^{(i)}\sigma_y^{(j)} + \sigma_z^{(i)}\sigma_z^{(j)}\\
    \chi_{ij}&\equiv \alpha_i^* g_j e^{-i\Delta t} + \alpha_j g_i e^{i\Delta t}\\
    \Omega_x^{i} &\equiv \sum_j \parea{(g_i\alpha_j-g_j\alpha_i) e^{i\Delta t} + (g_i\alpha_j^*-g_j\alpha_i^*) e^{-i\Delta t}}\\
    \Omega_y^{i} &\equiv i\sum_j \parea{(g_i\alpha_j+g_j\alpha_i) e^{i\Delta t} - (g_i\alpha_j^*+g_j\alpha_i^*) e^{-i\Delta t}}.\label{eq:omegay_noise}
\end{align}

Because $\text{Re}(\chi_{ij})$ is a symmetric matrix, the first term commutes with the Hamiltonian and, so, is a constant of motion. This term is analogous to $J^2$ in the limit of homogeneous coupling. In addition, because $\avg{\sigma^{2(ij)}}=1/2$ for $i\neq j$, this term scales linearly with $N$, and so corresponds to a small global frequency shift $(g\sqrt{N})^2/\Delta \ll \Delta$. Thus, we will ignore it in what follows.

The last two terms correspond to Rabi oscillations, whose stochasticity can lead to excess noise. These terms are zero both in the limit of \textit{constant} $g_i=\bar{g}_i$ and in the adiabatic limit where $\alpha_i e^{i\Delta t}\approx \alpha_i^* e^{-i\Delta t}$. 

All results will eventually depend on the correlation function 

\begin{eqnarray}
    \avg{\delta g_i(t)\delta g_j(t')}_\text{cl} = \bar{g}_i^2 \pare{\frac{\di B/\di y_i}{B_i}}^2 \bar{v}\ell \min\{t,t'\}\delta_{ij},
\end{eqnarray}
where $\di B_i/\di y_i$ is the gradient of the vacuum mode along the direction of spin (taken to be changing only along the solenoid axis for simplicity). This approximation is valid in the limit where the spin doesn't move too far away from its initial position and when the gradients are small. We also assume linear gradients, which is an overestimate, as the magnetic field profile at the center of a coil is a local maximum and, so, the spins near the center will experience slower changes, proportional to the second derivative of the magnetic field. The approximation of small displacements is appropriate for the gaseous and liquid systems that we have, since each spin diffuses a distance $\sqrt{\bar{v}\ell t_\text{exp}}\approx 0.5 - 0.05$cm for a protocol that lasts $t_\text{exp}\sim10$s, and for the range of densities assumed in this work. The magnetic field thus varies little over such distances. We will remove the $y_i$ dependence of the magnetic field gradients in what follows for simplicity, and assume a universal gradient of order $(\di B/\di y_i)/B_i\equiv f$.

First, it is straightforward to see that $\avg{\Omega_x^{i}}=0$ and $\avg{\Omega_y^{i}}=\bar{g}_i^2 f^2\bar{v}\ell\sin^2(\Delta t)/\Delta^2\approx \bar{g}_i^2 f^2\bar{v}\ell/(2\Delta^2)$, after averaging over timescales $\gg \Delta^{-1}$. This is a tiny Rabi frequency, on the order of $10^{-35}$Hz, even if the vacuum magnetic field varies $\mathcal{O}(1)$ over the sample. Therefore we can safely neglect the mean values of these Rabi terms. Nevertheless, these Rabi terms entail collectivity, in the sense that $\avg{\Omega_y^{(i)}\Omega_y^{(j)}}\neq 0$ for $i\neq j$. Therefore, to second order in perturbation theory, these could lead to $N$-enhanced, and thus appreciable, noise. We can therefore derive a master equation by going to second order in perturbation theory. The density matrix will then obey

\begin{equation}
    \begin{split}
    \dot{\rho}&=-\sum_{i,j}\int_0^t\di t'\, \avg{\Omega_y^{(i)}(t)\Omega_y^{(j)}(t')}_\text{cl}\pare{\sigma_y^{(i)}\sigma_y^{(j)}\rho(t')-\sigma_y^{(i)}\rho(t')\sigma_y^{(j)}-\sigma_y^{(j)}\rho(t')\sigma_y^{(i)}+\rho(t')\sigma_y^{(j)}\sigma_y^{(i)}}\\
    &=\sum_{i,j}\Gamma_{ij}(t)\pare{\sigma_y^{(i)}\rho(t)\sigma_y^{(j)}+\sigma_y^{(j)}\rho(t)\sigma_y^{(i)}-\sigma_y^{(i)}\sigma_y^{(j)}\rho(t)-\rho(t)\sigma_y^{(j)}\sigma_y^{(i)}},
    \end{split}
\end{equation}
and there will be a similar term with $y\to x$, for the last term in the effective Hamiltonian. In the second line we used perturbativity, which implies Markovian evolution~\cite{Mandel_Wolf_1995}, by setting $\rho(t')= \rho(t) +  (t'-t)\dot{\rho}(t)+...\approx\rho(t)$. We have also defined

\begin{equation}
    \begin{split}
            \Gamma_{ij}(t)&=\int_0^t\di t'\, \avg{\Omega_y^{(i)}(t)\Omega_y^{(j)}(t')}_\text{cl}\\
            & \approx \frac{2\sin^4(\Delta t/2)}{\Delta^4}f_if_j\bar{v}\ell\sum_{k,l}  \pare{\bar{g}_i\bar{g}_j\bar{g}_k^2\delta_{lk} + \bar{g}_i\bar{g}_l\bar{g}_k^2\delta_{jk}+\bar{g}_k\bar{g}_j\bar{g}_i^2\delta_{li}+\bar{g}_k\bar{g}_l\bar{g}_i^2\delta_{ij}},
            \label{eq:master_timedepg}
    \end{split}
\end{equation}
neglecting four-point correlations of the $\delta g_i$, which are both parametrically suppressed and come with one fewer power of $N$. While generally a cumbersome expression, we can estimate the size of this rate by setting all the $\bar{g}_i=\bar{g}$ and $f_i=f$, and taking $\avg{\avg{\sin^4(\Delta t/2)}}\approx3/8$, to be the mean value of this fast-oscillating term. Then the Lindblad equation~\eqref{eq:master_timedepg} takes the form of superradiant decay, with rate $\Gamma \sim N g^4 f^2\bar{v}\ell/\Delta^4$. 

For a gradient $f=1/$m and the aforementioned parameters for the properties of the gas or liquid, we find that this would place a limit of squeezing at $88$dB, much more than what system decay or thermalization allow for in our protocol. As such, these Rabi terms are insignificant. We note that the fact that the rate $\Gamma$ does not depend on time, after averaging over the very fast $\mathcal{O}(\Delta)$ oscillations, and scales as $\Delta^4$, explicitly shows that an adiabatic approximation would be justified, which immediately sets $\Omega_x^i=\Omega_y^i=0$

Thus, we apply this in what follows, so that $\chi_{ij}\approx g_i g_j$. We are then left with the Hamiltonian $H=-\sum_{i,j}\frac{g_i(t)g_j(t)}{\Delta}\sigma_z^{(i)}\sigma_z^{(j)}$. The observables now are $S_\alpha\equiv \sum_i\frac{g_i(t)}{\bar{g}} s_\alpha^{(i)}$, where $\bar{g}\equiv \sum_i\bar{g}_i^2/(\sum_i \bar{g}_i)$, since the observable tracks the spin-circuit coupling as a function of time. Then $\avg{S_z^2}=\avg{\bar{S}_z^2}+\frac{1}{4\bar{g}^2}\sum_i \avg{\delta g_i^2(t)}_\text{cl}$, where $\bar{S}_\alpha\equiv \sum_i\frac{\bar{g}_i}{\bar{g}} s_\alpha^{(i)}$. Similarly, $\avg{S_y^2}=\avg{\bar{S}_y^2}+\frac{1}{4\bar{g}^2}\sum_i \avg{\delta g_i^2(t)}_\text{cl}$ and $\avg{S_yS_z+S_zS_y}=\avg{\bar{S}_y\bar{S}_z+\bar{S}_z\bar{S}_y}$. It is straightforward to modify Eqs~\eqref{eq:inh_z2},~\eqref{eq:inh_y2}, and~\eqref{eq:inh_yz} for the time-\emph{independent} observables $\bar{S}_\alpha$, to include this time-dependence in the coupling, which simply amounts to changing Eq.~\eqref{eq:sminus_inh} to

\begin{equation}
    s_-^{(i)} = \exp\parea{-\frac{i}{\Delta}\int_0^tg_i(t')\pare{g_i(t')+\sum_j g_j(t') s_z^{(j)}}\di t'}s_-^{(i)}(0),
\end{equation}

Then we find for the time-\emph{dependent} operators $S_\alpha$:

\begin{align}
    \avg{S_z^2} &= \frac{N_\text{eff}}{4}\pare{1+f^2\bar{v}\ell t}\\
    \avg{S_y^2} &= \frac{N_\text{eff}}{4}\pare{1+f^2\bar{v}\ell t} +  \frac{1}{8}\sum_{i\neq j}\frac{\bar{g}_i}{\bar{g}}\frac{\bar{g}_j}{\bar{g}}
    \parea{\prod_{k\neq i,j} \cos\pare{\frac{1}{\Delta}\int_0^t g_k(g_i-g_j)\di t'}
    -\prod_{k\neq i,j} \cos\pare{\frac{1}{\Delta} \int_0^t g_k(g_i+g_j)\di t'}}\\
     &\approx \frac{N_\text{eff}}{4}\pare{1+f^2\bar{v}\ell t} + \frac{1}{4}N_\text{eff}^3 \bar{\chi}^2 t^2 \parea{1+ \mathcal{O}[(\sqrt{N}\bar{\chi}t)^2]}\\
     \avg{S_yS_z+S_zS_y} &= \frac{1}{2}\sum_{i\neq j} \frac{\bar{g}_i}{\bar{g}}\frac{\bar{g}_j}{\bar{g}}\sin\pare{\frac{1}{\Delta}\int_0^t g_i g_j \di t'}\prod_{k\neq i,j}\cos\pare{\frac{1}{\Delta}\int_0^t g_i g_k \di t'}\\
     &=\frac{1}{2}N_\text{eff}^2\chi t\parea{1+ \mathcal{O}[(\sqrt{N}\bar{\chi}t)]},
\end{align}
with $N_\text{eff}=(\sum_i\bar{g}_i)^2/(\sum_i\bar{g}_i^2)$. The time-dependent increase in the variance due to diffusion leads to an excess noise of

\begin{eqnarray}
    \xi^{-2}_\text{excess} = f^2\bar{v}\ell t,
\end{eqnarray}
which would kick in for squeezing of order of 100dB for $f=1/$m. Therefore, time-dependence in the coupling is largely irrelevant.

It is also straightforward to modify the above formalism to take into account Doppler shifts. Assuming now a homogeneous vacuum mode $g_i=\bar{g}$ and that the frequency shift is small, we may define an effective time-dependent coupling

\begin{eqnarray}
    g_i(t)\equiv\bar{g} e^{i\os\int_0^t v_i(t')\di t'}\approx \bar{g}\pare{1+i\os\int_0^t v_i(t')\di t'},
\end{eqnarray}
so the only modification of the formalism above would be to allow the $g_i$ to be complex. Again, all effects will be quantified by

\begin{eqnarray}
    \avg{\delta g_i(t)\delta g_j^*(t')}_\text{cl} = \bar{g}^2 \omega_0^2 \bar{v}\ell \min\{t,t'\}\delta_{ij}.
\end{eqnarray}

Here, effectively, the magnetic field gradient $f$ above is replaced by $\omega_0$. Given the very low frequencies we are working with in NMR systems, $\omega_0=2\pi/\lambda$, with $\lambda
\sim0.1-1$km. So, in fact, these Doppler shifts are even more subdominant than the effects of the Brownian motion in an inhomogeneous vacuum mode, setting a limit of about 105~dB to squeezing. This is related to the fact that NMR systems exhibit Doppler \emph{narrowing} because $\lambda/(2\pi)\gg\ell$, which is a well-known effect~\cite{Dicke_narrowing}.

%%%%%%%%%%%%%%%%%%%%%

\subsection{Coupling fluctuations}
\label{app:morenoise}

Coupling fluctuations can affect this protocol on several levels, from limiting squeezing and magnification, to inducing confusion with respect to a signal. They depend on mechanical fluctuations of the coil and the cavity, changes in the quality factor $Q$ of the circuit and frequency instabilities or pick-up of noise from electronics and nearby sources. Coupling fluctuations appear on many timescales and can encompass anything from short-timescale fluctuations to long drifts. 

These are effects arising from the large spin numbers needed to achieve significant sensitivity for new physics searches, and are tied to the large circuit coherent state produced during the initialization of the ECSS. To our knowledge, these are new effects, as previous studies of squeezing in atom-cavity systems are not in a regime to encounter them.

Therefore, here we derive for the first time some of the tolerances required, and we hope that this formalism is a first step towards understanding the rich physics of this new spin-circuit cross-talk at large $N$. As far as the noise details are concerned, our considerations are necessarily incomplete as mitigation will depend crucially on the concrete experimental setup. 

We work in the RWA to make the computations clearer, but this section can be extended to the full non-rotating Hamiltonian, Eq.~\eqref{eq:HnonRWA}. Instead of the Lindbladian approach of App.~\ref{app:decoherence}, here we work in the Heisenberg picture and couple the cavity operator to the environment, $H_\text{env}=\sum_k
\omega_k a_k^\dagger a_k+\sqrt{\kappa/(2L)}(a_k a^\dagger + a_k^\dagger a)$, where $\kappa$ is the cavity linewidth and $a_k$ are the environmental modes that lead to decay of the cavity current (resistors, cables etc). The equation of motion for the $a_k$ is

\begin{equation}
    a_k(t)=a_k(0)e^{-i\omega_k t}+i\sqrt{\frac{\kappa}{L}}\int_0^t a(t')e^{-i\omega_k(t-t')}\di t'.
    \label{eq:ak_sol}
\end{equation}

Substituting this into the equation for the cavity operator and taking the limit $\omega_k t\gg 1$, we get

\begin{equation}
\begin{split}
    \dot{a}&=-i\pare{\omega_c-\frac{i\kappa}{2}}a-igJ_-+f(t),
\end{split}
\end{equation}
where $f(t)\equiv i\sqrt{\frac{\kappa}{L}} \sum_k a_k(0)e^{-i\omega_k t}$ is a zero-mean noise term. Tracking the effects of $f(t)$ from these coupled equations to the adiabatic elimination of the cavity operator and the computation of $\avg{J_z}$ and $\text{var}(J_z)$, gives exactly the results computed through the Lindbladian approach outlined in App.~\ref{app:decoherence}, albeit in a more algebraically complicated way. Here we will largely ignore it as our focus is on the coupling and frequency fluctuations.

Fluctuations here can arise from frequency instabilities of the spin or circuit resonance, $\delta\omega(t)$, or from amplitude variations of the coupling (e.g. mechanical vibrations change the volume of the cavity, directly changing its size). These can be incorporated into an effective complex coupling $g(t)$, with a varying amplitude and frequency, as we show below.

Defining the cavity operator as $\bar{a}= e^{i\oc t+\kappa t/2}a$ leads to the formal solution

\begin{eqnarray}
    \bar{a}=\bar{a}_0-i\int_0^t g(t')e^{i\int_0^{t'}\delta \Delta (t'') \di t''} J_-(t')e^{-i\bar{\Delta} t'}e^{\kappa t'/2} + F(t),
    \label{eq:a_flucts_formal}
\end{eqnarray}
where $\delta\Delta \equiv \delta\omega_0(t)-\delta\omega_c(t)$ and $\bar{\Delta}$ is the central frequency and $F(t)$ the appropriate integral of $f(t)$. As such, we have the effective coupling $g_\text{eff}(t)=g(t)e^{i\int_0^{t}\delta \Delta (t'') \di t''}$. For small coupling fluctuations $g_\text{eff}(t)\approx \bar{g}[1+\epsilon(t)+i\phi(t)]$, where $\epsilon(t)$ is the amplitude fluctuation and $\phi(t)\equiv i\int_0^t\delta\Delta (t')\di t'$ is the phase fluctuation.

Focusing on the contribution from the fluctuating coupling, i.e. taking $\bar{J}_-\sim N/2$, neglecting its variance that gives a small Poissonian contribution of order $(g\sqrt{N}/\Delta)^2$ and neglecting thermal noise from $f(t)$, to leading order in small $\epsilon$ and $\phi$ we find excess variance in the number of circuit quanta

\begin{equation}
\begin{split}
    \text{var}_\text{excs}(a^\dagger a) =& \pare{\frac{\bar{g}N}{2\Delta}}^4 e^{-2\kappa t} \int_0^t \di t'\int_0^t\di \bar{t}'\int_0^t\di \tau'\int_0^t\di \bar{\tau}'e^{-i\Delta (t-t'+\tau'-\bar{\tau}')} e^{\kappa(t+t'+\tau'+\bar{\tau}')/2}\\
    &\times\left[\avg{[\phi(\bar{t}')-\phi(t')][\phi(\tau')-\phi(\bar{\tau}')]} + \avg{[\epsilon(t)+\epsilon(t')][\epsilon(\tau')+\epsilon(\bar{\tau}')]}- i \avg{[\epsilon(t')+\epsilon(\bar{t}')][\phi(\tau')-\phi(\bar{\tau}')]}\right]\\
    &\quad\left.-i\avg{[\epsilon(\tau')+\epsilon(\bar{\tau}')][\phi(t')-\phi(\bar{t}')]}\right].
\end{split}
\end{equation}

We may now assume that the amplitude fluctuations $\epsilon$ and the frequency fluctuations $\delta\Delta$ are zero-mean Gaussian processes with real power spectra $S_g$ and $S_\Delta$. For instance, $S_g$ can receive contributions from volume fluctuations (wall jitter), and $S_\Delta$ from fluctuations of the spin frequency or the LC circuit frequency coming from electronics, thermal effects etc. Because $g$ depends on $\oc$, the phase and amplitude fluctuations may, in principle, be correlated, which is encapsulated by the last two terms of the equations above.

Since a concrete study of these power spectra is beyond the scope of this work, we shall assume for simplicity in what follows that the amplitude and phase fluctuations are uncorrelated, and will derive some general bounds on frequency and coupling stability.

We have

\begin{align}
    \avg{\phi(t_1)\phi(t_2)} =&\int_0^{t_1}\int_0^{t_2}\avg{\delta\Delta(t_1')\delta\Delta(t_2')}\di t_1'\di t_2'=\int_0^{t_1}\int_0^{t_2}\int_{-\infty}^{+\infty}\frac{\di\omega}{2\pi}S_\Delta(\omega)\cos\parea{\omega (t_1'-t_2')}\di t_1'\di t_2'\\
\avg{\epsilon(t_1)\epsilon(t_2)}=&\int_{-\infty}^{+\infty}\frac{\di\omega}{2\pi}S_g(\omega) \cos\parea{\omega(t_1-t_2)}.
\end{align}

We then find

\begin{equation}
\begin{split}
    \text{var}_\text{excs}(a^\dagger a)=4\pare{\frac{\bar{g}N}{2\Delta}}^4\int_{-\infty}^{+\infty} \frac{\di \omega}{2\pi}\frac{\Delta^6\parea{S_\Delta(\omega) + \Delta^2 S_g(\omega)}}{(\Delta^2+\kappa^2/4)^2((\Delta^2-\omega^2)^2 + \kappa^2(\Delta^2 + \omega^2)/2 + \kappa^4/16)},
\end{split}
\end{equation}
assuming all the transients are eliminated, $e^{-\kappa t}\to 0$ and $\Delta\gg \omega,\kappa$ in the coefficient of $S_g(
\omega)$. As long as these power-spectra don't have support near $\omega\sim\Delta$, which would mediate a \emph{resonant} spin-circuit interaction, the excess circuit quanta variance is  $\text{var}_\text{excs}(a^\dagger a)\sim 4\pare{\frac{\bar{g}N}{2\Delta}}^4 (f_\text{rms}^2 + \epsilon_\text{rms}^2)$, where $ f_\text{rms}^2\equiv \int\frac{\di\omega}{2\pi} \frac{S_\Delta(\omega)}{\Delta^2}$ is the rms fractional frequency fluctuation, and $\epsilon_\text{rms}^2\equiv \int\frac{\di\omega}{2\pi}S_g(\omega)$ is the rms coupling size fluctuation. 

Denoting $\tilde{n}^2\equiv \text{var}_\text{excs}(a^\dagger a) $, this excess photon variance backreacts as noise onto the spins, because $\dot{J}_z=-\frac{\di}{\di t}(a^\dagger a)$, from the unitary RWA Hamiltonian. Therefore, it imprints its variance onto the energy variance of the spins. The requirement is then approximately $\tilde{n}^2\lesssim N/\xi^2$, which translates to $f_\text{rms},\epsilon_\text{rms}\lesssim 10^{-9}$ for the requirements of our protocol for new physics searches and $\sim 50$~dB of squeezing. This is the formula used in Sec.~\ref{sec:tolerances} of the main text. Such an amount of frequency and volume stability has been achieved in large volume cavities by DarkSRF~\cite{DarkSRF}.

While this is a quick argument for the approximate size of the induced decoherence on the spins, one should compute $\text{var}(J_z)$ directly, as some subtleties arise due to the adiabatic elimination of the cavity operator. Substituting Eq.~\eqref{eq:a_flucts_formal} into the equation for $\dot{J}_z$ and neglecting the small $\propto a(0)$ terms (we have shifted the cavity field and are perturbing around the mean), yields

\begin{eqnarray}
    \dot{J}_z = -2\parea{g(t)J_+(t)\int_0^t g(t')J_-(t')\cos\parea{\Delta (t-t')+\phi(t')-\phi(t)}e^{\frac{\kappa}{2} (t'-t)}\di t'} + \text{thermal terms},
    \label{eq:jz_fluctuations}
\end{eqnarray}
where we separated the phase and amplitude fluctuations of the coupling $g_\text{eff}$ and have neglected a small term $\propto J_z$, arising from the commutator of $J_+$ and $J_-$. We will further take $J_-(t')\approx J_-(t) + (t-t')\dot{\bar{J}}_-(t)$~\cite{Mandel_Wolf_1995}, where the second term is small as long as $\bar{g}\sqrt{N}/\Delta\ll1$.

The ``thermal terms'' arise from the environmental modes $a_k(0)$ and so are uncorrelated with the first term containing the fluctuating coupling. These give the usual increase in the variance of $J_z$ that scales as $\bar{g}^2J^2\kappa/\Delta^2(2\bar{n}+1)$, where $\bar{n}\equiv\avg{a_{\os}^\dagger a_{\os}}$, i.e. the thermal occupation number at frequency $k=\os$. While the mean shifts due to the fluctuations of $g$, this shift is suppressed by $\epsilon_\text{rms}^2,f_\text{rms}^2$ compared to the standard value. So we will neglect this term in what follows, since, as we have discussed in Sec.~\ref{sec:tolerances}, means have to measured and subtracted.

Taking $J_+J_-\approx J^2$, to leading order in the fluctuations we find 

\begin{equation}
\begin{split}
    J_z \approx J_z(0)-2\bar{g}^2J^2&\left[\frac{\Delta^2}{(\Delta^2+\kappa^2/4)^2}\pare{1+\frac{\kappa t}{2}} +\int_0^t\di t'\int_0^{t'}\di t'' [\epsilon(t') + \epsilon(t'')]\cos[\Delta(t'-t'')] e^{-\frac{\kappa}{2} (t'-t'')}\right.\\
    &\quad\left.- \int_0^t\di t'\int_0^{t'}\di t'' [\phi(t'') -\phi(t')]\sin[\Delta(t'-t'')]e^{-\frac{\kappa}{2} (t'-t'')}\right] + \text{ second order}.
\end{split}
\end{equation}
The first term is the mean collective decay due to Dicke superradiance. The ``second order'' terms induce a small change in the mean of $\avg{J_z}$, proportional to $f_\text{rms}^2$ and $\epsilon_\text{rms}^2$, and so we neglect them. They cancel in the variance to second order in the fluctuations. 

Then we find

\begin{equation}
    \begin{split}
            \text{var}(J_z)=&\frac{N}{4} +16\bar{g}^4J^4\int_{-\infty}^{+\infty}\frac{\di\omega}{2\pi}\left[S_\Delta(\omega)\frac{\Delta ^2 \left(\omega  \left(\Delta ^2-\frac{3 \kappa ^2}{4}-\omega ^2\right) \cos \left(\frac{t \omega }{2}\right)+\kappa  \left(\Delta ^2+\frac{\kappa ^2}{4}\right) \sin \left(\frac{t \omega }{2}\right)\right)^2}{\omega ^2 \left(\Delta ^2+\frac{\kappa ^2}{4}\right)^2 \left[(\Delta -\omega)^2+\frac{\kappa ^2}{4}\right]^2 \left[(\Delta +\omega)^2+\frac{\kappa ^2}{4}\right]^2}\right.\\
            &\left.+S_g(\omega)\frac{\left(\omega  \left(\Delta ^2+\frac{\kappa ^2}{4}\right) \left(-\Delta ^2+\frac{\kappa ^2}{4}+\omega ^2\right) \cos \left(\frac{t \omega }{2}\right)-\kappa  \left(\frac{1}{2} \omega ^2 \left(\frac{3 \kappa ^2}{4}-\Delta ^2\right)+\left(\Delta ^2+\frac{\kappa ^2}{4}\right)^2+\frac{\omega ^4}{2}\right) \sin \left(\frac{t \omega }{2}\right)\right)^2}{\omega ^2 \left(\Delta ^2+\frac{\kappa ^2}{4}\right)^2 \left((\Delta -\omega )^2+\frac{\kappa ^2}{4}\right)^2 \left((\Delta +\omega )^2+\frac{\kappa ^2}{4}\right)^2}\right]
    \end{split}
\end{equation}

As in the computation of the excess variance in the circuit quanta number, as long as the power spectra $S_\Delta$ and $S_g$ don't have power near the resonance $\Delta$ (or are very suppressed on that frequency), we may take the large $\Delta$ limit, which yields the simpler result

\begin{equation}
    \text{var}(J_z)= \frac{N}{4} + 16\pare{\frac{\bar{g}J}{\Delta}}^4 \int_{-\infty}^{+\infty}\frac{\di\omega}{2\pi}\pare{\frac{S_\Delta(\omega)}{\Delta^2}+S_g(\omega)}\pare{\cos^2\frac{\omega t}{2} + \frac{\kappa}{2}\frac{\sin\frac{\omega t}{2}}{\omega/2}+\frac{\kappa^2}{4}\frac{\sin^2\frac{\omega t}{2}}{\omega^2/4}}
    \label{eq:excessvarz}
\end{equation}

The requirement is then that the last term is $<N/\xi^2$, as any variance of $J_z$ is imprinted onto a variance along the squeezing axis. Because the contribution from the last term gets an enhancement at low frequencies, compared to the naive previous argument about $\tilde{n}$, the concrete power spectra are needed to set limits on their shape and size. In general, for the protocols discussed in this work, the frequency and coupling strength stability should be on the order of $10^{-9}$ -- $10^{-10}$.

We emphasize that effects discussed in this section set a limit on $N$ directly, rather than $n_\text{S}$ or $n_\text{eff}$, as they arise due to collective effects.

%%%%%%%%%%%%%%%%%%%%%%%%%%
\subsubsection{Mechanical fluctuations}
\label{app:mechfl}

As a concrete example, we can work out the effect of coil volume fluctuations, e.g. due to the walls jittering. Modeling the displacement of a linear dimension of the coil $\delta x(t)$ as a harmonic oscillator of frequency equal to the mechanical resonant frequency of the coil, $\omega_\text{m}$, with a quality factor $Q_\text{m}$, and assuming it is driven by noise $N(t)$ we have

\begin{eqnarray}
    \ddot{\delta x} + \frac{\omega_\text{m}}{Q_\text{m}}\dot{\delta x} + \omega_\text{m}^2\delta x=N(t).
\end{eqnarray}

We will take $N(t)$ to be white noise such that $\avg{N}=0$ and $\avg{N(t)N(t')}=D\delta(t-t')$, where $D$ is a constant related to the rms displacement, as thermal noise or other sources of wall jitter occur on very fast timescales. If the rms displacement is $\delta x_\text{rms}$, then the power spectrum of the coupling amplitude fluctuations is

\begin{eqnarray}
    S_{\delta x}(\omega)=\frac{2\omega_\text{m}^3 \delta x_\text{rms}^2/Q_\text{m}}{(\omega^2-\omega_\text{m}^2)^2 + (\omega\omega_\text{m}/Q_\text{m})^2}.
\end{eqnarray}

DarkSRF~\cite{DarkSRF} has achieved fractional stability of $10^{-10}$ for a $V_\text{c}^{1/3}\approx1$~m cavity, or $\delta x_\text{rms}\approx 0.1$~nm. This stability is worse compared the thermal noise limit, which even at room temperature would be $\approx 10^{-4}$~nm for a $M_\text{c}=1$~kg cavity with the fundamental resonant mode at $\omega_\text{m}=1$~kHz. Because $g\propto V^{-1/2}$, we have that $\delta g/g\approx (3/2)\delta x/V^{1/3}$, so that $S_g(\omega)=(9/4)S_{\delta x}/V^{2/3}$. Moreover, $\omega_\text{LC}\propto L^{-1/2}\propto (A/\ell)^{-1/2}$, so that $S_{\Delta}(\omega)/\Delta^2 = (9/4)S_{\delta x}$ as well, assuming all dimensions have the same $\delta x_\text{rms}$. Because both the phase and the amplitude of $g_\text{eff}$ depend on the same stochastic process, in principle we should have included the cross-correlations of $\epsilon(t)$ and $\phi(t)$ in the computation of $\text{var}(J_z)$, but this should simply amount to an extra factor of $\sim2$ as the spectra are the same.

Equation~\eqref{eq:excessvarz} can then be evaluated analytically, giving a numerically dominant contribution to excess variance of $144 (\bar{g}N/(2\Delta))^4 (\delta x_\text{rms}/V^{1/3})^2$ (including the factor of 2 due to cross-correlations discussed above), given the parameters of the circuit and assuming the same mechanical response as the DarkSRF setup. For a protocol achieving 48~dB of squeezing and the parameters of Table~\ref{tab:parameters}, this does not introduce excess noise as long as $N\lesssim 4\times 10^{26}$, which independently sets our upper limit on $N$ to around $10^{26}$ spins.

%%%%%%%%%%%%%%%%%%%%%%%%%%%
\subsection{Rabi rotations}
\label{app:rabipicture}
\begin{figure}[h!]
    \centering
    \includegraphics[width=0.44\textwidth]{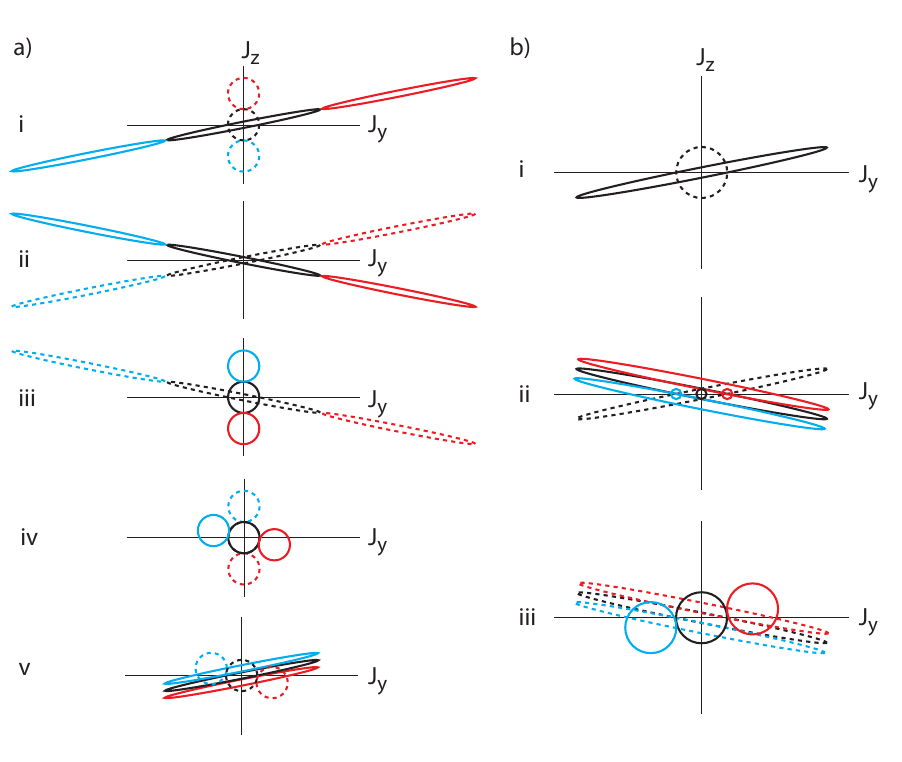}
    \caption{Effect of imperfections in the Rabi rotations on the protocol. In each graph, the dotted lines represent state prior to the action taken in that step, and the solid lines represent the situation afterwards. a) Insensitivity to initial mean $J_z$ values is evidenced by the identical $J_y$ distributions at the end of the protocol. Different colors represent states with different initial mean $J_z$ values. The sequence i-v of the protocol: i, squeeze; ii, rotate, iii, magnify; iv, rotate; v, magnify. b) Sensitivity to RF phase noise is evidenced by the final spread in the $J_y$ direction. The part of the sequence i-iii relevant for introduction of the phase noise into the protocol:  i, squeeze; ii; rotate; iii, magnify. In step ii, the small rotation is carried out around different axes due to RF phase jitter. The axes are marked with small circles of difference colors which correspond with the rest of the evolutions.}
    \label{fig:rabipicture}
\end{figure}

Here, we illustrate in a visual manner on the Bloch sphere the protocol's insensitivity to the initial mean $J_z$ values, as well as its sensitivity to RF phase jitter that accumulates throughout the sequence. The protocol starts with a CSS near the equator, aligned with the reference x-axis, and evolves as illustrated in Fig. \ref{fig:rabipicture} (see the caption for additional information).

\end{document}